\documentclass[11pt, a4paper]{article} 
\usepackage{jheppub}\makeatletter\gdef\@fpheader{~}\makeatother
\usepackage{todonotes}
\usepackage{dynkin-diagrams}
\usepackage{amsmath, amsthm, amssymb, amsbsy, amsfonts, mathrsfs, enumerate, float, wrapfig, mathtools, graphbox, framed, tcolorbox, bbm, microtype, adjustbox} 
\usepackage[section]{placeins}
\allowdisplaybreaks 
\usepackage{booktabs}
\usepackage{multirow}
\usepackage[dvipsnames]{xcolor}  
\usepackage{tikz}\usetikzlibrary{decorations.markings,arrows, calc, arrows.meta}  
\usepackage{orcidlink}\makeatletter\gdef\@OrigHeightRecip{0.005}\makeatother
\hypersetup{linkcolor=blue, urlcolor=blue, citecolor=blue}
\graphicspath{{Figures/}} 

\newcommand{\nn}{\nonumber}  
\newcommand{\Z}{\ensuremath{\mathbb{Z}}} 
\newcommand{\D}{\ensuremath{D}}  
\newcommand{\E}{\ensuremath{E}}    
\newcommand{\tF}{\ensuremath{F}}  
\newcommand{\F}{\textbf{F}}
  
\newcommand{\SO}{\text{SO}}
\newcommand{\U}{\text{U}}
\newcommand{\SU}{\text{SU}}

\title{5d \texorpdfstring{$\mathbb{Z}_2$}{Z2} S-folds and Seiberg–Witten curves}
\author[a]{Sung-Soo Kim,\,\orcidlink{0000-0001-7977-193X}}
\author[a]{Song Liu,\,\orcidlink{0009-0000-9824-7850}}
\author[a]{Mudassar Sabir,\,\orcidlink{0000-0002-8551-2608}}
\author[b]{Futoshi Yagi\,\orcidlink{0000-0002-6217-5151}}

\affiliation[a]{School of Physics, University of Electronic Science and Technology of China, \\
No.~2006 Xiyuan Ave, West Hi-Tech Zone, Chengdu, Sichuan 611731, China}
\affiliation[b]{School of Science, Huzhou Normal University, \\
No. 759, East 2nd Road, Huzhou, Zhejiang 313000, China}

\emailAdd{sungsoo.kim@uestc.edu.cn}
\emailAdd{sliu.susy@std.uestc.edu.cn} 
\emailAdd{mudassar.sabir@uestc.edu.cn} 
\emailAdd{60714@huznu.edu.cn}

\abstract{We compute the Seiberg–Witten curves for $\Z_2$ S-folds of five-dimensional $\mathcal{N}=1$ rank‑1 superconformal field theories. Two complementary constructions are presented. First, we implement the $\Z_2$ action by identifying pairs of points on the original curve, thereby mapping them to a single point on the S-folded curve. Second, motivated by~\cite{Kim:2021fxx}, we tune mass parameters to special values, corresponding to the insertion of $\D_4$ 7‑branes. As a nontrivial consistency check, we show that the two Seiberg–Witten curves agree with each other when expressed in the Weierstrass normal form. 
}

\begin{document}
\maketitle
\section{Introduction}\label{sec:intro}
Five-dimensional (5d) superconformal field theories (SCFTs) with eight supercharges admit powerful descriptions in terms of $(p,q)$ 5-brane webs in Type IIB string theory, which provide an explicit approach to the Coulomb branch physics of these SCFTs~\cite{Aharony:1997ju, Aharony:1997bh}. The areas of the compact faces of 5-brane webs are identified with the monopole string tensions of a given theory, which are the first derivatives of the cubic prepotential $\mathcal{F}$ with respect to the scalar in the vector multiplet~\cite{Aharony:1997bh,Intriligator:1997pq}. The prepotential can therefore be obtained by integrating the areas of the compact faces. The 5d cubic prepotential is 1-loop exact and receives no instanton corrections~\cite{Seiberg:1996bd, Intriligator:1997pq}. This should be distinguished from circle-compactified observables, which receive contributions from instanton particles. In suitable web realizations, the corresponding instanton partition functions~\cite{Nekrasov:2002qd} can be computed using the topological vertex or its refined version, yielding an expansion in the instanton fugacity~\cite{Aganagic:2003db,Iqbal:2007ii}.

For toric configurations, the dual diagrams  of local Calabi--Yau threefolds (CY3) used in M-theory geometric engineering~\cite{Leung:1997tw}. The associated Seiberg--Witten (SW) curve encodes the circle-compactified Coulomb branch geometry, including instanton effects~\cite{Aharony:1997bh,Kol:1997fv}. 
Upon compactifying a common worldvolume direction on a circle, T-dualizing, and lifting to M-theory, the 5-brane web is described by an M5-brane wrapping a holomorphic curve in $(\mathbb{C}^*)^2$, whose characteristic equation provides the Seiberg--Witten curve of the compactified theory~\cite{Witten:1997sc,Kol:1997fv,Brandhuber:1997ua,Aharony:1997bh}.

Introducing 7-branes into a 5-brane web enriches the description of 5d SCFTs~\cite{DeWolfe:1999hj}.
On the $(p,q)$-plane, 7-branes appear as points on which external 5-branes of the same charge are end. Their monodromy branch cuts change the $(p,q)$-charges of crossing 5-branes, and the 7-brane backgrounds also carry deficit angles~\cite{Gaberdiel:1997ud,Gaberdiel:1998mv,DeWolfe:1999hj}. Including 7-branes is, in fact, essential in many web descriptions with a large number of fundamental hypermultiplets or matter in higher-dimensional representations, leading to generalized web configurations that need not admit an ordinary toric description~\cite{DeWolfe:1999hj,Benini:2009gi,Kim:2014nqa}. Moreover, flavor symmetries can be read from the 7-brane configuration and its string-junction lattice. The $ADE$ singularities familiar from F-theory underlie the corresponding $ADE$ algebras~\cite{Vafa:1996xn,Gaberdiel:1997ud,DeWolfe:1999hj}. In particular, the exceptional flavor symmetries of rank-1 theories are made manifest by suitable 7-brane configurations, with an associated affine structure organizing the five-dimensional system~\cite{Gaberdiel:1997ud,Gaberdiel:1998mv,DeWolfe:1998zf,DeWolfe:1998eu,DeWolfe:1998pr,DeWolfe:1999hj}. The corresponding Seiberg--Witten curves can be obtained from generalized toric diagrams and compared with the known $E_n$ theory curves. In Weierstrass normal form, their coefficients can be organized in terms of $E_n$ characters~\cite{Eguchi:2002fc,Kim:2014nqa}.

Some 5-brane web diagrams admit discrete rotational symmetries on the $(p,q)$-plane. For example, 5d $\SU(2)_0$ theory can have 180$^\circ$ and 90$^\circ$ rotations, respectively corresponding to $\Z_2$ and $\Z_4$ discrete symmetries. When accompanied by an $\mathrm{SL}(2,\Z)$ transformation of the 5-brane charges, quotients by these combined discrete operations lead to five-dimensional $\Z_n$ S-fold constructions, where $n=2,3,4,6$~~\cite{Acharya:2021jsp,Kim:2021fxx}. Discrete gaugings and related quotient constructions have been studied from geometric, brane-web, holographic, and Seiberg--Witten perspectives~\cite{Tian:2021cif,Acharya:2021jsp,Kim:2021fxx,Xie:2022lcm,Apruzzi:2022nax,Closset:2023pmc,Furrer:2024zzu}. Related discrete gaugings of rank-1 $E_n$ SCFTs that fix the Coulomb branch coordinate while acting nontrivially on the Higgs branch have also been studied~\cite{Arias-Tamargo:2026rrf}. In particular, a systematic prescription for realizing S-folds on 5-brane webs is presented as the insertion of the proper 7-brane configurations at the fixed point of the $\Z_n$ discrete rotation. The inserted 7-brane configurations are those that have the correct monodromy and deficit angles associated with the S-foldings~\cite{Kim:2021fxx}.

The 5-brane web constructions of the S-folds suggest non-trivial relations between the parent theory $\mathcal{T}$ and the S-folded theory $\mathcal{T}/\Z_n$. In particular, the reduction of compact-face areas leads to a corresponding relation between their cubic prepotentials after identifying the parameters related by the discrete symmetry, $\mathcal{F}_{\mathcal{T}/\Z_n} = \frac{1}{n} \mathcal{F}_{\mathcal{T}}$. These relations provide important checks of the prescription, but they do not determine the complete circle-compactified Seiberg--Witten curve or the complexified mass data associated with the inserted 7-branes. An explicit curve construction can therefore test the prescription beyond the 5d cubic prepotential and clarify how the fixed-point data enter the compactified geometry.

In this paper, as the first explicit example, we consider the $\Z_2$ S-folds of rank-1 theories and compute their Seiberg–Witten curves. Their Laurent polynomial presentations can be obtained from toric or generalized toric diagrams. These presentations provide a natural setting in which to implement $\Z_2$ discrete actions directly on the curve coordinates. We also provide a different method for computing the Seiberg–Witten curve to account for the 7-brane insertion, where the inserted 7-branes are located at the fixed point of the $\Z_2$ symmetry, corresponding to imposing special holonomies for the masses of the inserted 7-branes. For all four examples of the $\Z_2$ S-folds of rank-1 theories, we demonstrate agreement of the two constructions in Weierstrass normal form after specifying the required coordinate and parameter transformations. The resulting coefficients organize into characters of $\mathfrak{su}(2)$, $\mathfrak{su}(3)$, $\mathfrak{so}(7)$, and $\mathfrak{f}_4$, respectively. Our result is thus an explicit quotient–tuning correspondence for these rank-1 theories, providing a Seiberg–Witten curve level test of the proposed $\Z_2$ S-fold prescription.

The paper is organized as follows. In section~\ref{sec:s-folds}, we review rank-1 SCFTs based on the Type IIB 5-brane web diagram and how S-folds are realized from discrete symmetry of 5-brane webs.  In section~\ref{sec:SWcurvesforS-folds}, we compute the Seiberg–Witten curves for the $\Z_2$ S-folds in two different methods. We then summarize the results and discuss possible physical interpretations and generalizations in section~\ref{sec:conclusion}. Appendix \ref{app:WNF} provides the transformations to the Weierstrass normal form for the Seiberg–Witten curve obtained from (non-)toric diagrams. In Appendix \ref{sec:phy-mass-para}, we develop the physical mass parameterization and its implementation in the curves obtained from a 5-brane web. In Appendix~\ref{app:Characters}, we collect the characters for flavor symmetries and their identities used in the comparison.

\section{\texorpdfstring{$\Z_2$}{Z2} S-folding of 5d rank-1 SCFTs on 5-brane webs}\label{sec:s-folds}
We review the salient features of Type IIB 5-brane constructions for 5d $\mathcal{N}=1$ superconformal theories of rank-1 as well as the 5-brane realization of $\Z_2$ S-folding of such theories. 

\subsection{5-brane construction of 5d rank-1 SCFTs}\label{sec:5-branes}
\begin{figure}
    \begin{adjustbox}{max width=0.93\linewidth, center}
    \begin{tikzpicture}[
    thick line/.style={line width=1.25pt},
    dashed line/.style={line width=1.25pt, dashed},
    blue dot/.style={circle, fill=blue, inner sep=1.8pt},
    red dot/.style={circle, fill=red, inner sep=1.8pt},
    axis arrow/.style={line width=1.25pt, -{Stealth[length=3mm, width=2.5mm]}}
]

    \begin{scope}[shift={(0,0)}]
        \draw[thick line] (-1.2, -0.8) rectangle (1.2, 0.8);
        
        \draw[thick line] (-1.2, 0.8) -- (-2.2, 1.8);
        \draw[thick line] (1.2, 0.8) -- (2.2, 1.8);
        \draw[thick line] (-1.2, -0.8) -- (-2.2, -1.8);
        \draw[thick line] (1.2, -0.8) -- (2.2, -1.8);
        
        \draw[axis arrow] (-2.8, -2.2) -- (-1.4, -2.2) node[right, font=\boldmath] {$x^6$};
        \draw[axis arrow] (-2.6, -2.4) -- (-2.6, -1.0) node[above, font=\boldmath] {$x^5$};

        \node at (0, -2.8) {\large (a)};
    \end{scope}

    \begin{scope}[shift={(6.5,0)}]
        \draw[thick line] (-1.2, -0.8) rectangle (1.2, 0.8);
        
        \draw[thick line] (-1.2, 0.8)  -- (-1.9, 1.5)  node[blue dot] (tl) {};
        \draw[thick line] (1.2, 0.8)   -- (1.9, 1.5)   node[red dot] (tr) {};
        \draw[thick line] (-1.2, -0.8) -- (-1.9, -1.5) node[red dot] (bl) {};
        \draw[thick line] (1.2, -0.8)  -- (1.9, -1.5)  node[blue dot] (br) {};
        
        \draw[dashed line] (tl) -- (-2.5, 2.1);
        \draw[dashed line] (tr) -- (2.5, 2.1);
        \draw[dashed line] (bl) -- (-2.5, -2.1);
        \draw[dashed line] (br) -- (2.5, -2.1);
        
        \node[below left, font=\bfseries, xshift=2pt]  at (tl) {$[1,-1]$};
        \node[above left, font=\bfseries, xshift=2pt]  at (bl) {$[1,1]$};
        \node[below right, font=\bfseries, xshift=-2pt] at (tr) {$[1,1]$};
        \node[above right, font=\bfseries, xshift=-2pt] at (br) {$[1,-1]$};
        
        \node at (0, -2.8) {\large (b)};
    \end{scope}

    \begin{scope}[shift={(13,0)}]
        \draw[thick line] (-1.2, -0.8) rectangle (1.2, 0.8);
        
        \draw[dashed line] (-1.2, 0.8)  -- (-2.0, 1.6);
        \draw[dashed line] (1.2, 0.8)   -- (2.0, 1.6);
        \draw[dashed line] (-1.2, -0.8) -- (-2.0, -1.6);
        \draw[dashed line] (1.2, -0.8)  -- (2.0, -1.6);
        
        \draw[dashed line] (-1.2, 0.8)  -- (-0.5, 0.1)  node[blue dot] (itl) {};
        \draw[dashed line] (-1.2, -0.8) -- (-0.5, -0.1) node[red dot] (ibl) {};
        \draw[dashed line] (1.2, 0.8)   -- (0.5, 0.1)   node[red dot] (itr) {};
        \draw[dashed line] (1.2, -0.8)  -- (0.5, -0.1)  node[blue dot] (ibr) {};
        
        \node[above=2pt, font=\bfseries\small] at (itl) {$\mathbf{B}$};
        \node[below=2pt, font=\bfseries\small] at (ibl) {$\mathbf{C}$};
        \node[above=2pt, font=\bfseries\small] at (itr) {$\mathbf{C}$};
        \node[below=2pt, font=\bfseries\small] at (ibr) {$\mathbf{B}$};
        
        \node at (0, -2.8) {\large (c)};
    \end{scope}
\end{tikzpicture}
\end{adjustbox}
     \caption{5-brane webs for the SU(2)$_0$ theory. (a) A 5-brane web without 7-branes attached. (b) A 5-brane web where 7-branes are attached to external 5-branes. (c) A 5-brane configuration where the external 7-branes are taken to the Coulomb branch by the Hanany-Witten moves.}
     \label{fig:su2web}
 \end{figure}
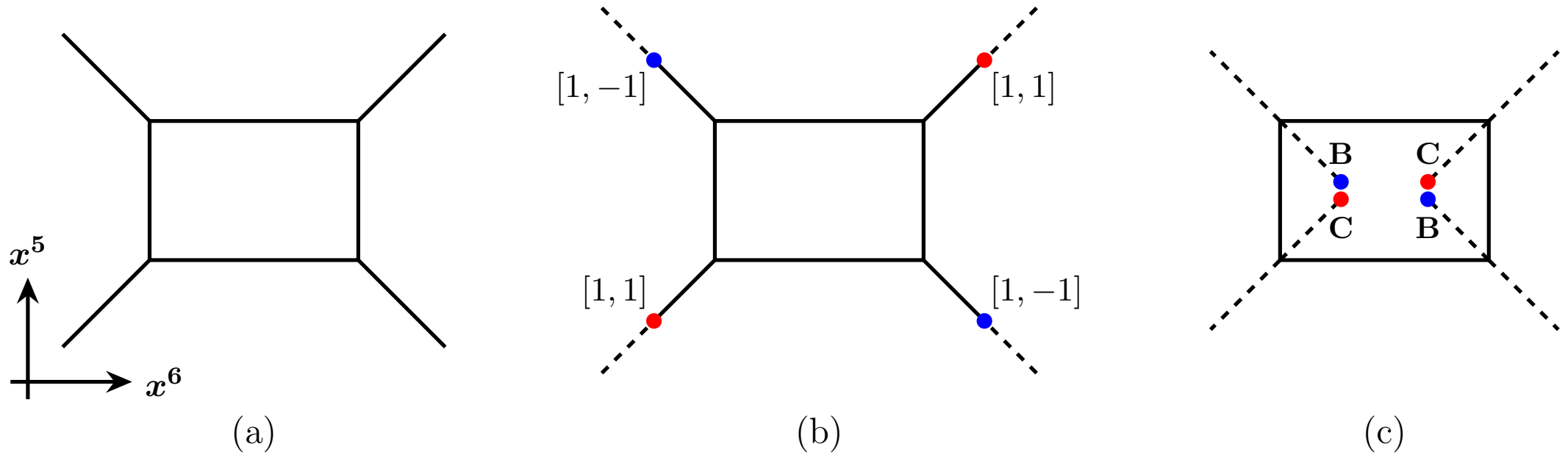

The 5-brane web in Type IIB string theory is a powerful tool that can engineer 5d $\mathcal{N}=1$ supersymmetric gauge theories based on intersecting $(p,q)$ 5-branes,  where $p,q$ are co-prime. These 5-branes form a charge conserving configuration, making a 5-brane web diagram~\cite{Aharony:1997bh}. 
For example, a representative 5-brane web diagram for the 5d $\mathcal{N}=1$ SU(2) theory is shown in figure~\ref{fig:su2web}(a), where the $(p,q)$ 5-brane web is depicted in the $x^5$–$x^6$ plane, called the $(p,q)$-plane. We use the convention that a D5-brane is denoted by a $(1,0)$ 5-brane, while an NS5-brane is represented by a $(0,1)$ 5-brane. Their brane configurations are summarized in Table~\ref{tab:brane-config}.

\begin{table}[htbp]
\centering
\begin{tabular}{c|ccccc|cc|ccc}
\hline
5-/7-brane & 0 & 1 & 2 & 3 & 4 & 5 & 6 & 7 & 8 & 9 \\
\hline
NS5~(0,1)  & $\times$ & $\times$ & $\times$ & $\times$ & $\times$ & $\times$ &  &  &  &  \\
D5~(1,0)   & $\times$ & $\times$ & $\times$ & $\times$ & $\times$ &  & $\times$ &  &  &  \\
$(p,q)$    & $\times$ & $\times$ & $\times$ & $\times$ & $\times$ & $\theta$ & $\theta$ &  &  &  \\
7-brane    & $\times$ & $\times$ & $\times$ & $\times$ & $\times$ &  &  & $\times$ & $\times$ & $\times$ \\
\hline
\end{tabular}
\caption{Type IIB brane configuration for 5d $\mathcal{N}=1$ supersymmetric gauge theories. The symbol `$\times$' indicates the directions that the brane extends. A $(p,q)$ 5-brane is represented by a line making an angle $\theta = \tan^{-1}(q/p)$.}
\label{tab:brane-config}
\end{table}

One can introduce 7-branes such that the external 5-branes terminate on 7-branes of the same charges. As depicted in figure~\ref{fig:su2web}(b), 7-branes (including O7-planes), which are extended along $x^{7,8,9}$-directions, are realized as a dot in the ($p,q$)-plane with the monodromy cut associated with their charge, which is depicted as a dashed line. When an $[r,s]$ 7-brane crosses the monodromy cut of a $[p, q]$ 7-brane in the counterclockwise direction, its charge is transformed to $[r', s']$ according to the monodromy matrix $K_{[p,q]}$ associated with the $[p, q]$ 7-brane\footnote{We use $[p,q]$ to denote the charge of a 7-brane and $(p,q)$ to denote the charge of a 5-brane.},
\begin{align}\label{eq:monodromyMatrix}
\Big[\begin{smallmatrix} r' \\ s' \end{smallmatrix}\Big]
=K_{[p,q]}\Big[\begin{smallmatrix} r \\ s \end{smallmatrix}\Big]\ , 
\quad \mathrm{where}\quad 
K_{[p,q]} =\begin{pmatrix}
    1+pq & -p^2\\
     q^2 & 1- p q
    \end{pmatrix}\ . 
\end{align}
If an $[r,s]$ 7-brane crosses the monodromy cut of a $[p, q]$ 7-brane in the clockwise direction, the inverse of the monodromy matrix $K_{[p,q]}^{-1}$ is used .

\begin{figure}[htbp]
    \begin{adjustbox}{max width=0.93\linewidth, center}
    \begin{tikzpicture}[
        x=0.75pt,
        y=0.75pt,
        yscale=-1,
        xscale=1,
        line/.style={line width=1.3pt, line cap=round, line join=round},
        dot/.style={circle, fill=black, inner sep=1.7pt},
        rdot/.style={circle, fill=red, draw=none, inner sep=2.1pt},
        lab/.style={font=\large, inner sep=1pt},
        redlab/.style={font=\large, red, inner sep=1pt}
    ]

    \coordinate (A1) at (71,122);
    \coordinate (A2) at (141,122);
    \coordinate (A3) at (141,162);
    \coordinate (A4) at (71,162);

    \coordinate (Btop) at (39,91);
    \coordinate (Ctop) at (171,92);
    \coordinate (Cbot) at (40,191);
    \coordinate (Bbot) at (174,193);

    \draw[line] (A1)--(A2)--(A3)--(A4)--cycle;
    \draw[line] (A1)--(Btop);
    \draw[line] (A2)--(Ctop);
    \draw[line] (A4)--(Cbot);
    \draw[line] (A3)--(Bbot);

    \node[dot] at (Btop) {};
    \node[dot] at (Ctop) {};
    \node[dot] at (Cbot) {};
    \node[dot] at (Bbot) {};

  \node[lab, above left=2pt] at (Btop) {$\mathbf{B}$};
\node[lab, above right=2pt] at (Ctop) {$\mathbf{C}$};
\node[lab, below left=2pt] at (Cbot) {$\mathbf{C}$};
\node[lab, below right=2pt] at (Bbot) {$\mathbf{B}$};

    \node[lab] at (106,240) {(a)};

    \node[font=\Large] at (201,142) {$\Longleftrightarrow$};

    \coordinate (M1) at (303,122);
    \coordinate (M2) at (373,122);
    \coordinate (M3) at (334.35,162);
    \coordinate (M4) at (264.35,162);

    \coordinate (Ntop) at (303,89);
    \coordinate (Nbot) at (334.35,194);
    \coordinate (Xleft) at (223.35,181.96);
    \coordinate (Xright) at (414,102.04);

    \draw[line] (M1)--(M2)--(M3)--(M4)--cycle;
    \draw[line] (M1)--(Ntop);
    \draw[line] (M3)--(Nbot);
    \draw[line] (M4)--(Xleft);
    \draw[line] (M2)--(Xright);

    \node[rdot] at (Ntop) {};
    \node[dot] at (Nbot) {};
    \node[dot] at (Xleft) {};
    \node[dot] at (Xright) {};

    \draw[line, red, ->, >=stealth] (313,88) -- (313,116);

    \node[lab] at (295,75) {$\mathbf{X}_{[0,1]}$};
    \node[redlab] at (330,98) {\small \sf HW};
    \node[lab] at (344,208) {$\mathbf{X}_{[0,1]}$};
    \node[lab] at (221,195) {$\mathbf{X}_{[2,1]}$};
    \node[lab] at (402,86) {$\mathbf{X}_{[2,1]}$};

    \node[lab] at (319,240) {(b)};

    \node[font=\Large] at (427,142) {$\Longleftrightarrow$};
    \coordinate (R1) at (492,121);
    \coordinate (R2) at (592,121);
    \coordinate (R3) at (551,162.2);
    \coordinate (R4) at (531.83,162.11);

    \coordinate (RXleft) at (451,100);
    \coordinate (RXright) at (629,101);
    \coordinate (RNleft) at (532,190);
    \coordinate (RNright) at (551,190.2);

    \draw[line] (R1)--(R2);
    \draw[line] (R1)--(R4);
    \draw[line] (R2)--(R3);
    \draw[line] (R4)--(R3);
    \draw[line] (R4)--(RNleft);
    \draw[line] (R3)--(RNright);
    \draw[line] (R1)--(RXleft);
    \draw[line] (R2)--(RXright);

    \node[dot] at (RXleft) {};
    \node[dot] at (RXright) {};
    \node[rdot] at (RNleft) {};
    \node[dot] at (RNright) {};

    \node[lab] at (474,84) {$\mathbf{X}_{[2,-1]}$};
    \node[lab] at (617,84) {$\mathbf{X}_{[2,1]}$};
    \node[lab] at (526,205) {$\mathbf{X}_{[0,1]}$};
    \node[lab] at (565,205) {$\mathbf{X}_{[0,1]}$};
    
    \node[lab] at (542,240) {(c)};
    \end{tikzpicture}
   \end{adjustbox}
 \caption{Three equivalent 5-brane webs for the SU(2)$_0$ theory. An SL(2,\Z) T-transformation on (a) yields (b). A Hanany-Witten ({\color{red}\sf HW}) move on the (red dot) 7-brane $\mathbf{X}_{[0,1]}$ on (b) leads to (c). }
    \label{fig:su2HW}   
\end{figure}
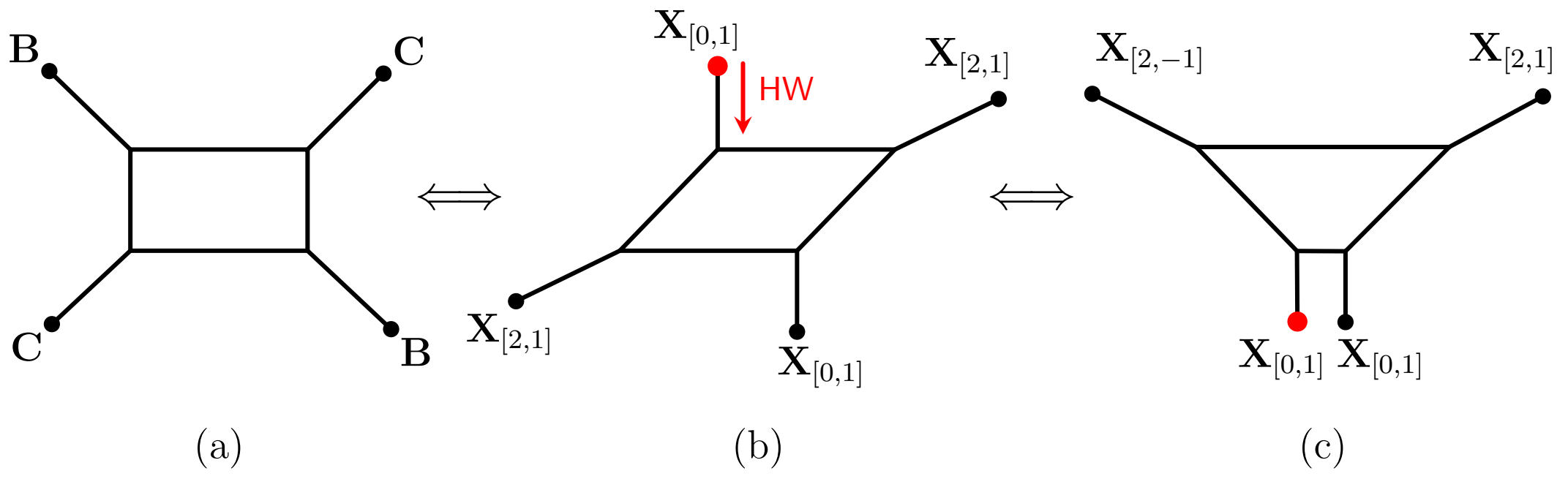

The presence of 7-branes on 5-brane webs enriches the Type IIB construction of five-dimensional supersymmetric gauge theories. One notable example is the enhancement of global symmetry in the SU(2) gauge theory with $N_f\le 7$ hypermultiplets in the fundamental representation (SU$(2)+N_f\F)$, which provides an explicit realization of $\E_{N_f+1}$ flavor symmetry at the fixed point~\cite{Seiberg:1996bd}. In recognition of its flavor symmetry at the fixed point, the superconformal theory for the SU$(2)+N_f\F$ theory is often referred to as the rank-1 $\E_{N_f+1}$ theory. Reading off the flavor symmetry of such rank-1 $\E_{N_f+1}$ theories based on 5-brane webs with 7-branes is as follows: First, one brings all the external 7-branes inside the Coulomb branch, which corresponds to the compact face of the web diagram. It then follows from the F-theory that 7-branes collapse to an $ADE$ Kodaira singularity, giving rise to the global symmetry of a 5d SCFT~\cite{Gaberdiel:1997ud, Gaberdiel:1998mv, DeWolfe:1998zf, DeWolfe:1998eu, DeWolfe:1998pr, DeWolfe:1999hj}. For convenience, we denote frequently used 7-branes as
\begin{align}
\mathbf{A}=[1,0],\qquad  \mathbf{B}=[1,-1],\qquad \mathbf{C}=[1,1]\ . 
\end{align}
The corresponding $ADE$-type global symmetries are given as
\begin{align}\label{eq:7-braneconfi}
    \mathrm{SU}(N):& \quad \mathbf{A}^N \cr
    \mathrm{SO}(2N):& \quad \mathbf{A}^N\mathbf{B}\mathbf{C} \qquad (N\ge4) \cr
    \E_{N}:&\quad \mathbf{A}^{N-1}\mathbf{B}\mathbf{C}\mathbf{B} \quad  ( N\ge 6)\ , 
\end{align}
up to SL(2,\Z) transformations\footnote{See \cite{Kim:2021fxx} for equivalent 7-brane configurations up to SL(2,\Z) transformations.}. It follows from \eqref{eq:monodromyMatrix} that the 7-brane configuration shown in figure~\ref{fig:su2web}(c) leads to
\begin{align}\label{eq:7-braneMani}
    \mathbf{BCBC}\quad \overset{\text{SL}(2,\Z)}{\Longleftrightarrow}\quad\mathbf{X}_{[0,1]} \mathbf{X}_{[2,1]}\mathbf{X}_{[0,1]}\mathbf{X}_{[2,1]} \quad\overset{\text{Eq.}\eqref{eq:monodromyMatrix}}{\Longleftrightarrow} \quad\mathbf{X}_{[2,-1]}\mathbf{X}_{[0,1]}^2\mathbf{X}_{[2,1]}\ , 
\end{align}
where the first equivalence is due to an SL(2,\Z) T-transformation, and the second equivalence is due to $[\begin{smallmatrix} 2 \\ -1 \end{smallmatrix}]
=K_{[0,1]}^{-1}[\begin{smallmatrix} 2 \\ 1 \end{smallmatrix}]$. One can easily see that $\mathbf{X}_{[0,1]}^2$ is transformed to $\mathbf{AA}$ by an SL(2,\Z) S-transformation, yielding an SU($2$) flavor symmetry\footnote{The true flavor symmetry is $\SU(2)/\Z_2=\SO(3)$~\cite{BenettiGenolini:2020doj, Apruzzi:2021vcu}.}. We note that the manipulation of the 7-brane configurations in \eqref{eq:7-braneMani} can also be understood as the Hanany-Witten moves of 5-brane webs, as shown in figure~\ref{fig:su2HW}.

\begin{figure}[htbp]
    \centering
    \begin{tikzpicture}[
        x=0.75pt,
        y=0.75pt,
        yscale=-1,
        xscale=1,
        line/.style={line width=1.2pt, line cap=round, line join=round},
        dot/.style={circle, fill=black, draw=none, inner sep=1.7pt},
        eqarrow/.style={font=\Large}
    ]

    \begin{scope}[shift={(105,90)}, scale=0.85, shift={(-105,-90)}]
        \coordinate (A1) at (71,71);
        \coordinate (A2) at (141,71);
        \coordinate (A3) at (141,111);
        \coordinate (A4) at (71,111);

        \coordinate (L1) at (41,40);
        \coordinate (L2) at (169,41);
        \coordinate (L3) at (169,141);
        \coordinate (L4) at (41,140);

        \draw[line] (A1)--(A2)--(A3)--(A4)--cycle;
        \draw[line] (L1)--(A1);
        \draw[line] (A2)--(L2);
        \draw[line] (A3)--(L3);
        \draw[line] (L4)--(A4);
    \end{scope}

    \node[eqarrow] at (176,91.5) {$\leftrightarrow$};
    \node[font=\large] at (174,170) {(a)};

    \begin{scope}[shift={(237,90.7)}, scale=1.32, shift={(-237,-90.7)}]
        \coordinate (B1) at (236.69,71.00);
        \coordinate (B2) at (257.00,90.69);
        \coordinate (B3) at (237.31,111.00);
        \coordinate (B4) at (217.00,91.31);

        \coordinate (Bleft)   at (217.57,91.31);
        \coordinate (Bright)  at (256.43,90.69);
        \coordinate (Btop)    at (236.69,71.85);
        \coordinate (Bbottom) at (237.59,110.43);
        \coordinate (Bcenter) at (237.00,90.72);

        \draw[line] (B1)--(B2)--(B3)--(B4)--cycle;
        \draw[line] (Bleft)--(Bright);
        \draw[line] (B1)--(B3);

        \node[dot] at (Bleft) {};
        \node[dot] at (Bright) {};
        \node[dot] at (Btop) {};
        \node[dot] at (Bbottom) {};
        \node[dot] at (Bcenter) {};
    \end{scope}

    \begin{scope}[xshift=-15pt]

        \begin{scope}[shift={(426,90)}, scale=0.85, shift={(-426,-90)}]
            \coordinate (C1) at (387,71);
            \coordinate (C2) at (426,71);
            \coordinate (C3) at (465,111);
            \coordinate (C4) at (387,111);

            \coordinate (Cup)    at (426,50);
            \coordinate (Cleft)  at (365,50);
            \coordinate (Cdown)  at (366,131);
            \coordinate (Cright) at (506,131);

            \draw[line] (C1)--(C4);
            \draw[line] (C1)--(C2);
            \draw[line] (C2)--(C3);
            \draw[line] (C4)--(C3);
            \draw[line] (Cup)--(C2);
            \draw[line] (Cleft)--(C1);
            \draw[line] (C4)--(Cdown);
            \draw[line] (C3)--(Cright);
        \end{scope}

        \node[eqarrow] at (500,90.5) {$\leftrightarrow$};
        \node[font=\large] at (502,170) {(b)};

        \begin{scope}[shift={(565,90)}, scale=1.32, shift={(-565,-90)}]
            \coordinate (Dtop)       at (565.00,70.00);
            \coordinate (Dbottom)    at (565.00,110.00);
            \coordinate (Dleft)      at (545.00,90.00);
            \coordinate (Dmid)       at (565.00,90.00);
            \coordinate (DdiagBot)   at (564.33,109.33);
            \coordinate (DrightTopA) at (584.33,70.00);
            \coordinate (DrightTopB) at (585.00,70.67);
            \coordinate (DrightTopC) at (585.00,70.00);

            \draw[line] (Dtop)--(Dbottom);
            \draw[line] (Dleft)--(DdiagBot);
            \draw[line] (Dleft)--(Dmid);
            \draw[line] (Dleft)--(Dtop);
            \draw[line] (Dtop)--(DrightTopA);
            \draw[line] (Dbottom)--(DrightTopB);
            \draw[line] (Dmid)--(DrightTopC);

            \node[dot] at (Dmid) {};
            \node[dot] at (Dleft) {};
            \node[dot] at (Dtop) {};
            \node[dot] at (Dbottom) {};
            \node[dot] at (DrightTopA) {};
        \end{scope}

    \end{scope}

    \end{tikzpicture}
    \caption{Toric diagrams as the dual diagram of 5-brane webs for (a) the SU(2)$_0$ theory and (b) the SU(2)${}_\pi$ theory. The vertices are denoted by fat dots, which are associated with monomials for the Seiberg–Witten curves.}
    \label{fig:toric}
\end{figure}
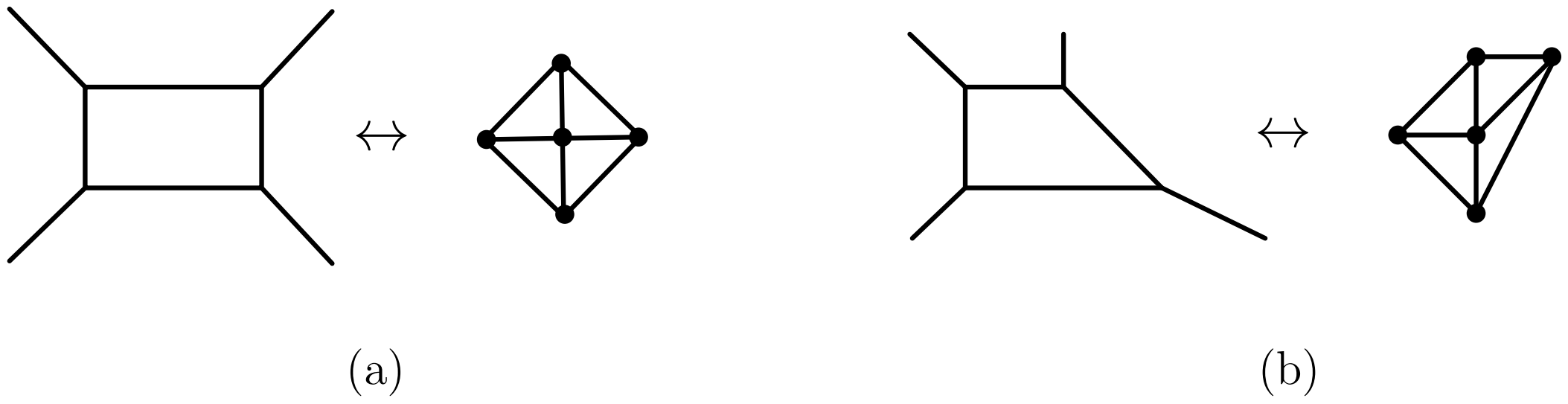

\paragraph{Seiberg–Witten curves.} 
Given a 5-brane web diagram, one can consider a dual diagram, known as a toric diagram for the associated CY3 \cite{Leung:1997tw}, as depicted in figure~\ref{fig:toric}. This dual diagram serves a dual representation of M5-brane configurations wrapping a Seiberg–Witten curve in a CY3. Through this duality, the Seiberg–Witten curve can be obtained by the following procedure. First, given a Type IIB 5-brane web, one takes the circle compactification along $x^4$, followed by the T-duality along this circle $x^4=x^4+2\pi R_A$ and then uplifts it to the M-theory along the M-circle $x^{11}=x^{11}+2\pi R_M$. The curve is embedded in $\mathbb{R}^2 \times \mathbb{T}^2$ where $\mathbb{R}^2$ is along the 5-brane $(x^5, x^6)$-directions and the torus $\mathbb{T}^2$ is along the $(x^{4},x^{11})$-directions. The coordinates on $\mathbb{R}^2 \times \mathbb{T}^2$ are given by two complex coordinates $t$ and $w$:
\begin{align}\label{eq:tw-coord}
    t=e^{-\frac{1}{R_{M}}(x^6+ix^{11})},\qquad w=e^{-\frac{1}{R_A}(x^5+ix^4)}\ .
\end{align}
Secondly, one constructs the Newton polygon corresponding to the dual toric diagram, which takes the form
\begin{align}
    F(t, w) = \sum_{m,n} c_{mn}t^m w^n = 0\ , 
\end{align}
where $c_{mn}$ are the non-vanishing coefficients associated with the $(m,n)$ coordinate in the toric diagram. The Seiberg–Witten curve is then obtained in a straightforward manner by determining the values of $c_{mn}$ in terms of physical parameters of the theory, the Coulomb branch parameters and the mass parameters. To this end, one associates the external $(p,q)$ 5-brane charges to asymptotic behaviors of the curve, and relates the boundary conditions to mass parameters including the instanton factor, and finally identifies the internal vertices to the Coulomb branch parameters~\cite{Kim:2014nqa}. The resulting Seiberg–Witten curve is a holomorphic curve on $\mathbb{R}^2\times \mathbb{T}^2\simeq(\mathbb{C}^*)^2$ that an M5-brane is wrapping.  
The Seiberg–Witten differential is
\begin{align}\label{eq:SWdifferential}
    \lambda_\mathrm{SW}= \mathrm{d}(\log w)\log t \ . 
\end{align}

As a representative example, consider toric diagrams for the SU(2)$_\theta$ theories, depicted in figure~\ref{fig:toric}. The corresponding Seiberg–Witten curves for different discrete $\theta$-angles are given as follows:
\begin{align}
    &\mathrm{SU}(2)_0~:~  \frac{1}{w}+w+U+\sqrt{q}\Big(\frac{1}{t}+t\Big)=0\ ,\label{eq:swcurve-su0} \\
    &\mathrm{SU}(2)_\pi~:~ \frac{1}{w}+w+U+\sqrt{q}\Big(\frac{1}{t}+tw\Big)=0\ .\label{eq:swcurve-su0pi}
\end{align}
Here, the powers of monomials are chosen with the help of rescaling degrees of freedom such that the Coulomb branch parameter $U$ is identified as the coefficient of the monomial $t^0w^0$, which corresponds to the inner vertex of each toric diagram for rank-1 theories. The instanton factor is chosen as $q=e^{-m_0}$, where the instanton mass~$m_0$ is the inverse coupling~$1/g_0^2$, which is diagrammatically identified with the separation between two pairs of the 7-branes $\mathbf{BC}$ as shown in figure~\ref{fig:su2web}(c), measured from the left pair to the right pair along the center of the Coulomb branch. The infinite coupling limit $g_0\to\infty$ then corresponds to the zero distance between two pairs $\mathbf{BC}$ in figure~\ref{fig:su2web} ($m_0\to0$, or equivalently $q\to1$).

For a rank-1 5d $\mathcal{N}=1$ theory compactified on a circle, the Seiberg–Witten geometry is an elliptic fibration over the one complex dimensional Coulomb branch, locally parameterized by $u$. At a generic point $u$, the Seiberg–Witten curve is a smooth elliptic curve. The fibration admits a Weierstrass
description,
\begin{align}
    y^2 = 4 x^3 - g_2(u, M_i) x - g_3 (u,M_i)\ ,
\end{align}
where $g_2$ and $g_3$ are polynomials in $u$ of degrees at most four and six, respectively, with coefficients depending on the mass parameters $m_i$, revealing the flavor symmetry of theory.
For example, the Seiberg–Witten curves for the two SU(2) theories, \eqref{eq:swcurve-su0} and \eqref{eq:swcurve-su0pi}, can be expressed as the Weierstrass normal form as:
\begin{align}\label{eq:Weier-su0}
    &\mathrm{SU}(2)_0~:~  &&g^{\E_1}_2 = \frac1{12} u^4 -\frac23 \chi_1^{\E_1} u^2 +\frac43 \big(\chi_1^{\E_1}\big)^2 -4\ , \nonumber\\
    &&&g^{\E_1}_3= \frac1{216}u^6 -\frac1{18} \chi_1^{\E_1} u^4 + \Big(\frac29\big(\chi_1^{\E_1}\big)^2 -\frac13\Big)u^2 -\frac8{27}\big(\chi_1^{\E_1}\big)^3 +\frac43 \chi_1^{\E_1}\ ,\\
    &\mathrm{SU}(2)_\pi~:~ &&g^{\widetilde{E}_1}_2 = \frac1{12} u^4 -\frac23  u^2 -2\chi_2^{\widetilde{E}_1} u+\frac43 \ , \nonumber\\
    &&&g^{\widetilde{E}_1}_3=  \frac1{216}u^6 -\frac1{18}  u^4 -\frac16 \chi_2^{\widetilde{E}_1} u^3 +\frac29u^2 +\frac23 \chi_2^{\widetilde{E}_1} u + (\chi^{\widetilde{E}_1}_2)^2-\frac8{27},
\end{align}
where we rescaled $U$ as $U=q^{1/4}u$ in the $\SU(2)_0$ theory, and
\begin{align}\label{eq:su2-char}
\chi_1^{\E_1}=\sqrt{q}+\frac1{\sqrt{q}}
\end{align}
is the character of $\SU(2)$ symmetry showing that the global symmetry for the $\SU(2)_0$ theory is enhanced to $\E_1=\SU(2)$, whereas that for SU(2)$_\pi$ remains as $\U(1)_I$, with $\chi_2^{\widetilde{\E_1}}=q$.

\subsection{\texorpdfstring{$\Z_2$}{Z2} S-folding of 5d rank-1 SCFTs}\label{sec:Z2-S-folds}

Some 5-brane web diagrams have a discrete $\Z_{n}$ symmetry, which is generated by a rotation of the web by an angle ${2\pi}/{n}$ (${n=2,3,4,6}$)\footnote{Here, the axio-dilation $\tau$ takes any value for $n=2$, while $e^{\frac{2\pi}{3}i}, i,  e^{\frac{\pi}{3}i}$ for $n=3,4,6$, respectively.}. These $\Z_{n}$ symmetries are a subgroup of the SL(2,\Z) transformation that transform the charges of 5-branes or 7-branes. 
Now, one can imagine such discrete $\Z_{n}$ rotating of the web by ${2\pi}/{n}$, combined with the $\Z_{n}$ SL(2,\Z) transformation. For example, recall figure~\ref{fig:toric}, and one clearly sees that the SU(2)$_0$ theory has a $\Z_2$ symmetry, while the SU(2)$_\pi$ theory does not have such discrete symmetry. It is also easy to see that the SU(2)$_0$ theory also has a $\Z_4$ symmetry in the infinite coupling limit. The S-folds~\cite{Acharya:2021jsp, Kim:2021fxx} is the gauging of this discrete $\Z_{n=2,3,4,6}$  symmetry. In other words, an S-fold for the $\Z_{n}$ symmetry is obtained by a quotient with respect to the $\Z_{n}$ group in the original theory.

\paragraph{Prescription.}
Explicit constructions of $\Z_{n}$ S-folds in terms of  Type IIB 5-brane webs were discussed in \cite{Acharya:2021jsp, Kim:2021fxx}. The prescription proposed in \cite{Kim:2021fxx} can be summarized as follows: 
\begin{enumerate}
    \item[(I)] Given a 5-brane web invariant under the S-fold action, one restricts the web to a {\it fundamental domain} whose images under this $\Z_{n}$ S-fold action reproduce the original configuration.
    \item[(II)] To realize the $\Z_{n}$ S-fold for $n=2,3,4,6$, one places the corresponding 7-brane configuration $\mathcal{S}_{\D_4}$, $\mathcal{S}_{\E_6}$, $\mathcal{S}_{\E_7}$, or 
    $\mathcal{S}_{\E_8}$, respectively, at the fixed point in the fundamental domain.
\end{enumerate}
 Here, ${\cal{S}}_{\D_4,\E_6,\E_7,\E_8}$ are the 7-brane configurations associated with, respectively, $\D_4, \E_6,\E_7,\E_8$ symmetries, as discussed in \eqref{eq:7-braneconfi}:
\begin{align}
{\cal{S}}_{\D_4}= \mathbf{A}^4\mathbf{BC}, \quad     
{\cal{S}}_{\E_{6}}= \mathbf{A}^{5}\mathbf{BCB}, \quad     
{\cal{S}}_{\E_{7}}= \mathbf{A}^{6}\mathbf{BCB}, \quad     \mathrm{and}\quad
{\cal{S}}_{\E_{8}}= \mathbf{A}^{7}\mathbf{BCB}\ . 
\end{align}
The inserted 7-brane configurations ${\cal{S}}$ have the deficit angles and monodromies required to consistently identify the boundary 5-branes of the fundamental domain, thereby providing a 5-brane-web realization of the $\Z_{n}$ S-folds. Moreover, the appearance of the 7-brane configurations $\mathcal{S}_{\E_{6,7,8}}$ suggests that the resulting SCFTs can be identified from the total 7-brane content at the fixed point.

\begin{figure}
    \centering
    \newcommand{\tikzfigscale}{1.25}

\begin{tikzpicture}[
  scale=\tikzfigscale,
  transform shape,
  line cap=round,
  line join=round,
  line width=1.25pt,
  font=\small,
  guide/.style={
    draw=blue,
    dash pattern=on 2pt off 4.5pt
  },
  redline/.style={ draw=red
  },
  reddash/.style={
    redline,
    dash pattern=on 2pt off 4.5pt
  },
  redarrow/.style={
    redline,
    line width=1.25pt,
    -{Stealth[length=5pt,width=5pt]}
  },
  doublearrow/.style={
    line width=0.9pt,
    {Stealth[length=5pt,width=5pt]}-{Stealth[length=5pt,width=5pt]}
  },
  innerbox/.style={
    draw=blue,
    inner xsep=4pt,
    inner ysep=2pt,
    font=\scriptsize
  },
  label/.style={
    font=\scriptsize,
    inner sep=1pt
  }
]

\pgfmathsetmacro{\xone}{0.00}
\pgfmathsetmacro{\xtwo}{2.80}
\pgfmathsetmacro{\xthree}{4.50}
\pgfmathsetmacro{\xfour}{6.55}
\pgfmathsetmacro{\xfive}{8.95}

\newcommand{\redcrossat}[1]{%
  \draw[redline,line width=1pt] ($(#1)+(-0.085,-0.085)$) -- ($(#1)+(0.085,0.085)$);
  \draw[redline,line width=1pt] ($(#1)+(-0.085,0.085)$) -- ($(#1)+(0.085,-0.085)$);
}

\newcommand{\reddotat}[1]{%
  \fill[red] (#1) circle[radius=0.055];
}

\begin{scope}[shift={(\xone,0)}]
  \draw (-0.6,-0.42) rectangle (0.6,0.42);
  \draw (-1.05, 0.92) -- (-0.6, 0.42);
  \draw ( 0.6, 0.42) -- ( 1.05, 0.92);
  \draw (-0.6,-0.42) -- (-1.05,-0.92);
  \draw ( 0.6,-0.42) -- ( 1.05,-0.92);
  \draw[redarrow] (0.94,0.38) .. controls (1.20,0.12) and (1.20,-0.12) .. (0.94,-0.38);
  \draw[redarrow] (-0.94,-0.38) .. controls (-1.20,-0.12) and (-1.20,0.12) .. (-0.94,0.38);
  \redcrossat{0,0}
  \node[label] at (0,-1.08) {$\Z_{2}$};
\end{scope}

\begin{scope}[shift={(\xtwo,0)}]
  \draw[guide] (0,-0.96) -- (0,0.96);
  \draw (-0.6, 0.42) -- (0, 0.42);
  \draw (-0.6, 0.42) -- (-0.6,-0.42);
  \draw (-0.6,-0.42) -- (0,-0.42);
  \draw (-1.05, 0.92) -- (-0.6, 0.42);
  \draw (-0.6,-0.42) -- (-1.05,-0.92);

  \redcrossat{0,0}
\end{scope}

\begin{scope}[shift={(\xthree,0)}]
  \draw[guide] (0,-0.96) -- (0,0.96);
  \draw (-0.6, 0.42) -- (0, 0.42);
  \draw (-0.6, 0.42) -- (-0.6,-0.42);
  \draw (-0.6,-0.42) -- (0,-0.42);
  \draw (-1.05, 0.92) -- (-0.6, 0.42);
  \draw (-0.6,-0.42) -- (-1.05,-0.92);

  \reddotat{0,0}
  \node[label,red,anchor=west] at (0.10,0.03) {$\mathbf{A}^{4}\mathbf{BC}$};
\end{scope}

\begin{scope}[shift={(\xfour,0)}]
  \draw (-0.6,-0.42) rectangle (0.6,0.42);
  \draw[dash pattern=on 3pt off 3pt] (-0.52, 0.34) -- (-1.05, 0.92);
  \draw[reddash] ( 0.52, 0.34) -- ( 1.05, 0.92);
  \draw[dash pattern=on 3pt off 3pt] (-0.52,-0.34) -- (-1.05,-0.92);
  \draw[reddash] ( 0.52,-0.34) -- ( 1.05,-0.92);

  \foreach \x in {-0.08,0.00,0.08,0.16} {
    \draw[reddash] (\x,-0.34) -- (\x,-0.92);
  }

  \node[label] at (-0.37, 0.24) {$\mathbf{B}$};
  \node[label,red] at ( 0.38, 0.17) {$\mathbf{C}$};
  \node[label] at (-0.37,-0.2) {$\mathbf{C}$};
  \node[label,red] at ( 0.38,-0.17) {$\mathbf{B}$};
  \node[label,red] at (0.05,-0.17) {$\mathbf{A^{4}}$};
\end{scope}

\begin{scope}[shift={(\xfive,0)}]
  \draw (0,0) ellipse[x radius=1,y radius=0.55];
  \node[innerbox] (ebox)at (-0.09,0) {{\color{red}$\mathbf{A^{4}BC}$}{\color{black}$\mathbf{B}$}};
\node[font=\scriptsize,inner sep=0pt,anchor=west] at ([xshift=1pt]ebox.east) {$\mathbf{C}$};
  \node[label,blue] at (0,-0.78) {$\E_{5}$};
\end{scope}

 \node[font=\small]  at(7.65,0){$\Leftrightarrow$};
\end{tikzpicture}
\caption{The $\Z_{2}$ S-fold of the SU(2)$_0$ theory and 5-brane realization. The resulting 7-brane configuration in a box of the last figure shows that  $\E_5$ symmetry.}
    \label{fig:pureSU2-web-E5}
\end{figure}
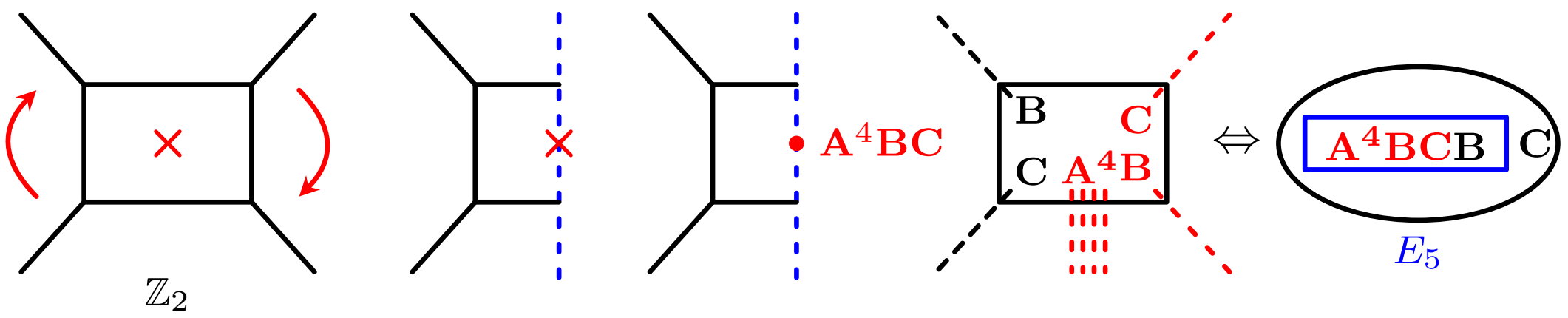

Among the rank-1 theories, four admit a $\Z_{2}$ S-fold
construction:
\begin{align}\label{eq:fourZ2}
\mathrm{SU}(2)_0,\qquad
\mathrm{SU}(2)+2\F,\qquad
\mathrm{SU}(2)+4\F,\qquad
\mathrm{SU}(2)+6\F\ .    
\end{align}
The simplest example, obtained by S-folding the $\mathrm{SU}(2)_0$ theory, is illustrated in figure~\ref{fig:pureSU2-web-E5}. The first panel displays the $\Z_{2}$ symmetry of the $\mathrm{SU}(2)_0$ 5-brane web, realized as a rotation by $180^\circ$ about the fixed point marked by a red $\times$. The second panel shows a choice of fundamental domain, given by the left half of the web. In the third panel, the 7-brane configuration associated with the $\D_4$ singularity, $\mathbf{A}^4\mathbf{B}\mathbf{C},$ is inserted at the fixed point according to prescription~(II). The resulting configuration therefore provides a 5-brane-web realization of the $\Z_{2}$ S-fold of the $\mathrm{SU}(2)_0$ theory and is closely related to the $\mathrm{SU}(2)+4\F$ theory. Indeed, as shown in the final panel of figure~\ref{fig:pureSU2-web-E5}, the total 7-brane configuration  $\mathbf{A}^4\mathbf{B}\mathbf{C}\mathbf{B}\mathbf{C}$ exhibits the $\E_5$ flavor symmetry associated with the UV fixed point of the $\mathrm{SU}(2)+4\F$ theory.

\begin{figure}
    \centering
    \newcommand{\tikzfigscale}{1.25}

\begin{tikzpicture}[
  scale=\tikzfigscale,
  transform shape,
  line cap=round,
  line join=round,
  line width=1.25pt,
  font=\small,
  guide/.style={
    draw=blue,
    dash pattern=on 2pt off 4.5pt
  },
  label/.style={
    font=\scriptsize,
    inner sep=1pt
  },
  dot/.style={
    fill=red,
    draw=red
  }
]

\pgfmathsetmacro{\figsep}{3.00}
\pgfmathsetmacro{\xone}{0*\figsep}
\pgfmathsetmacro{\xtwo}{1*\figsep}
\pgfmathsetmacro{\xthree}{2*\figsep}

\newcommand{\reddotat}[1]{%
  \fill[dot] (#1) circle[radius=0.055];
}

\newcommand{\halfrectangle}{%
  \draw[guide] (0,-1.42) -- (0,1.02);
  \draw (-0.6, 0.42) -- (0, 0.42);
  \draw (-0.6, 0.42) -- (-0.6,-0.42);
  \draw (-0.6,-0.42) -- (0,-0.42);
  \draw (-1.05, 0.92) -- (-0.6, 0.42);
  \draw (-0.6,-0.42) -- (-1.05,-0.92);
  \reddotat{0,0}
  \node[label,red,anchor=west] at (0.10,0.03) {$\mathbf{A^{4}BC}$};
  \node[label] at (-1.22,1.04) {$\mathbf{B}$};
}

\begin{scope}[shift={(\xone,0)}]
  \halfrectangle
  \draw (-1.45,-0.92) -- (-1.05,-0.92);
  \draw (-1.05,-0.92) -- (-1.05,-1.30);

  \node[label] at (-1.65,-0.84) {$\mathbf{A}$};
  \node[label] at (-1.12,-1.5) {$\mathbf{X}_{[0,1]}$};
  \node[label] at (0,-2.22) {$\E_{6}$};
\end{scope}

\begin{scope}[shift={(\xtwo,0)}]
  \halfrectangle
  \draw (-1.45,-0.92) -- (-1.05,-0.92);
  \draw (-1.05,-0.92) -- (-1.05,-1.30);
  \draw (-1.45,-1.30) -- (-1.05,-1.30);
  \draw (-1.05,-1.30) -- (-0.68,-1.66);

  \node[label] at (-1.62,-0.84) {$\mathbf{A}$};
  \node[label] at (-1.62,-1.22) {$\mathbf{A}$};
  \node[label] at (-0.72,-1.82) {$\mathbf{B}$};
  \node[label] at (0,-2.22) {$\E_{7}$};
\end{scope}

\begin{scope}[shift={(\xthree,0)}]
  \halfrectangle
  \draw (-1.45,-0.92) -- (-1.05,-0.92);
  \draw (-1.05,-0.92) -- (-1.05,-1.30);
  \draw (-1.45,-1.30) -- (-1.05,-1.30);
  \draw (-1.05,-1.30) -- (-0.68,-1.66);
  \draw (-1.45,-1.66) -- (-0.68,-1.66);
  \draw (-0.68,-1.66) -- (-0.30,-1.88);

  \node[label] at (-1.62,-0.84) {$\mathbf{A}$};
  \node[label] at (-1.62,-1.22) {$\mathbf{A}$};
  \node[label] at (-1.62,-1.58) {$\mathbf{A}$};
  \node[label] at (-0.70,-2.02) {$\mathbf{X}_{[2,-1]}$};
  \node[label] at (0,-2.22) {$\E_{8}$};
\end{scope}
\end{tikzpicture}
\caption{$\Z_2$ S-folds of the SU(2)$+2n\F$ theories ($n=1,2,3$) and their SCFTs}\label{fig:SU2-E679-web}
\end{figure}
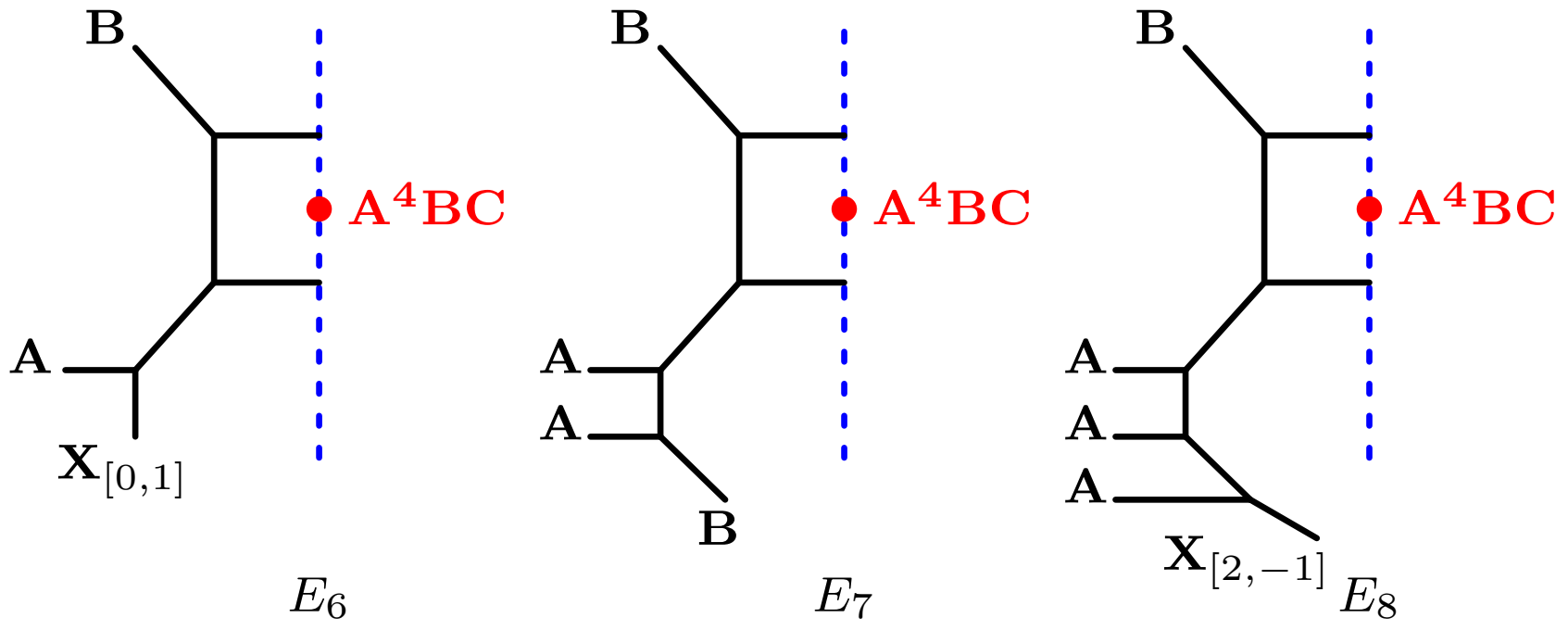

For the remaining $\Z_2$ S-folds of SU(2)$+2n\F$ theories,  with $n=1,2,3$, the
corresponding 5-brane webs are obtained by adding $n$ flavor branes to the fundamental domain, as illustrated in
Figure~\ref{fig:SU2-E679-web}. The resulting SCFTs are identified with the $\E_{5+n}$ theories, whose total 7-brane configurations can be brought to the canonical form
\begin{align}
\mathbf{A}^{n+4}\mathbf{B}\mathbf{C}\mathbf{B}\mathbf{C}\, .
\end{align}
These configurations displayed in figure~\ref{fig:SU2-E679-web} can be recast in this form by rearranging the 7-branes using their monodromy relations, following~\cite{Gaberdiel:1997ud, Gaberdiel:1998mv, DeWolfe:1998zf, DeWolfe:1998eu, DeWolfe:1998pr, DeWolfe:1999hj, Kim:2015jba}. For example, reading the
7-branes in the first panel counterclockwise, the $\E_6$ configuration given in the first panel in figure~\ref{fig:SU2-E679-web} can be rewritten as 
\begin{align}
\mathbf{A}^{4}\mathbf{B}\mathbf{C}\mathbf{B}\mathbf{A}\mathbf{X}_{[0,1]}=\mathbf{A}^{4}\mathbf{B}\mathbf{C}\mathbf{B}\mathbf{C}\mathbf{A} = \mathbf{A}^{5}\mathbf{B}\mathbf{C}\mathbf{B}\mathbf{C}\ ,
\end{align}
where we have used
$\mathbf{A}\mathbf{X}_{[0,1]}=\mathbf{C}\mathbf{A}$.
Similarly, the configurations $\E_7$ and $\E_8$ are brought to their
canonical forms using $
\mathbf{A}^{2}\mathbf{B}=\mathbf{C}\mathbf{A}^{2}$ and $    \mathbf{A}^{3}\mathbf{X}_{[2,-1]}
    =\mathbf{C}\mathbf{A}^{3}$, 
respectively.
These brane constructions make manifest a close relation between the $\Z_2$ S-folds of the $\mathrm{SU}(2)+2n\F$ theories and the $\E_{5+n}$ SCFTs for $n=0,1,2,3$. 

\paragraph{Prepotential.} Beyond the 5-brane web considerations, the prescription of the S-folds is also tested at the level of the prepotential~\cite{Kim:2021fxx}. On the Coulomb branch where the gauge group is broken to abelian $\U(1)^\mathrm{rk_{G}}$ where $\mathrm{rk_{G}}$ is the rank of gauge group $G$, the prepotential characterizing the low-energy abelian theory is given by~\cite{Seiberg:1996bd, Intriligator:1997pq}
\begin{align}
    \mathcal{F}(\phi) = \frac{m_0}{2}h_{ij}\phi_i\phi_j+\frac{\kappa}{6}d_{ijk}\phi_i\phi_j\phi_k +\frac{1}{12}\bigg(\sum_{\mathsf{r}\in\mathbf{ R}}|\mathsf{r}\cdot\phi|^3 - \sum_f\sum_{\mathsf{w}\in \mathbf{W}_f} |\mathsf{w}\cdot\phi+m_f|^3\bigg),
\end{align}
where $\phi$ is the vacuum expectation value of the scalar field in the vector multiplet, $m_0$ is the instanton mass, and $\kappa$ is the classical Chern-Simons level, and $m_f$ are the mass of the matter $f$. Here, $\mathsf{r}$ is a root of the Lie algebra $\mathfrak{g}$ of the gauge group $G$, $\mathsf{w}$ is a weight associated with the matter in the representation $\mathbf{W}_f$ of $\mathfrak{g}$, and the remaining factors are defined as $h_{ij}=\mathrm{Tr}(T_iT_j)$ and $d_{ijk}=\frac12\mathrm{Tr}\big(T_i\{T_j,T_k\}\big)$ with the Cartan generators $T_i$.

Derivatives of the prepotential with respect to $\phi$ give physical observables that can be read off from 5-brane webs. The first derivative yields the magnetic monopole string tension $\partial_i\mathcal{F}$ which corresponds to the areas of compact faces of 5-brane webs, and the second derivative determines the effective coupling $\big(g^{-2}_\mathrm{eff}\big)_{ij}=\partial_i\partial_j\mathcal{F}$ which corresponds to the perimeters of the compact phases. The prepotential for the $\SU(2)_0$ theory is a good and simple illustrative example as it is rank-1 and does not possess the Chern-Simons term. Given an $\SU(2)_0$ 5-brane web of a rectangular shape shown in figure~\ref{fig:toric}, the vertical length is $2\phi$, corresponding to the separation of two color D5-branes and the horizontal length is $2\phi+m_0$, corresponding to the separation of two NS5-brane taking into account the bare coupling $m_0$. Together, the area of the compact phase yields $(2\phi)(2\phi+m_0)$ which corresponds to the monopole string tension, and its integration with respect to $\phi$ yields the prepotential $\mathcal{F}$:
\begin{align}
    \partial_\phi\mathcal{F}_{\SU(2)_0}=(2\phi)(2\phi+m_0) \quad\longrightarrow \quad \mathcal{F}_{\SU(2)_0} = m_0\phi^2+ \frac{8}{6}\phi^3\ .
\end{align}

Based on 5-brane descriptions of the $\Z_n$ S-folds where the area of the compact phases is reduced by $1/n$, it was proposed~\cite{Kim:2021fxx} that the prepotential of the S-folded theory $\mathcal{T}/\Z_n$ is given by 
\begin{align}\label{eq:1/nF}
    \mathcal{F}(\mathcal{T}/\Z_n) = \frac{1}{n}\mathcal{F}(\mathcal{T})\Big|_{\Z_n\text{-identification}}\ .
\end{align}
Here, the right-hand side denotes the prepotential of the parent theory $\mathcal{T}$, evaluated after imposing the $\Z_n$ identification on its K\"ahler parameters and rescaled by a factor of $1/n$. For the $\Z_2$ S-folds of rank-1 theories, \eqref{eq:1/nF} supports the insertion of $\mathcal{S}_{\D_4}$. This can be shown more explicitly as follows: The prepotential of the parent theory $\SU(2)+2n\F$ takes the form~\cite{Seiberg:1996bd, Intriligator:1997pq}
\begin{align}
    \mathcal{F}_{\mathcal{T}} ( m_0, m_{i=1,\dots, 2n}) = m_0 \phi^2 +\bigg(\,\frac{8}{6} |\phi|^3 -\frac1{12}\sum_{i=1}^{2n} \big(|\phi-m_i|^3+|\phi+m_i|^3 \big)\bigg)\ , 
\end{align}
where $m_0$ denotes the instanton mass of the parent theory $\mathcal{T}=\SU(2)+2n\F$,  the flavor index $i$ runs from $1$ to $2n$, and $m_i$ are the flavor masses. It follows from \eqref{eq:1/nF} that the prepotential for the S-folded theory $\mathcal{F_T}/\Z_2$ is then given by
\begin{align}\label{eq:FtoF/2}
    \mathcal{F}_{\mathcal{T}/\Z_2}(m_0, m_{i=1,\dots, n}) 
     &=\frac12 m_0 \phi^2 +\bigg(\,\frac{4}{6} |\phi|^3 -\frac1{12}\sum_{i}^{n} \big(|\phi-m_i|^3+|\phi+m_i|^3 \big)\bigg)\cr
    &= \Big(\frac{ m_0}{2}\Big) \phi^2 + \bigg(\,\frac{8}{6} |\phi|^3-\frac{4}{6} |\phi|^3 -\frac1{12}\sum_{i}^{n} \big(|\phi-m_i|^3+|\phi+m_i|^3 \big)\bigg)\cr
    &\Leftrightarrow\, \mathcal{F}_{\mathcal{T}^\mathcal{S}}(m^\mathcal{S}_0, m^\mathcal{S}_{i=1,\dots, n+4})\ , 
\end{align}
 where $\mathcal{F}_{\mathcal{T}^\mathcal{S}}$ denotes the prepotential for the theory with the 7-brane insertion $\mathcal{S}_{\D_4}$ to the parent theory, $\mathcal{T}^\mathcal{S}=\SU(2)+(4+n)\F$  or $\E_{5+n}$. Here, the mass parameters of $\mathcal{T}^\mathcal{S}$,  $\{m^\mathcal{S}\}$,  are \emph{identified} with those of $\mathcal{T}$ as
 \begin{align}\label{eq:m0/2andmi}
     m^\mathcal{S}_0  = \frac{m_0}{2}, \qquad \{  m^\mathcal{S}_{i=1,\dots, n+4} \} = \{ m_{i=1,\dots, n};\, 0,\,0,\,0,\,0\}\ , 
 \end{align}
 where the term $-\frac{4}{6}|a|^3$ in the second line of \eqref{eq:FtoF/2} is interpreted as the contribution of four massless hypermultiplets in the fundamental representation, so that  $\mathcal{T}^\mathcal{S}$ is consistent with the $\SU(2)+(4+n)\F$ theory with four extra flavors that originated from the 7-branes  $\mathcal{S}_{\D_4} = \mathbf{A}^4\mathbf{BC}$, whose masses are zero because they are inserted at the fixed point of $\Z_2$ symmetry. In short, the prescription of the $\Z_2$ S-folds for rank-1 theories, $\SU(2)+2n\F$ $(n=0,1,2,3)$, is summarized as 
 \begin{align}
      \mathcal{F}_{\SU(2)+2n\F/\Z_2}\big(m_0, \{m_{i=1,\dots, n}\}\big)  = \mathcal{F}_{\SU(2)+(n+4)\F}\big({m_0}/{2}, \{m_{i=1,\dots, n}; 0,0,0,0\}\big)\ .
 \end{align}

In the following, we examine this relation for all four rank-1 theories by explicitly deriving their Seiberg–Witten curves and comparing them with the Seiberg–Witten curves for the corresponding 7-brane-inserted $\E_5$, $\E_6$, $\E_7$, and $\E_8$ theories.


\section{Seiberg–Witten curves for 5d rank-1 \texorpdfstring{$\Z_2$}{Z2} S-folds}\label{sec:SWcurvesforS-folds}
In this section, we compute the Seiberg–Witten curves for rank-1 $\Z_2$ S-folds in two different ways. One way is that we start from the Seiberg–Witten curves for the original theories and then we implement the $\Z_2$ action explicitly by introducing a two-to-one map between curve coordinates and physical mass parameters. The other way is that we first insert  $\mathcal{S}_{\D_4}=\mathbf{A}^4\mathbf{BC}$ to the fixed point and then specially tune the masses associated with the inserted four D7-branes that are stuck at the fixed point. We will show that two Seiberg–Witten curves obtained from these two different ways are equivalent, which can be shown explicitly when the curves are expressed in Weierstrass normal form.

\subsection{SW curves from two-to-one map}\label{sec:2-1}

The 5d Seiberg–Witten curve is a holomorphic curve $F(t,w)=0$ embedded in $(\mathbb{C}^*)^2$. Via the duality between Type IIB string theory on a circle and M-theory on a torus, the mapping-class group of the M-theory torus is identified with the SL(2,\Z) duality group of Type IIB string theory. Accordingly, the same $\Z_n$ S-fold action that acts on the $(x^5,x^6)$-plane together with the $(p,q)$-charge lattice admits a natural lift to the Seiberg–Witten geometry, acting on the torus coordinates $(x^4,x^{11})$. For the $\Z_2$ S-fold, this action takes the form 
\begin{align}
(x^5, x^6; x^4, x^{11})~~ \mapsto~~ (-x^5, -x^6; -x^4, -x^{11})\ ,    
\end{align}
or, equivalently, 
\begin{align}\label{eq:involution}
    (t,w)~~\mapsto~~ (t^{-1},w^{-1})\ .
\end{align}
This is a holomorphic involution of $(\mathbb{C}^*)^2$. It leaves the Seiberg–Witten differential $\lambda_\mathrm{SW}$ \eqref{eq:SWdifferential} invariant and preserves the holomorphic two-form $\Omega = d\log t \wedge d \log w$. The $\Z_2$ S-fold therefore acts on the Seiberg–Witten curves through the involution \eqref{eq:involution}, requiring 
\begin{align}\label{eq:twZ2}
 F(t,w)=F(t^{-1},w^{-1})   \ .
\end{align}
This invariance ensures that the curve descends to the quotient $(\mathbb{C}^*)^2/\Z_2$ and 
constrains the mass parameters $M_i$ and the Coulomb modulus $U$ appearing in the original curve $F(t,w)=0$. 

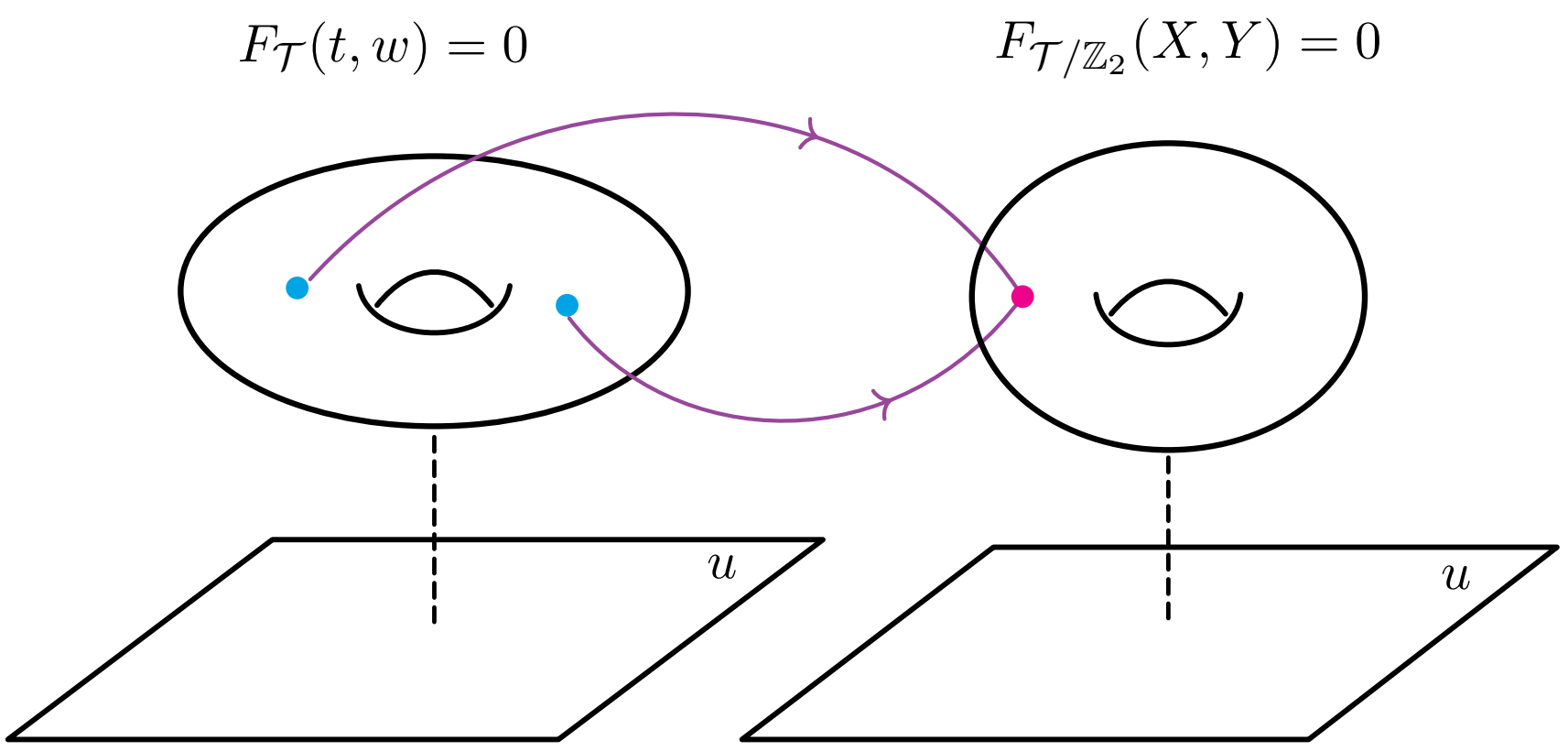
\begin{figure}
    \centering
    \begin{tikzpicture}[
    scale=0.78,
    line cap=round,
    line join=round
]

\draw[line width=1.1pt]
(-6.0,-3.0) -- (-0.9,-3.0) -- (1.55,-1.15) --
(-3.55,-1.15) -- cycle;

\draw[densely dashed,line width=0.9pt]
(-2.05,-0.20) -- (-2.05,-1.95);

\draw[line width=1.2pt]
(-2.05,1.15) ellipse (2.35 and 1.25);

\draw[line width=1.1pt]
(-2.75,1.20)
.. controls (-2.65,0.62) and (-1.45,0.62) ..
(-1.35,1.20);

\draw[line width=1.1pt]
(-2.58,1.02)
.. controls (-2.25,1.43) and (-1.85,1.43) ..
(-1.52,1.02);

\fill[Cerulean] (-3.32,1.18) circle (0.105);
\fill[Cerulean] (-0.82,1.02) circle (0.105);


\node[anchor=west] at (0.3,-1.40) {$u$};
\node[anchor=east] at (-1.0,3.40) {$F_{\mathcal{T}}(t,w)=0$};

\draw[
    Purple,
    thick,
    postaction={decorate},
    decoration={
        markings,
        mark=at position 0.68 with {\arrow{>}}
    }
]
(-3.20,1.26)
.. controls (-1.30,3.35) and (1.95,3.30) ..
(3.40,1.10);

\draw[
    Purple,
    thick,
    postaction={decorate},
    decoration={
        markings,
        mark=at position 0.68 with {\arrow{>}}
    }
]
(-0.80,0.90)
.. controls (0.15,-0.35) and (2.25,-0.45) ..
(3.40,1.10);
\draw[line width=1.1pt]
(0.80,-3.0) -- (6.05,-3.0) -- (8.35,-1.22) --
(3.13,-1.22) -- cycle;

\draw[densely dashed,line width=0.9pt]
(4.75,-0.39) -- (4.75,-1.95);

\draw[line width=1.2pt]
(4.75,1.10) ellipse (1.82 and 1.42);

\draw[line width=1.1pt]
(4.08,1.12)
.. controls (4.15,0.50) and (5.35,0.50) ..
(5.42,1.12);

\draw[line width=1.1pt]
(4.22,0.94)
.. controls (4.55,1.34) and (4.95,1.34) ..
(5.28,0.94);

\fill[magenta] (3.4,1.10) circle (0.105);

\node[anchor=west] at (7.1,-1.50) {$u$};
\node[anchor=east] at (6.9,3.40) {$F_{\mathcal{T}/\Z_2}(X,Y)=0$};

\end{tikzpicture}
\caption{Two-to-one map. 
Under the $\Z_2$ S-fold action, pairs of points on the torus (fiber) of the original theory ($F_{\mathcal{T}}(t,w)=0)$ are mapped to a single point on a new torus (fiber) of the $\Z_2$ S-folded theory, while the Coulomb branch (base) remains unaltered, yielding the Seiberg–Witten curves for the $\Z_2$ S-fold,  $F_{\mathcal{T}/\Z_2}(X,Y)=0.$}\label{fig:2-to-1 map}
\end{figure}

As the rank-1 SW geometry is a family of tori parametrized by the Coulomb-branch coordinate $u$, the $\Z_2$ S-fold action \eqref{eq:twZ2} corresponds to a map that sends two generic points on a torus to a point on a new torus, which acts trivially on the base Coulomb branch, as depicted in figure \ref{fig:2-to-1 map}.
It is a $\Z_2$-invariant two-to-one map realizing the S-fold.

For an explicit construction of such a map, consider the following system of two $\Z_2$-invariant algebraic equations in the coordinates $t$ and $w$:%
\footnote{We note that such a two-to-one map is not unique. For instance, one may take the following two-to-one map:
\begin{align*}
    \tilde{X}=t+\frac1t, \qquad \tilde{Y} = \frac{1+tw}{t+w}\ , 
\end{align*}
which looks simpler than \eqref{eq:2-to-1} but the corresponding $t(\tilde{X},\tilde{Y})$ and $w(\tilde{X},\tilde{Y})$ are more complicated than \eqref{eq:21maptw}. Yet this two-to-one map leads to the same Weierstrass form as the one obtained from \eqref{eq:2-to-1}.}
\begin{align}\label{eq:2-to-1}
X=\frac{t-1}{t+1}\frac{w-1}{w+1},
    \qquad Y=\frac{t-1}{t+1}\frac{w+1}{w-1}\ .
\end{align}
For generic points in the new coordinates, 
this system admits two solutions, which lead to a two-to-one map.
One can choose: 
\begin{align}\label{eq:21maptw}
    t=\frac{1+\sqrt{X} \sqrt{Y}}{1-\sqrt{X}\sqrt{Y}},\qquad w=-\frac{\sqrt{X}+\sqrt{Y}}{\sqrt{X}-\sqrt{Y}}\ .
\end{align}
By plugging $t(X,Y),w(X,Y)$ into the Seiberg–Witten curve $F_{\mathcal{T}}(t,w)=0$ for the original theory $\mathcal{T}$, we find the Seiberg–Witten curve for the $\Z_2$ S-folded theory $\mathcal{T}/\Z_2$, 
\begin{align}
    F_{\mathcal{T}/\Z_2}(X,Y)=0 \ .
\end{align}

We now consider the four S-folds, $\mathcal{T}/\Z_2=\SU(2)_0+2n\F/\Z_2~(n\!=\!0,1,2,3)$ and compute their Seiberg–Witten curves. The curves are expressed in terms of the characters of a symmetry group that the remaining mass parameters after the $\Z_2$ S-fold organize themselves. To study the resulting symmetry structure, we analyze the Kodaira fiber configuration after expressing the Seiberg--Witten curves in the Weierstrass normal form
\begin{align}
    y^2 = 4 x^3 - g_2(u)\, x- g_3(u).
\end{align}
For generic values of the mass parameters, we find that the relatively minimal elliptic surface
associated with the Seiberg--Witten curve has Kodaira fiber
configuration $(I_{4-{N_f}/{2}};\,I_2^4I_1^{N_f/2})$, where the distinguished fiber at
$u=\infty$ is of type $I_{4-{N_f}/{2}}$ and the four singular fibers of type $I_2$ and $N_f/2$ singular fibers of type $I_1$ are at finite $u$.
The fiber types are determined by the vanishing orders of
$g_2$, $g_3$, and
$\Delta(u)=g_2(u)^3-27g_3(u)^2$
in local minimal Weierstrass models, including the model near
$u=\infty$, using the Kodaira criteria reviewed
in~\cite{Closset:2023pmc}.
Applying the flavor-lattice prescription
of~\cite{Caorsi:2018ahl,Oguiso:1991mw} to the associated
Mordell--Weil lattice, as discussed
in~\cite{Xie:2022lcm}, yields a flavor root system of type $G$
and hence the flavor algebra $\mathfrak{g}$.

\subsubsection{\texorpdfstring{$\mathrm{SU}(2)_0/\Z_2$}{SW curve for SU(2)0Z2 theory}}
The Seiberg–Witten curve for the SU(2)$_0$ theory \eqref{eq:swcurve-su0} takes the form invariant under $t \to t^{-1}$ and $w \to w^{-1}$, 
\begin{align}\label{eq:sw-su2-z2}
     F_{\SU(2)_0}(t,w) = 0 \quad \Longleftrightarrow \quad U+w+\frac1w+\sqrt{q}\Big(t+\frac1t\Big)=0 \ .
\end{align}
As explained, by substituting \eqref{eq:21maptw} into the curve \eqref{eq:sw-su2-z2}, one finds the Seiberg–Witten curve of the SU(2)$_0/\Z_2$ theory, 
\begin{align}\label{eq:swz2inv}
 & F_{\SU(2)_0/\Z_2} (X,Y)=0  \quad \Longleftrightarrow   \\
 &  X^2 Y (-2 \sqrt{q}+U-2)+X Y^2 (2 \sqrt{q}-U-2)+X (2
   -2\sqrt{q}-U)+Y (2 \sqrt{q}+U+2)=0\ .\nn
\end{align}
The corresponding Weierstrass normal form reads
\begin{align}\label{eq:WSformSU20}
    y^2=4x^3-g_2^{\SU(2)_0/\Z_2}x-g_3^{\SU(2)_0/\Z_2}\ ,
\end{align}
where
\begin{align}\label{eq:g2-su2-0}
 g_2^{\SU(2)_0/\Z_2}&=\frac1{12}u^4-\frac23\chi_1^{\SU(2)}u^2+\frac43\big(\chi_1^{\SU(2)}\big)^2+16\ ,\\
    g_3^{\SU(2)_0/\Z_2}&=\frac1{216}u^6-\frac1{18}u^4\chi_1^{\SU(2)}+\bigg(\frac29\big(\chi_1^{\SU(2)}\big)^2-\frac83\bigg)u^2-\frac8{27}\big(\chi_1^{\SU(2)}\big)^3+\frac{32}3\chi_1^{\SU(2)}\ .\nn
\end{align}
Here, $\chi_1^{\SU(2)}=\sqrt{q}+1/\sqrt{q}$ is the character for the fundamental representation of SU(2), and $U$ and $u$ are related by the rescaling $U\to q^{1/4}u$.  Converting a cubic curve \eqref{eq:swz2inv} into the Weierstrass normal form \eqref{eq:WSformSU20} is explained in Appendix \ref{app:WNF2}, or also see \cite{Huang:2013yta}.  Flavor symmetry $\SU(2)$ is obtained from the corresponding Kodaira fiber configuration $(I_4;I_2^4)$.

\subsubsection{\texorpdfstring{$\mathrm{SU}(2)+2\F/\Z_2$}{SW curve for SU(2)+2FZ2 theory}}

The Seiberg–Witten curves for the $\SU(2)+N_f\F$ theory based on a 5-brane web are presented in \cite{Kim:2014nqa}. We here sketch the procedure in a way that the $\Z_2$ identification can be easily applied. The curve takes the following form 
\begin{align}
    F(t,w)=c_{00}+c_{10}(t+t^{-1})+c_{11}(tw+t^{-1}w^{-1})+c_{01}(w+w^{-1})=0\ , 
\end{align}
where the coefficients $c_{mn}$ are fixed by the parameters of the theory, the boundary conditions of the external 5-branes, and the symmetry of the theory. By imposing the $\Z_2$ symmetry for the S-folds, we first identify the mass parameters $M_1=M_2$. It follows from the boundary conditions that we can take $c_{10}=-c_{11}M_1$. By a rescaling degree of freedom, we choose $c_{10} = 1$. The remaining coefficients, $c_{00}$ and $c_{01}$, are then identified as the Coulomb parameter $U$ and $\sqrt{q/M_1}$, with $q$ being instanton parameter, respectively. The Seiberg–Witten curve for the $\SU(2)+2\F$ theory of the $\Z_2$ symmetry is given by
\begin{align}\label{eq:Ftwsu22}
   F_{\SU(2)+2\F}(t,w)=U+\left(t+t^{-1}\right)-\frac1{M_1}\left(tw+t^{-1}w^{-1}\right)+\sqrt{\frac{q}{M_1}}\left(w+w^{-1}\right)=0 \ .
\end{align}

Applying the two-to-one map, as explained in the $\mathrm{SU}(2)_0/\Z_2$ case, to \eqref{eq:Ftwsu22}, one finds the Seiberg–Witten curve of the $\Z_2$ S-folded $\mathrm{SU}(2) + 2\F$ theory, 
\begin{align}\label{eq:su22fz2}
& F_{\SU(2)+2\F/\Z_2}(X,Y)=0 \quad \Longleftrightarrow\cr
&X^2 Y (2 \sqrt{M_1 q}+2 M_1-U'+2)+X Y^2 (2 \sqrt{M_1 q}-2
   M_1+U'+2) \\
 & +X (-2 \sqrt{M_1 q}+2 M_1+U'+2)  +Y (-2 \sqrt{M_1 q}-2M_1-U'+2)+8 X Y=0\ ,\nonumber
\end{align}
where we have rescaled $U\to U'/M_1$. Using Appendix \ref{app:WNF2} for \eqref{eq:su22fz2}, the corresponding Weierstrass normal form reads
\begin{align*}
    y^2=4x^3-g_2^{2\F/\Z_2}x-g_3^{2\F/\Z_2}\ ,
\end{align*}
with
\begin{align}\label{eq:su2+2Fz2main}
    g_2^{2\F/\Z_2}&=\frac1{12} u^4 - \frac{2}3 \chi_1^{\SU(3)} u^2 + 16 u + \frac43 (\chi_1^{\SU(3)})^2 + 16 \chi_2^{\SU(3)},\\
    g_3^{2\F/\Z_2}&=\frac{1}{216}u^6-\frac{1}{18}u^4\chi_1^{\SU(3)}-\frac83u^3+\Big(\frac29\big(\chi_1^{\SU(3)}\big)^2-\frac83\chi_2^{\SU(3)}\Big)u^2+\frac{32}3\chi_1^{\SU(3)}u\notag\\
    &\quad +\frac{32}3\chi_1^{\SU(3)}\chi_2^{\SU(3)}-\frac8{27}\big(\chi_1^{\SU(3)}\big)^3,\nonumber
\end{align}
where $U'\to \sqrt{(M_1q^{1/3})}u$ has been used, and $\chi_1^{\SU(3)}$ and $\chi_2^{\SU(3)}$ are the characters of the fundamental and anti-fundamental representations of $\SU(3)$, respectively:
\begin{align}\label{eq:su3-cha}
    \chi_1^{\SU(3)}=y_1+\frac1{y_2}+\frac{y_2}{y_1},\qquad
    \chi_2^{\SU(3)}=\frac1{y_1}+y_2+\frac {y_1}{y_2},
\end{align}
with 
\begin{align}
    y_1=\frac {M_1}{q^{1/3}},\qquad y_2=M_1 q^{1/3}\ .
\end{align}
The flavor symmetry $\SU(3)$ is obtained from the corresponding Kodaira fiber configuration~$(I_3;I_2^4I_1)$.

\subsubsection{\texorpdfstring{$\mathrm{SU}(2)+4\F/\Z_2$}{SW curve for SU(2)+4FZ2 theory}}

After imposing the $\Z_2$ invariance, the Seiberg–Witten curve for the $\SU(2)+4\F/\Z_2$ theory takes the form
\begin{align}\label{eq:sw-su2-4f}
    F(t,w)=c_{00}+c_{01}(w+w^{-1})+c_{11}(tw+t^{-1}w^{-1})+c_{1-1}(tw^{-1}+t^{-1}w)+c_{10}(t+t^{-1})=0\ .
\end{align}
The mass parameters are identified so that only two independent masses $M_1,M_2$ remain, and we choose $c_{11}=1$, by the rescaling symmetry. This yields the Seiberg–Witten curve for the $\SU(2)+4\F/\Z_2$ theory of the $\Z_2$ invariance:
\begin{align}
     F_{\SU(2)+4\bf{F}}(t,w)&=U+\sqrt{\frac{{M_1M_2}}{q}}(q+1)(w+w^{-1})+(tw+t^{-1}w^{-1})+M_1M_2(tw^{-1}+t^{-1}w)\notag\\
     &\quad -(M_1+M_2)(t+t^{-1})=0\ ,
\end{align}
where $q$ is the instanton fugacity and $c_{00}=U$ is the Coulomb parameter.

By substituting the two-to-one map \eqref{eq:21maptw} into \eqref{eq:sw-su2-4f}, one finds
\begin{align}\label{eq:su24fz2}
    &F_{\SU(2)+4\bf{F}/\Z_2}(X,Y)=0\qquad \Longleftrightarrow\cr
    &X^2Y\bigg(-2 \sqrt{\frac{M_1 M_2}{q}}-2 \sqrt{M_1 M_2 q}+2 M_1+2 M_1 M_2+2 M_2+U+2\bigg)\notag\\
    &-XY^2\bigg(2 \sqrt{\frac{M_1 M_2}{q}}+2 \sqrt{M_1 M_2 q}+2 M_1-2 M_1 M_2+2 M_2+U-2\bigg)+8XY(1-M_1M_2)\notag\\
    &+X\bigg(2 \sqrt{\frac{M_1 M_2}{q}}+2 \sqrt{M_1 M_2 q}+2 M_1+2 M_1 M_2+2 M_2-U+2\bigg)\notag\\
    &+Y\bigg(2 \sqrt{\frac{M_1 M_2}{q}}+2 \sqrt{M_1 M_2 q}-2 M_1+2 M_1 M_2-2 M_2+U+2\bigg)=0\ .
\end{align}
Using Appendix \ref{app:WNF2} for \eqref{eq:su24fz2}, the corresponding Weierstrass normal form reads 
\begin{align*}
    y^2=4x^3-g_2^{4\F/\Z_2}x-g_3^{4\F/\Z_2}\ ,
\end{align*}
with
\begin{align}\label{eq:SU24F-Z2}
    g_2^{4\F/\Z_2}&=\frac1{12}u^4-\frac23\Big(\chi_1^{\SO(7)}\!-27\Big)u^2+16\chi_2^{\SO(7)}u+\frac43\big(\chi_1^{\SO(7)}\big)^2\!-24\chi_1^{\SO(7)}\!+16\chi_3^{\SO(7)}\!+44 \,, \notag\\
    g_3^{4\F/\Z_2}&=\frac1{216}u^6-\frac1{18}\Big(\chi_1^{\SO(7)}+45\Big)u^4-\frac83\chi_2^{\SO(7)}u^3\\
     &~~+\Big(12\chi_1^{\SO(7)}\!+\frac29\big(\chi_1^{\SO(7)}\big)^2\!-\frac83\chi_3^{\SO(7)}\!-\frac{106}3\Big)u^2 +\frac{32}3\Big(\chi_1^{\SO(7)}\chi_2^{\SO(7)}\!-3\chi_2^{\SO(7)}\big)u\cr
    &~~-\frac8{27}\big(\chi_1^{\SO(7)}\big)^3-8\big(\chi_1^{\SO(7)}\big)^2+\frac{32}3\chi_1^{\SO(7)}\chi_3^{\SO(7)}+\frac{136}3\chi_1^{\SO(7)}-32\chi_3^{\SO(7)}-56 \,,\nn
\end{align}
where $U\to\sqrt{M_1M_2}u$ has been used, and $\chi_1^{\SO(7)}$, $\chi_2^{\SO(7)}$ and $\chi_3^{\SO(7)}$ are the characters of the 7-, 8- and 21-dimensional representations of $\SO(7)$, respectively:
\begin{align}\label{eq:cha-B3}
    \chi_1^{\SO(7)}&= \sum_{i=1}^3 \left( y_i + \frac{1}{y_i} \right) + 1 \ ,\notag\\
    \chi_2^{\SO(7)}&= \prod_{i=1}^3 
   \left(\sqrt{y_i}+\frac1{\sqrt{y_i}}\right) \ ,\notag\\
   \chi_3^{\SO(7)}&= \sum_{1 \le i < j \le 3} 
   \left( y_i + \frac{1}{y_i} \right)
   \left( y_j + \frac{1}{y_j} \right) 
   + \sum_{i=1}^3 \left( y_i + \frac{1}{y_i} \right)
   + 3 \ ,
\end{align}
with
\begin{align}
    y_1=\frac1{M_1M_2},\qquad y_2=\frac{M_1}{M_2},\qquad y_3=q \ .
\end{align}
The flavor symmetry $\SO(7)$ is obtained from the corresponding Kodaira fiber configuration~$(I_2;I_2^4I^2_1)$.

\subsubsection{\texorpdfstring{$\mathrm{SU}(2)+6\F/\Z_2$}{SW curve for SU(2)+6FZ2 theory}}\label{sec:SWSU2+6F/Z2}

For the $\SU(2)+6\F/\Z_2$ theory, the Seiberg–Witten curve   can be computed in a similar way, although the dual diagram is non-toric, as explicitly worked out in \cite{Kim:2014nqa}. A $\Z_2$ invariant curve takes the form:
\begin{align}\label{eq:sw-su2-6f-z2}
    F_{\SU(2)+6\bf{F}}(t,w)&=U+\frac1{tw^2}+tw^2+M^2\chi^{\SU(4)}_6\Big(t+\frac1t\Big)\nonumber\\
    & +\big(M\chi^{\SU(4)}_{\bar{4}}+M^3\chi^{\SU(4)}_4\big)\Big(w+\frac1w\Big)-M^3\chi^{\SU(4)}_{\bar4}\Big(\frac tw+\frac wt\Big) \\
    &-M\chi^{\SU(4)}_4\Big(tw+\frac1{tw}\Big)-2M^2\Big(w^2+\frac1{w^2}\Big)+M^4\Big(\frac t{w^2}+\frac{w^2}t\Big)\!=0,\nn
\end{align}
where $M=(qM_1M_2M_3)^{1/4}$ is the fugacity for U(1) and the characters of SU(4) are 
\begin{align}
    \chi_4^{\SU(4)}=\sum_{i=0}^3\widehat{M_i},\quad
    \chi_6^{\SU(4)}=\!\!\sum_{0\leq i<j\leq3}\widehat{M_i}\widehat{M_j},\quad
    \chi_{\bar4}^{\SU(4)}=\sum_{i=0}^3\widehat{M_i}^{-1}\ ,
\end{align}
with $\widehat{M_0}\equiv q/M$ and $\widehat{M_i}\equiv{M_i}/{M}, ~ i=1,2,3$.
By applying the two-to-one map \eqref{eq:21maptw} into \eqref{eq:sw-su2-6f-z2}, one finds
\begin{align}\label{eq:SW6f-reduction-maintext}
F_{\SU(2)+6\bf{F}/\Z_2}(X,Y)=0\quad\Longleftrightarrow\quad \sum_{m,n} C_{m,n} X^m Y^n=0
\end{align}
where non-vanish coefficients are given by
\begin{align}
&C_{0,2} =  - (M-M^{-1})
\left( \chi_4^{\SU(4)} - \chi_{\bar 4}^{\SU(4)} + M - M^{-1} \right) - \chi_6^{\SU(4)} - \mathcal{U},
\cr
&C_{1,1} = 2 \chi_6^{\SU(4)} - 6 (M-M^{-1})^2 +2\mathcal{U},
\cr
&C_{1,2} = 8(M^2-M^{-2})+4(M^{-1} \chi_4^{\SU(4)} - M \chi_{\bar 4}^{\SU(4)}),
\cr
&C_{1,3} = 
(M+M^{-1})
\left( \chi_4^{\SU(4)}+\chi_{\bar 4}^{\SU(4)} - M -M^{-1} \right) - \chi_6^{\SU(4)} + \mathcal{U},
\cr
&C_{2,0} = 
(M-M^{-1})
\left( \chi_4^{\SU(4)} - \chi_{\bar 4}^{\SU(4)} - M+M^{-1} \right) - \chi_6^{\SU(4)} - \mathcal{U},
\cr
&C_{2,1} = 8(M^2-M^{-2})-4(M^{-1} \chi_4^{\SU(4)} - M \chi_{\bar 4}^{\SU(4)}),
\cr
&C_{2,2} = 2 \chi_6^{\SU(4)} - 6 (M+M^{-1})^2 -2\mathcal{U},
\cr
&C_{3,1} = 
- (M+M^{-1})
\left( \chi_4^{\SU(4)}+\chi_{\bar 4}^{\SU(4)} + M+M^{-1} \right) - \chi_6^{\SU(4)} + \mathcal{U}\ ,
\end{align}
with
\begin{align}
    \mathcal{U}=\frac{U}{2M^2}\ .
\end{align}
See also Appendix \ref{app:coodTrans} for more details.

The corresponding Weierstrass normal form is given by 
\begin{align*}
    y^2=4x^3-g^{6\F/\Z_2}_2x-g_3^{6\F/\Z_2}\ ,
\end{align*}
with
\begin{align}\label{eq:sw-su26f-f4}
g_2^{6\F/\Z_2}\!=\,&\frac1{12}u^4-\frac{2}{3} \bigg(\big(\chi _4^{\tF_4}\big)^2+26 \chi^{\tF_4} _4-28 \chi^{\tF_4} _1-2 \chi^{\tF_4} _3+2378\bigg)u^2\\
   &+\bigg(32 \big(\chi_4^{\tF_4}\big)^2
-16 \chi_1^{\tF_4}\chi_4^{\tF_4}
+1952 \chi_4^{\tF_4}-1984 \chi_1^{\tF_4}+16 \chi_2^{\tF_4}-48 \chi_3^{\tF_4}+118864\bigg)u\notag\\
&+\frac{4}{3}\big(\chi_4^{\tF_4}\big)^4
+\frac{16}{3}\big(\chi_4^{\tF_4}\big)^3
-\frac{128}{3}\chi_1^{\tF_4}\big(\chi_4^{\tF_4}\big)^2
-\frac{16}{3}\chi_3^{\tF_4}\big(\chi_4^{\tF_4}\big)^2
+\frac{1232}{3}\big(\chi_4^{\tF_4}\big)^2
\notag\\
&
+\frac{2624}{3}\chi_1^{\tF_4}\chi_4^{\tF_4}
-32\chi_2^{\tF_4}\chi_4^{\tF_4}
-\frac{32}{3}\chi_3^{\tF_4}\chi_4^{\tF_4}
-\frac{164704}{3}\chi_4^{\tF_4}
+\frac{256}{3}\big(\chi_1^{\tF_4}\big)^2
+\frac{64}{3}\big(\chi_3^{\tF_4}\big)^2
\notag\\
&
+\frac{156800}{3}\chi_1^{\tF_4}
-960\chi_2^{\tF_4}+\frac{256}{3}\chi_1^{\tF_4}\chi_3^{\tF_4}
-\frac{5120}{3}\chi_3^{\tF_4}
-\frac{7512320}{3} \ ,\notag\\
& ~\cr
g_3^{6\F/\Z_2}\!=&\,\frac1{216}u^6-4u^5+\bigg(-\frac{1}{18}\left(\chi_4^{\tF_4}\right)^2
+\frac{23}{9}\chi_4^{\tF_4}
-\frac{22}{9}\chi_1^{\tF_4}
+\frac{1}{9}\chi_3^{\tF_4}
+\frac{7811}{9}\bigg)u^4\notag\\
&+\bigg(\frac{80}{3}\left(\chi_4^{\tF_4}\right)^2
+\frac{8}{3}\chi_1^{\tF_4}\chi_4^{\tF_4}
-\frac{1552}{3}\chi_4^{\tF_4}
+\frac{1376}{3}\chi_1^{\tF_4}
-\frac{8}{3}\chi_2^{\tF_4}
-56\chi_3^{\tF_4}
-\frac{255080}{3}\bigg)u^3\notag\\
&+\bigg(\frac{2}{9}\big(\chi_4^{\tF_4}\big)^4
-\frac{88}{9}\big(\chi_4^{\tF_4}\big)^3
+\frac{128}{9}\chi_1^{\tF_4}\big(\chi_4^{\tF_4}\big)^2
-\frac{8}{9}\chi_3^{\tF_4}\big(\chi_4^{\tF_4}\big)^2
-\frac{30728}{9}\big(\chi_4^{\tF_4}\big)^2\notag\\
&
\qquad-\frac{3104}{9}\chi_1^{\tF_4}\chi_4^{\tF_4}
+\frac{16}{3}\chi_2^{\tF_4}\chi_4^{\tF_4}
+\frac{176}{9}\chi_3^{\tF_4}\chi_4^{\tF_4}
+\frac{347152}{9}\chi_4^{\tF_4}
-\frac{448}{9}\big(\chi_1^{\tF_4}\big)^2
\notag\\
&
\qquad -\frac{16}{9}\big(\chi_3^{\tF_4}\big)^2\!-\!\frac{278720}{9}\chi_1^{\tF_4}
\!+416\chi_2^{\tF_4}
\!-\frac{256}{9}\chi_1^{\tF_4}\chi_3^{\tF_4}
\!+\frac{64736}{9}\chi_3^{\tF_4}\!+\frac{38661776}{9}\bigg)u^2\notag\\
&+\bigg(-\frac{128}{3}\big(\chi_4^{\tF_4}\big)^4
-\frac{32}{3}\chi_1^{\tF_4}\big(\chi_4^{\tF_4}\big)^3
+1088\big(\chi_4^{\tF_4}\big)^3
-\frac{4288}{3}\chi_1^{\tF_4}\big(\chi_4^{\tF_4}\big)^2
\notag\\
&\qquad +\frac{32}{3}\chi_2^{\tF_4}\big(\chi_4^{\tF_4}\big)^2 +\frac{544}{3}\chi_3^{\tF_4}\big(\chi_4^{\tF_4}\big)^2
+170208\big(\chi_4^{\tF_4}\big)^2
+\frac{128}{3}\big(\chi_1^{\tF_4}\big)^2\chi_4^{\tF_4}
\notag\\
&\qquad +13376\chi_1^{\tF_4}\chi_4^{\tF_4}
-\frac{1472}{3}\chi_2^{\tF_4}\chi_4^{\tF_4}
 +\frac{64}{3}\chi_1^{\tF_4}\chi_3^{\tF_4}\chi_4^{\tF_4}
-\frac{6464}{3}\chi_3^{\tF_4}\chi_4^{\tF_4}
-1247936\chi_4^{\tF_4}
\notag\\
&\qquad +\frac{15872}{3}\big(\chi_1^{\tF_4}\big)^2
+64\big(\chi_3^{\tF_4}\big)^2
+871296\chi_1^{\tF_4}
-\frac{128}{3}\chi_1^{\tF_4}\chi_2^{\tF_4}
-\frac{63680}{3}\chi_2^{\tF_4}
\notag\\
&\qquad +\frac{8320}{3}\chi_1^{\tF_4}\chi_3^{\tF_4}
-\frac{64}{3}\chi_2^{\tF_4}\chi_3^{\tF_4}
-\frac{1067776}{3}\chi_3^{\tF_4}
-\frac{328387520}{3}\bigg)u\cr
&
-\frac{8}{27}\big(\chi_4^{\tF_4}\big)^6\!
-\!\frac{16}{9}\big(\chi_4^{\tF_4}\big)^5\!
-\!\frac{160}{9}\chi_1^{\tF_4}\big(\chi_4^{\tF_4}\big)^4\!
+\!\frac{16}{9}\chi_3^{\tF_4}\big(\chi_4^{\tF_4}\big)^4\!
+\frac{5200}{3}\big(\chi_4^{\tF_4}\big)^4\!
\notag\\
&
+\frac{5120}{9}\chi_1^{\tF_4}\big(\chi_4^{\tF_4}\big)^3
-\frac{64}{3}\chi_2^{\tF_4}\big(\chi_4^{\tF_4}\big)^3
+\frac{64}{9}\chi_3^{\tF_4}\big(\chi_4^{\tF_4}\big)^3
+\frac{1024}{9}\big(\chi_1^{\tF_4}\big)^2\big(\chi_4^{\tF_4}\big)^2
\notag\\
&
+\frac{64}{9}\big(\chi_3^{\tF_4}\big)^2\big(\chi_4^{\tF_4}\big)^2 +\frac{314752}{9}\chi_1^{\tF_4}\big(\chi_4^{\tF_4}\big)^2
-\frac{2048}{3}\chi_2^{\tF_4}\big(\chi_4^{\tF_4}\big)^2
+\frac{640}{9}\chi_1^{\tF_4}\chi_3^{\tF_4}\big(\chi_4^{\tF_4}\big)^2
\notag\\
&
-\frac{67840}{9}\chi_3^{\tF_4}\big(\chi_4^{\tF_4}\big)^2
-\frac{8882752}{3}\big(\chi_4^{\tF_4}\big)^2
-\frac{20992}{9}\big(\chi_1^{\tF_4}\big)^2\chi_4^{\tF_4}
+\frac{128}{9}\big(\chi_3^{\tF_4}\big)^2\chi_4^{\tF_4}
\notag\\
&
+\frac{256}{3}\chi_1^{\tF_4}\chi_2^{\tF_4}\chi_4^{\tF_4}
+\frac{31360}{3}\chi_2^{\tF_4}\chi_4^{\tF_4}
-\frac{10240}{9}\chi_1^{\tF_4}\chi_3^{\tF_4}\chi_4^{\tF_4}
+\frac{128}{3}\chi_2^{\tF_4}\chi_3^{\tF_4}\chi_4^{\tF_4}
\notag\\
&
-\frac{815296}{27}\big(\chi_4^{\tF_4}\big)^3+\frac{532736}{9}\chi_3^{\tF_4}\chi_4^{\tF_4}
-\frac{4096}{27}\big(\chi_1^{\tF_4}\big)^3
-\frac{512}{27}\big(\chi_3^{\tF_4}\big)^3
-\frac{1254400}{9}\big(\chi_1^{\tF_4}\big)^2
\notag\\
&
-\frac{1024}{9}\chi_1^{\tF_4}\big(\chi_3^{\tF_4}\big)^2
+\frac{20480}{9}\big(\chi_3^{\tF_4}\big)^2 
-\frac{73661440}{9}\chi_1^{\tF_4}
+2560\chi_1^{\tF_4}\chi_2^{\tF_4}
+352000\chi_2^{\tF_4}
\notag\\
&-\frac{2048}{9}\big(\chi_1^{\tF_4}\big)^2\chi_3^{\tF_4}
-\frac{586240}{9}\chi_1^{\tF_4}\chi_3^{\tF_4}
-\frac{1223168}{9}\chi_1^{\tF_4}\chi_4^{\tF_4}+1280\chi_2^{\tF_4}\chi_3^{\tF_4}\cr
&
+\frac{131765120}{9}\chi_4^{\tF_4}+\frac{55041280}{9}\chi_3^{\tF_4}
+\frac{30114656000}{27}\ ,\nn
\end{align}
where 
$\chi_i^{\tF_4}=\chi_i^{\tF_4}(q,M_1,M_2,M_3)$, $i=1,2,3,4$, are the characters of 52-,1274-, 273- and 26-dimensional representations of $\tF_4$, respectively, and
$\mathcal{U}\to \frac12u$. 
The flavor symmetry $F_4$ is obtained from the corresponding Kodaira fiber configuration~$(I_1;I_2^4I^3_1)$. 

\vspace{0.5cm}


\subsection{SW curves from the \texorpdfstring{$\D_4$}{D4} tuning of the inserted 7-branes \texorpdfstring{$\mathcal{S}_{\D_4}$}{SD4}}\label{sec:SW-EN-SCFT}

As discussed earlier, the $\Z_2$ S-folds of $\SU(2)+2n\F$ ($n=0,1,2,3$) are realized as the 7-brane insertion associated with $\mathcal{S}_{\D_4}=\mathbf{A}^4\mathbf{BC}$, yielding the $\E_{5+n}$ fixed point configurations. They contain the masses associated with four $\mathbf{A}~[1,0]$ 7-branes. These masses are confined to the position of $\mathbf{BC}$ or an O7$^-$-plane that is located at the center of the compact face of the 5-brane webs. 
However, it does not mean that the corresponding K\"ahler parameters $\mathsf{M}_k~(k=1,2,3,4)$ represent massless ($\mathsf{M}_k=1$), rather they should be responsible for the twisted sectors arising due to the S-folding, implying that the K\"ahler parameters are tuned to be special values other than the massless case. Inspired by \cite{Hayashi:2023boy, Kim:2024ufq}, where a D-brane charge neutral 7-brane combination $\mathrm{O7}^-+4\mathrm{D7}$ is tuned to effectively represent a $\Z_2$-reflection, we propose that four masses $\mathsf{M}_i$ are stuck at 
\begin{align}\label{eq:tuning-condi}
    \mathsf{M}_1=-\mathsf{M}_2=\mathsf{M}_3=-\mathsf{M}_4=1\ ,
\end{align}
up to the Weyl symmetry exchanging $\mathsf{M}_k$.

This locus can be understood from the T-dual IIA picture where two O6$^-$-planes sit at $w=\pm1$ and each carries two D6-branes so that the total D-brane charges are locally canceled, which translates under T-duality to the flavor branes at $ w=\mathsf{M}_k=\pm1$.

As described based on the cubic prepotential in section~\ref{sec:Z2-S-folds}, the instanton factor $M_0$ of the $\SU(2)+(4+n)\F$ theories is also tuned as \eqref{eq:m0/2andmi}:
\begin{align}\label{eq:instTuning}
m_0^{\SU(2)+(4+n)\F} = \frac{1}{2}m_0^{\SU(2)+2n\F}\quad \Longrightarrow\quad  M_0 = \sqrt{q}\ , 
\end{align}
where $M_0 = e^{-m_0^{\SU(2)+(4+n)\F}}$ and $q=e^{-m_0^{\SU(2)+2n\F}}$.
From the perspective of the Seiberg–Witten curve for the $\Z_2$ S-folds, it then amounts to the special tuning of four masses \eqref{eq:tuning-condi} of the Seiberg–Witten curves for $\E_{5+n}$ SCFTs (the $\D_4$-tuning):
\begin{align}\label{eq:proposal}
\text{SW}_{\E_{5+n}}~~ \xrightarrow{\text{$\D_4$-tuning:}~\eqref{eq:tuning-condi} ~ \& ~ \eqref{eq:instTuning}} ~~\text{SW}_{\SU(2)+2n\F/\Z_2}\ .
\end{align}
In other words, we claim that one obtains 
the Seiberg–Witten curves for the $\Z_2$ S-folds, $\SU(2)+2n\F/\Z_2$, from
the Seiberg–Witten curves for the $\E_{5+n}$ SCFTs by tuning the four masses 
$\mathsf{M}_k$ and the instanton factor $M_0$ as \eqref{eq:tuning-condi}
and \eqref{eq:instTuning}, respectively.
As the $\E_{5+n}$ theories have $(n+5)$ flavor fugacities which include mass parameters and instanton, the special tuning, \eqref{eq:tuning-condi} and \eqref{eq:instTuning}, is to set the flavor fugacities to~be 
\begin{align}\label{eq:special tuning}
 \left\{\, \sqrt{q}, \, 1 ,-1,\, 1, -1 ;\, M_{i=1,\dots, n}\,\right\}  \ , 
\end{align}
leaving $n$ mass parameters $M_{i}$ untuned, which will be identified with the independent mass parameters of the $\SU(2)+2n\F/\Z_2$ theories. The resulting flavor symmetries are then expected to be expressed in terms of the characters of a $\text{rank-}n$ subgroup of $\E_{5+n}$, as five of $5+n$ are tuned. We note that for the mass parameters, one needs to use physical masses that are measured from the position of $\mathbf{BC}=\mathrm{O7}^-$. See Appendix \ref{sec:phy-mass-para} for more details. 

In what follows, given the Seiberg–Witten curves for the $\E_{5+n}$ theory \cite{Minahan:1997ch, Eguchi:2002fc, Eguchi:2002nx, Huang:2013yta, Kim:2014nqa,Closset:2021lhd} expressed in the Weierstrass normal form, we tune four mass parameters \eqref{eq:tuning-condi} associated with the 7-brane insertion $\mathcal{S}_{\D_4}=\mathbf{A}^4\mathbf{BC}$ to obtain the Seiberg–Witten curves for the $\SU(2)+2n\F/\Z_2$ theories, and compare them with those obtained by two-to-one map in the preceding section \ref{sec:2-1}.

\subsubsection{\texorpdfstring{$\E_5$ to $\SU(2)_0/\Z_2$}{E5 to SU20/Z2}}
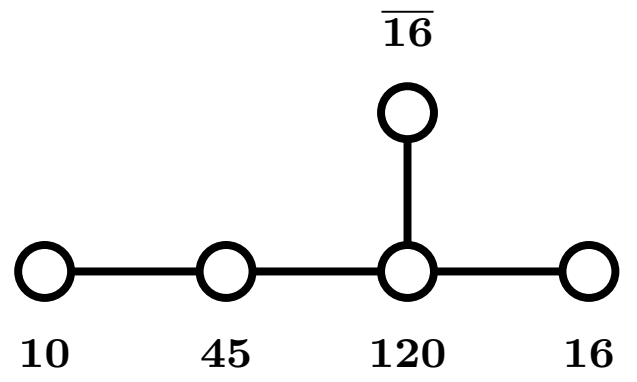
\begin{figure}
    \centering
\begin{tikzpicture}[
  x=10mm,
  y=10mm,
  dynkin edge/.style={
    draw,
    line width=1.5pt,
    line cap=round,
    line join=round
  },
  dynkin node/.style={
    circle,
    draw,
    fill=white,
    minimum size=3.5mm,
    inner sep=0pt,
    outer sep=0pt,
    line width=1.5pt
  },
  root label/.style={
    font=\scriptsize,
    inner sep=1pt,
    label distance=1.5mm
  },
  diagram title/.style={
    font=\normalsize,
    inner sep=1pt
  }
]

\coordinate (a1) at (-2.4,0);
\coordinate (a3) at (-1.2,0);
\coordinate (a4) at (0,0);
\coordinate (a5) at (1.2,0);
\coordinate (a2) at (0,1.05);

\draw[dynkin edge] (a1) --(a3) -- (a4) -- (a5) ;
\draw[dynkin edge] (a4) -- (a2) ;

\node[dynkin node, label={[root label]below:$\bf 10$}] at (a1) {};
\node[dynkin node, label={[root label]above:$\bf \overline{16}$}] at (a2) {};
\node[dynkin node, label={[root label]below:$\bf 45$}] at (a3) {};
\node[dynkin node, label={[root label]below:$\bf 120$}] at (a4) {};
\node[dynkin node, label={[root label]below:$\bf16$}] at (a5) {};

\end{tikzpicture}
\caption{Dynkin diagram for $\E_5$. The numbers denote the dimension of the fundamental weights associated with each node.}
\label{fig:E5-Dynkin}
\end{figure}

The Seiberg–Witten curve for the rank-1 $\E_5$ theory in the Weierstrass normal form is given by
\begin{align*}
    y^2=4x^3-g_2^{\E_5}x-g_3^{\E_5}\ ,
\end{align*}
with
\begin{align}\label{eq:sw-e5-curve}
    g_2^{\E_5}=&\, \frac1{12}u^4- \frac{2}{3}\chi_1^{\E_5}u^2
- 2\chi_2^{\E_5}u
+\frac{4}{3}\chi_1^{\E_5}\chi_1^{\E_5}-4 \chi_3^{\E_5}+4 {}\ , \notag\\
g_3^{\E_5}=& \, \frac{1}{216} u^6  -\frac{1}{18}\chi_1^{\E_5}u^4
    - \frac{1}{6}\chi_2^{\E_5}u^3
    + \bigg(\frac{2}{9}\big(\chi_1^{\E_5}\big)^2  -\frac{1}{3} \chi_3^{\E_5}
       -\frac{11}{3}\bigg)u^2   + \bigg(\frac{2}{3}\chi_1^{\E_5}\chi_2^{\E_5} -4 \chi_5^{\E_5} \bigg)u\notag\\
 &- \frac{8}{27}\chi_1^{\E_5}\chi_1^{\E_5}\chi_1^{\E_5}
   +\frac{4}{3}\chi_1^{\E_5}\chi_3^{\E_5}
   -\frac{4 }{3}\chi_1^{\E_5}
 +\chi_2^{\E_5}\chi_2^{\E_5}
   -4\chi_4^{\E_5} \ ,
\end{align}
where $\chi_1^{\E_5}, \chi_2^{\E_5}, \dots, \chi_5^{\E_5}$ are the characters of $10$-, $\overline{16}$-, $45$-, $120$-, and $16$-dimensional representations of the $\E_5=\text{SO}(10)$ group, respectively. The node associated with each
fundamental weight is denoted in the Dynkin diagram of $\E_5$ in figure \ref{fig:E5-Dynkin}. 
The characters for the fundamental weights of $\E_5$ can be expressed in terms of those of $\SO(8)\times \U(1)$ by the embedding\footnote{Note that throughout this paper, we use the convention $M_0=e^{-m_0}$, where $M_0$ carries $\U(1)_I$ charge $+1$.}:
\begin{align}\label{eq:E5decompositions}
\E_5=\SO(10) &\supset \SO(8)\times \U(1)\\
\mathbf{10} & =
\mathbf8_{\bf v}{}_{(0)}+\mathbf1_{(+1)}+\mathbf1_{(-1)}\cr
\mathbf{\overline{16}}& = \mathbf8_{\bf s}{}_{(-\frac12)}+\mathbf8_{\bf c}{}_{(+\frac12)}\cr
\mathbf{45}& =\mathbf{28}_{(0)}+\mathbf1_{(0)}
+\mathbf8_{\bf v}{}_{(+1)}+\mathbf8_{\bf v}{}_{(-1)}\cr
%
\mathbf{120} & =\mathbf8_{\bf s}{}_{(0)}
\times\mathbf8_{\bf c}{}_{(0)}
+\mathbf{28}_{(+1)}+\mathbf{28}_{(-1)}\cr
\mathbf{16}& =
\mathbf8_{\bf s}{}_{(+\frac12)}+\mathbf8_{\bf c}{}_{(-\frac12)}\, \nn
\end{align}
where $\SO(8)$ is the symmetry for the four hypermultiplets of $\SU(2)+4\F$ with physical masses $\mathsf{M}_i$ and the U(1) instanton charge is denoted by the subscript $(n)$. The explicit form of the characters $\chi_i^{\E_5}$ is given in~\eqref{eq:sw-e5-cha}, up to the triality of SO(8) with $\mathbf{v}\to\mathbf{s}\to\mathbf{c}\to \mathbf{v}$. 
See Appendix~\ref{app:Characters} for more details.

Upon the application of the special tuning \eqref{eq:special tuning}, or more explicitly, 
\begin{align}\label{eq:stforSU2+4F}
    \{\, M_0,\, \mathsf{M}_{i=1,2,3,4} \}=\{\sqrt{q}, \, 1,-1,\,1,-1\, \}\ ,
\end{align}
where we identify the instanton factor $M_0$ of $\SU(2)+4\F$ with the instanton factor $q$ of SU(2)$_0$ theory as $M_0=\sqrt{q}$.  These $\E_5$ characters become the SU(2) characters: 
\begin{align}\label{eq:e5-tuned-cha}
    \chi_1^{\E_5}\to\chi_1^{\SU(2)},\quad\chi_2^{\E_5}\to 0,\quad\chi_3^{\E_5}\to-3,\quad \chi_4^{\E_5}\to-4\chi_1^{\SU(2)},\quad \chi_5^{\E_5}\to0 \ ,
\end{align}
where $\chi_1^{\SU(2)}=\sqrt{q}+1/\sqrt{q}$. This can be readily understood as follows. As the special tuning \eqref{eq:tuning-condi} is imposed on the $\SO(8)$ characters with the alternating values $\pm1$, the characters of $\SO(8)$ of degree-odd, $\mathbf{8}_{\bf v, s, c}$, vanish and only the characters of degree-even, $\mathbf{28}, \mathbf{1}$, survive, yielding that only remaining parameter is the SU(2) character associated with the instanton factor $M_0$ for $\SU(2)+4\F$, which becomes that of the SU(2) character for the $\SU(2)_0/\Z_2$ theory, given in \eqref{eq:su2-char}:
\begin{align}\label{eq:SU(2)M5}
    \chi_1^{\SU(2)}(M_0){}\Big|_{M_0\to \sqrt{q}} = \sqrt{q}+\frac{1}{\sqrt{q}}= \chi_1^{\SU(2)/\Z_2}(q)\ .
\end{align}  
It then follows from \eqref{eq:e5-tuned-cha} and \eqref{eq:SU(2)M5} that $g_2^{\E_5}$ and $g_3^{\E_5}$ given in \eqref{eq:sw-e5-curve} become exactly the same as $g_2^{\SU(2)_0/\Z_2}$ and $g_3^{\SU(2)_0/\Z_2}$ in \eqref{eq:g2-su2-0}:
\begin{align}
    g_{2,3}^{\E_5}({M}_i; u) ~\xrightarrow{\eqref{eq:stforSU2+4F}}  ~  g_{2,3}^{\SU(2)_0/\Z_2} (q; u)\ ,
\end{align}
hence, confirming our proposal \eqref{eq:proposal},
\begin{align}
    \text{SW}_{\E_5} ~\xrightarrow{\text{$\D_4$-tuning}:\,
    \eqref{eq:stforSU2+4F}}  ~  \text{SW}_{\SU(2)_0/\Z_2}\ 
.
\end{align}

\subsubsection{\texorpdfstring{$\E_6$ to $\SU(2)+2\F/\Z_2$}{E6 to SU(2)+2F/{Z}2}}
The Seiberg–Witten curve for the rank-1 $\E_6$ theory 
is given~by
\begin{align*}
    y^2=4x^3-g_2^{\E_6}x-g_3^{\E_6}\ ,
\end{align*}
with
\begin{align}\label{eq:E6-Weierstrass}
     g_2^{\E_6}&=\frac{1}{12}u^4 - \frac{2}{3}\chi_1^{\E_6}u^2
- \big(2 \chi_2^{\E_6}-12  \big)u
+\frac{4} {3}\chi_1^{\E_6}\chi_1^{\E_6}-4 \chi_3^{\E_6}+4 {\chi_6^{\E_6}}, \\
g_3^{\E_6}&= \frac{1}{216} u^6  -\frac{1}{18}\chi_1^{\E_6}u^4
    - \bigg(\frac{1}{6}\chi_2^{\E_6}+3 \bigg)u^3
    + \bigg(\frac{2}{9}\chi_1^{\E_6}\chi_1^{\E_6}  -\frac{1}{3} \chi_3^{\E_6}
       -\frac{11}{3}\chi_6^{\E_6}\bigg)u^2 \notag\\
       &\quad + \bigg(\frac{2}{3}\chi_1^{\E_6}\chi_2^{\E_6} -4 \chi_5^{\E_6} \bigg)u-\frac{8}{27}\chi_1^{\E_6}\chi_1^{\E_6}\chi_1^{\E_6}+\frac{4}{3}\chi_1^{\E_6}\chi_3^{\E_6}
   -\frac{4}{3}\chi_1^{\E_6}\chi_6^{\E_6} +\chi_2^{\E_6}\chi_2^{\E_6}-4\chi_4^{\E_6},
   \notag
\end{align}
where $\chi_1^{\E_6}, \dots, \chi_6^{\E_6}$ are  the characters of the $27$-, $78$-, $\overline{351}$-, $2925$-, ${351}$-, and $\overline{27}$-dimensional representations of $\E_6$, respectively.
Here, as stated in Appendix \ref{app:Characters}, the $\E_6$ characters are expressed in 
terms of the characters of the fundamental weights of $\SO(10)\times \U(1)_I$ where $\chi_i^{\SO(10)}$ are respectively the $10$-, $\overline{16}$-, $45$-, $120$-, and $16$-dimensional representations, and the $\U(1)_I$ charges are denoted by the subscript numbers:  
\begin{align}
\E_6&\supset \SO(10)\times \U(1)_I\cr
\mathbf{27}&=
\mathbf1_{(\frac43)}
+\mathbf{10}_{(-\frac23)}
+\mathbf{16}_{(\frac13)}
\cr
\mathbf{78}&=
\mathbf1_{(0)}
+\mathbf{16}_{(1)}
+\overline{\mathbf{16}}_{(-1)}
+\mathbf{45}_{(0)}
\cr
\overline{\mathbf{351}}&=
\mathbf{10}_{(\frac23)}
+(\mathbf{10}\times{\mathbf{16}})_{(-\frac13)}
+\mathbf{16}_{(\frac53)}
+\mathbf{45}_{(-\frac43)}
+\mathbf{120}_{(\frac23)}
\cr
\mathbf{2925}&= (\mathbf{10}\times \mathbf{120})_{(0)}+
({\mathbf{16}}\times\mathbf{45})_{(-1)}
+\mathbf{45}_{(0)}
+(\mathbf{45}\times\overline{\mathbf{16}})_{(1)}
+\mathbf{120}_{(2)}
+\mathbf{120}_{(-2)}
\cr
\mathbf{351}&=
\mathbf{10}_{(-\frac23)}
+(\mathbf{10}\times\overline{\mathbf{16}})_{(\frac13)}
+\overline{\mathbf{16}}_{(-\frac53)}
+\mathbf{45}_{(\frac43)}
+\mathbf{120}_{(-\frac23)}
\cr
\overline{\mathbf{27}}&=
\mathbf1_{(-\frac43)}
+\mathbf{10}_{(\frac23)}
+\overline{\mathbf{16}}_{(-\frac13)}
\ ,
\end{align}

\begin{figure}
    \centering
\begin{tikzpicture}[
  x=10mm,
  y=10mm,
  dynkin edge/.style={
    draw,
    line width=1.5pt,
    line cap=round,
    line join=round
  },
  dynkin node/.style={
    circle,
    draw,
    fill=white,
    minimum size=3.5mm,
    inner sep=0pt,
    outer sep=0pt,
    line width=1.5pt
  },
  root label/.style={
    font=\scriptsize,
    inner sep=1pt,
    label distance=1.5mm
  },
  diagram title/.style={
    font=\normalsize,
    inner sep=1pt
  }
]

\coordinate (a1) at (-2.4,0);
\coordinate (a3) at (-1.2,0);
\coordinate (a4) at (0,0);
\coordinate (a5) at (1.2,0);
\coordinate (a6) at (2.4,0);
\coordinate (a2) at (0,1.05);

\draw[dynkin edge] (a1) --(a3) -- (a4) -- (a5) -- (a6);
\draw[dynkin edge] (a4) -- (a2) ;

\node[dynkin node, label={[root label]below:$\mathbf{27}$}] at (a1) {};
\node[dynkin node, label={[root label]above:$\mathbf{78}$}] at (a2) {};
\node[dynkin node, label={[root label]below:$\mathbf{\overline{351}}$}] at (a3) {};
\node[dynkin node, label={[root label]below:$\mathbf{2925}$}] at (a4) {};
\node[dynkin node, label={[root label]below:$\mathbf{351}$}] at (a5) {};
\node[dynkin node, label={[root label]below:$\mathbf{\overline{27}}$}] at (a6) {};

\end{tikzpicture}
\caption{Dynkin diagram for $\E_6$. The numbers denote the dimension of the fundamental weights associated with each node.}
\label{fig:E6-Dynkin}
\end{figure}
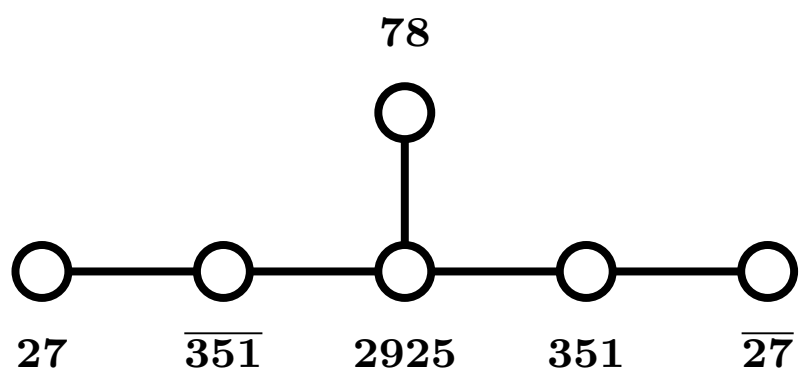
As shown in the Dynkin diagram of $\E_6$ in figure~\ref{fig:E6-Dynkin}, the nodes associated with $\alpha_{2,3,4,5,6}$ represent $\SO(10)$ characters.

The special tuning 
\eqref{eq:special tuning} 
\begin{align}
\big\{\sqrt{q},\, 1, -1, 1, -1;\, M_1\big\}\ ,
\end{align}
where the instanton factor $M_0$ and four masses for the $\SU(2)+5\F$ theory are respectively tuned to be $M_0=\sqrt{q}$ and $\mathsf{M}_i=\pm1$, 
yields 
\begin{align}\label{eq:e6-after-char} 
    \chi_1^{\E_6}&\to\chi_1^{\SU(3)}\ ,\cr
    \chi_2^{\E_6}&\to-2\ ,\cr 
    \chi_3^{\E_6}&\to-3\chi_2^{\SU(3)}\ ,\cr
    \chi_4^{\E_6}&\to-4\chi_1^{\SU(3)}\chi_2^{\SU(3)}+1\ , \cr
   \chi_5^{\E_6}&\to-3\chi_1^{\SU(3)}\ ,\cr
   \chi_6^{\E_6}&\to\chi_2^{\SU(3)}\ ,
\end{align}
where $\chi_1^{\SU(3)}$ and $\chi_2^{\SU(3)}$ are the characters of the two fundamental weights of SU(3) in terms of untuned mass $M_1$ and the instanton factor $M_0=\sqrt{q}$ of $\SU(2)+(4+1)\F$, 
\begin{align}\label{eq:enhancedSU3}
    \chi_1^{\SU(3)} &= 
    q^{2/3} +M_1q^{-1/3} + M_1^{-1}q^{-1/3}, 
    \cr
    \chi_2^{\SU(3)} &  =  
    q^{-2/3} +M_1^{-1}q^{1/3} + M_1q^{1/3}\ , 
\end{align}
which are exactly the same as the flavor SU(3) characters of the $\SU(2)+2\F/\Z_2$ theory, given in \eqref{eq:su3-cha}.
 It follows that $g_2^{\E_6}$ and $g_3^{\E_6}$ in \eqref{eq:E6-Weierstrass} reduce to $g_2^{2\F/\Z_2}$ and $g_3^{2\F/\Z_2}$ for the $\SU(2)+2\F/\Z_2$ theory in \eqref{eq:su2+2Fz2main}. This, therefore, confirms our claim \eqref{eq:proposal}.

 We note that although it is obvious from \eqref{eq:enhancedSU3} that the resulting flavor symmetry is an $\SU(3)$, it is worth discussing how $\SU(3)$ is obtained. The special tuning freezes $\SO(8)$ inside $\SO(10)$, and the remaining symmetry gives rise to  an $\SU(2)$ such that together with $\U(1)_I$ it forms an $\SU(3)$, 
 \begin{align}
     \E_6 \supset \SO(10)\times \U(1)_I ~\xrightarrow{\text{$\D_4$-tuning}}  ~  \SU(2)\times \U(1)_I \subset \SU(3)\ .
 \end{align}
\subsubsection{\texorpdfstring{$\E_7$ to $\SU(2)+4\F/\Z_2$}{E7 to SU(2)+4F/{Z}2}}
\begin{figure}
    \centering
\begin{tikzpicture}[
  x=10mm,
  y=10mm,
  dynkin edge/.style={
    draw,
    line width=1.5pt,
    line cap=round,
    line join=round
  },
  dynkin node/.style={
    circle,
    draw,
    fill=white,
    minimum size=3.5mm,
    inner sep=0pt,
    outer sep=0pt,
    line width=1.5pt
  },
  root label/.style={
    font=\scriptsize,
    inner sep=1pt,
    label distance=1.5mm
  },
  diagram title/.style={
    font=\normalsize,
    inner sep=1pt
  }
]

\coordinate (a1) at (-2.4,0);
\coordinate (a3) at (-1.2,0);
\coordinate (a4) at (0,0);
\coordinate (a5) at (1.2,0);
\coordinate (a6) at (2.4,0);
\coordinate (a2) at (0,1.05);
\coordinate (a7) at (3.6,0);

\draw[dynkin edge] (a1) --(a3) -- (a4) -- (a5) -- (a6) -- (a7);
\draw[dynkin edge] (a4) -- (a2);

\node[dynkin node, label={[root label]below:$\bf 133$}] at (a1) {};
\node[dynkin node, label={[root label]above:$\bf 912$}] at (a2) {};
\node[dynkin node, label={[root label]below:$\bf 8645$}] at (a3) {};
\node[dynkin node, label={[root label]below:$\bf 365750$}] at (a4) {};
\node[dynkin node, label={[root label]below:$\bf 27664$}] at (a5) {};
\node[dynkin node, label={[root label]below:$\bf 1539$}] at (a6) {};
\node[dynkin node, label={[root label]below:$\bf 56$}] at (a7) {};

\end{tikzpicture}
\caption{Dynkin diagram for $\E_7$. The numbers denote the dimension of the fundamental weights associated with each node.}
\label{fig:E7-Dynkin}
\end{figure}
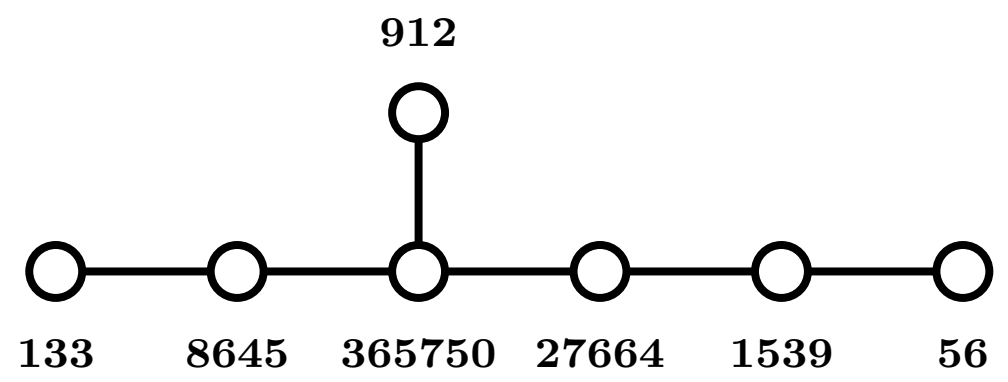

The Seiberg–Witten curve for the rank-1 $\E_7$ theory is given by
\begin{align*}
    y^2=4x^3-g_2^{\E_7}x-g_3^{\E_7}\ ,
\end{align*}
with
\begin{align}\label{eq:E7}
    g_2^{\E_7}=&~\frac{1}{12}u^4 - \bigg(\frac{2}{3}\chi_1^{\E_7}-\frac{50}{3}\bigg)u^2
- \big(2 \chi_2^{\E_7}-12 \chi_7^{\E_7} \big)u
+\frac{4} {3}\big(\chi_1^{\E_7}\big)^2-\frac83\chi_1^{\E_7} -4 {\chi_3^{\E_7}}+4 {\chi_6^{\E_7}}+ \frac{28}{3} \,, \nonumber\\
g_3^{\E_7}=&~\frac{1}{216} u^6  -\bigg(\frac{1}{18}\chi_1^{\E_7}+\frac{47}{18}\bigg)u^4
    - \bigg(\frac{1}{6}\chi_2^{\E_7}+3\chi_7^{\E_7} \bigg)u^3
    \notag\\
    &+\left(\frac{2}{9}\chi_1^{\E_7}\chi_1^{\E_7} +\frac{20}{9}\chi_1^{\E_7}  -\frac{1}{3} \chi_3^{\E_7}
       -\frac{11}{3}\chi_6^{\E_7}
       -\frac{196}{9}\right)u^2\notag\\
    &+ \bigg(\frac{2}{3}\chi_1^{\E_7}\chi_2^{\E_7} +\frac{10}{3}\chi_2^{\E_7}-4 \chi_5^{\E_7} - 16\chi_7^{\E_7}\bigg)u
  \notag\\
 & - \frac{8}{27}\big(\chi_1^{\E_7}\big)^3+\big(\chi_2^{\E_7}\big)^2-\frac{28}{9}\big(\chi_1^{\E_7}\big)^2+\frac{4}{3}\chi_1^{\E_7}\chi_3^{\E_7}
   -\frac{4 }{3}\chi_1^{\E_7}\chi_6^{\E_7}+\frac{40}{9}\chi_1^{\E_7}\cr
   &+\frac{20}{3}\chi_3^{\E_7}
   -4\chi_4^{\E_7} -\frac83\chi_6^{\E_7} -\frac{136}{27} \ ,
   \end{align}
 where $\chi_i^{\E_7}$ are the characters of the fundamental weights of $\E_7$ which are of dimension 133, 912, 8645, 365750, 27664, 1539, and 56, respectively. The node associated with each fundamental weight is denoted in the Dynkin diagram of $\E_7$ in figure~\ref{fig:E7-Dynkin}.
Their explicit forms are given in Appendix \ref{app:Characters}, which are expressed in terms of the characters of the fundamental weights of $\SO(12)\times \SU(2) \supset \SO(12)\times \U(1)_I$:
\begin{align}
\E_7&\supset \SO(12)\times \SU(2)\cr
\mathbf{133}&= \bf (1,2^2)+(32,2)+(66,1)-(1,1)\cr
\mathbf{912}&= \bf (32,3)+(12,2)+(220,2)+(12\times\overline{32},1)-\mathrm{2}(32,1)\cr
\mathbf{8645}&= \bf
(\overline{32},2^3)
+(66,2^2)+(495,2^2)+(1,2^2)+(66\times\overline{32},2)-\mathrm{2}(\overline{32},2)
\cr
&\quad \bf
+(\overline{32}^2,1)-\mathrm{2}(66,1)+(12\times 220,1)-\mathrm{3}(495,1)-\mathrm{2}(1,1)\cr
\mathbf{365750}&= \bf
(495,2^4)
+(220\times 32,2^3)+(12\times 220,2^2)+(66\times 495,2^2)\cr
&\quad \bf
-\mathrm{4}(495,2^2)
+(12\times 220\times\overline{32},2)
-\mathrm{3}(220\times 32,2)
+(12^2,1)-(66^2,1)\cr
&\quad \bf
+(220^2,1)
+(66\times\overline{32}^2,1)
+(32^2,1)-\mathrm{2}(66,1)+(12^2\times 495,1)\cr
&\quad \bf
-\mathrm{4}(66\times 495,1)
+(495,1)-(12\times\overline{32}\times 32,1)-(1,1)\cr
\mathbf{27664}&= \bf (220,2^3)+ (66\times 32, 2^2)+ (12\times 66, 2)-\mathrm{3}(220, 2)+(12\times 495, 2)\cr
&\quad \bf  + (220\times\overline{32}, 1)+(12^2\times32, 1)-\mathrm{3}(66\times 32,1 )\cr
\mathbf{1539}&= \bf (12^2, 1)+(12\times 32, 2)+ (66, 2^2)-\mathrm{2}(66, 1)+(495, 1) \cr
\mathbf{56}&= \bf (32,1) + (12,2)\ ,
\end{align}
where the fundamental weights of $\SO(12)$ are $\{ \bf 12, 66, 220, 495, \overline{32}, 32 \}$.

The special tuning \eqref{eq:special tuning} for this case is
\begin{align}
\{ M_0, \mathsf{M}_{i};\, M_1, M_2
\} = \{ \sqrt{q}, 1,\, -1, 1, \, -1;\, M_1, M_2\}\ .
\end{align}
where the instanton factor $M_0$ and four masses for the $\SU(2)+6\F$ theory are respectively tuned to be $M_0=\sqrt{q}$ where $q$ is the instanton factor for $\SU(2)+4\F$. 
The tuning then yields 
\begin{align}\label{eq:E7-tun-cha}
    \chi_1^{\E_7}&\to\chi_1^{\SO(7)}-2 \ ,\cr
    \chi_2^{\E_7}&\to-2\chi_2^{\SO(7)} \ ,\cr
    \chi_3^{\E_7}&\to-3\chi_3^{\SO(7)}+\chi_1^{\SO(7)}-3\ , \notag\\
\chi_4^{\E_7}&\to3\big(\chi_1^{\SO(7)}\big)^2+\big(\chi_2^{\SO(7)}\big)^2-4\chi_1^{\SO(7)}\chi_3^{\SO(7)}-9\chi_1^{\SO(7)}+5\chi_3^{\SO(7)}+5\ ,\notag\\
    \chi_5^{\E_7}&\to-3\chi_1^{\SO(7)}\chi_2^{\SO(7)}+3\chi_2^{\SO(7)},\cr
    \chi_6^{\E_7}&\to-3\chi_1^{\SO(7)}+\chi_3^{\SO(7)}+3\ ,\cr
    \chi_7^{\E_7}&\to\chi_2^{\SO(7)}\ ,
    \end{align}
where $\chi_i^{\SO(7)},i=1,2,3$, are the same as those given by \eqref{eq:cha-B3}. It follows that $g_2^{\E_7}$ and $g_3^{\E_7}$ in \eqref{eq:E7} reduce to $g_2^{4\F/\Z_2}$ and $g_3^{4\F/\Z_2}$ for the $\SU(2)+4\F/\Z_2$ theory in \eqref{eq:SU24F-Z2}. This, therefore, confirms our claim \eqref{eq:proposal}.

We note that the special tuning freezes $\SO(8)$ inside $\SO(12)$, and the remaining symmetry gives rise to  an $\SU(2)\times \SU(2)$ such that together with $\SU(2)_I$ the symmetry is further enhanced to an $\SO(7)$, 
\begin{align}
      \E_7 \supset \SO(12)\times \SU(2)_I ~\xrightarrow{\text{$\D_4$-tuning}}  ~  \SU(2)\times\SU(2) \times \SU(2)_I \subset \SO(7)\ .
\end{align}

\subsubsection{\texorpdfstring{$\E_8$ to $\SU(2)+6\F/\Z_2$}{E8 to SU(2)+6F/{Z}2}}

The Seiberg–Witten curve for the rank-1 $\E_8$ theory is given by 
\begin{align*}
    y^2=4x^3-g_2^{\E_8}x-g_3^{\E_8}\ ,
\end{align*}
with
\begin{align}\label{eq:E8}
g_2^{\E_8}=&~\frac{1}{12}u^4 - \Big(\frac{2}{3}\chi_1^{\E_8}-\frac{50}{3}\chi_8^{\E_8}+1550\Big)u^2 
\notag\\
&+\big(70 \chi_1^{\E_8}-2 \chi_2^{\E_8}+12 \chi_7^{\E_8} -1840 {\chi_8^{\E_8}}+115010\big)u\nonumber\\
  &+\frac{4} {3}{\chi_1^{\E_8}}{\chi_1^{\E_8}}-\frac{8 }{3}{\chi_1^{\E_8}}
   {\chi_8^{\E_8}}-1824 {\chi_1^{\E_8}}+112
   {\chi_2^{\E_8}}-4 {\chi_3^{\E_8}}+4 {\chi_6^{\E_8}}
    -680 \chi_7^{\E_8}\nonumber\\
    &+\frac{28}{3}\chi_8^{\E_8} \chi_8^{\E_8} +50744 \chi_8^{\E_8}-2399276 \ , \nonumber\\
g_3^{\E_8}=&~
 \frac{1}{216} u^6 - 4u^5 - \Big(\frac{1}{18}\chi_1^{\E_8}+\frac{47}{18}\chi_8^{\E_8}-\frac{5177}{6} \Big)u^4 \nonumber\\
&
- \Big(-\frac{107}{6}\chi_1^{\E_8}+\frac{1}{6}\chi_2^{\E_8}+3\chi_7^{\E_8} -\frac{1580}{3}\chi_8^{\E_8}+\frac{504215}{6}\Big)u^3 \nonumber\\
&
   - \bigg(-\frac{2}{9}\chi_1^{\E_8}\chi_1^{\E_8}-\frac{20}{9}\chi_1^{\E_8} \chi_8^{\E_8}+\frac{5866}{3}\chi_1^{\E_8}-\frac{112}{3}\chi_2^{\E_8}+\frac{1}{3} \chi_3^{\E_8} \nonumber\\
   &\qquad +\frac
   {11}{3}\chi_6^{\E_8}-\frac{1450}{3}\chi_7^{\E_8}+\frac{196}{9}\chi_8^{\E_8}\chi_8^{\E_8}+39296\chi_8^{\E_8}-\frac{12673792}{3} \bigg)u^2\nonumber\\
&
    - \bigg(\frac{94}{3}\big(\chi_1^{\E_8}\big)^2-\frac{2 }{3}\chi_1^{\E_8}\chi_2^{\E_8}+\frac{718}{3}\chi_1^{\E_8}\chi_8^{\E_8}-\frac{270736}{3}\chi_1^{\E_8}-\frac{10 }{3}\chi_2^{\E_8}\chi_8^{\E_8}\notag\\
    &\qquad+2630 \chi_2^{\E_8}-52\chi_3^{\E_8}+4 \chi_5^{\E_8} -416 \chi_6^{\E_8}+16
   \chi_7^{\E_8} \chi_8^{\E_8}+25880
   \chi_7^{\E_8}\notag\\
   &\qquad-\frac{7328
   }{3}\big(\chi_8^{\E_8}\big)^2-\frac{3841382
   }{3}\chi_8^{\E_8}+107263286 \bigg)u \nonumber\\
   &- \frac{8}{27}\big(\chi_1^{\E_8}\big)^3-\frac{28
   }{9}\big(\chi_1^{\E_8}\big)^2 \chi_8^{\E_8}+1065
   \big(\chi_1^{\E_8}\big)^2-\frac{118 }{3}\chi_1^{\E_8}
   \chi_2^{\E_8}+\frac{4 }{3}\chi_1^{\E_8}
   \chi_3^{\E_8}\nonumber\\
   &
   -\frac{4 }{3}\chi_1^{\E_8}\chi_6^{\E_8} +\frac{8 }{3}\chi_1^{\E_8}
   \chi_7^{\E_8}+\frac{40}{9}\chi_1^{\E_8}
   \big(\chi_8^{\E_8}\big)^2+\frac{19264 }{3}\chi_1^{\E_8}
   \chi_8^{\E_8}-\frac{4521802
   }{3}\chi_1^{\E_8}\nonumber\\
   &-\frac{572}{3}\chi_2^{\E_8} \chi_8^{\E_8}+232 \chi_5^{\E_8} +59482
   \chi_2^{\E_8}+\frac{20 }{3}\chi_3^{\E_8}
   \chi_8^{\E_8}-1880 \chi_3^{\E_8}-4
   \chi_4^{\E_8} \nonumber\\
   &-\frac{8
   }{3}\chi_6^{\E_8} \chi_8^{\E_8}-11808
   \chi_6^{\E_8}+\frac{2740 }{3}\chi_7^{\E_8}
   \chi_8^{\E_8}+460388
   \chi_7^{\E_8}-\frac{136}{27}\big(\chi_8^{\E_8}\big)^3\cr
   &-\frac{205492
   }{3}\big(\chi_8^{\E_8}\big)^2 -\frac{45856940}{3}\chi_8^{\E_8}+\big(\chi_2^{\E_8}\big)^2+1091057493 \ ,
\end{align}
where $\chi_i^{\E_8}$ are the characters of the fundamental weights of $\E_8$ which are of dimension, $3875, 147250, 6696000, 6899079264, 146325270, 2450240, 30380, 248$, respectively. The node associated with each fundamental weight is denoted in the Dynkin diagram of $\E_8$ in figure~\ref{fig:E8-Dynkin}. Their explicit
forms are given in Appendix \ref{app:Characters}, which are expressed in terms of the characters of the
fundamental weights of $\SO(16)\supset\SO(14)\times \U(1)_I$:
\begin{align}
\E_8&\supset \SO(16)\cr
\bf 3875 &= \bf 16^2+128 \times 16-120+1820-\overline{128}-1,\cr
&\vdots\cr
\bf 30380 &= \bf 
 16\times 560 - 1820 + 8008 + 120\times\overline{128}-\overline{128}  ,
\cr
\bf 248 &= \bf 120+\overline{128}\ ,
\end{align}
where we only list the three lowest dimensional representations, as listing other higher dimensional weights is not very illuminating.

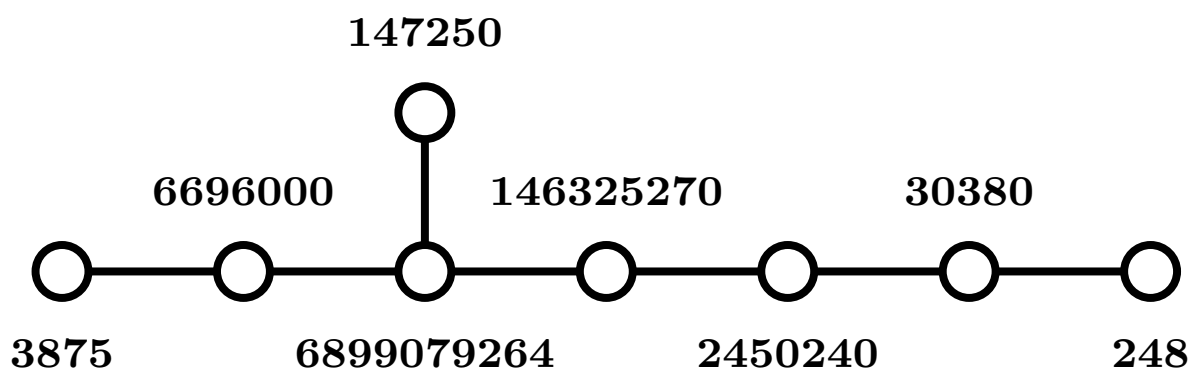
\begin{figure}
    \centering
\begin{tikzpicture}[
  x=10mm,
  y=10mm,
  dynkin edge/.style={
    draw,
    line width=1.5pt,
    line cap=round,
    line join=round
  },
  dynkin node/.style={
    circle,
    draw,
    fill=white,
    minimum size=3.5mm,
    inner sep=0pt,
    outer sep=0pt,
    line width=1.5pt
  },
  root label/.style={
    font=\scriptsize,
    inner sep=1pt,
    label distance=1.5mm
  },
  diagram title/.style={
    font=\normalsize,
    inner sep=1pt
  }
]

\coordinate (a1) at (-2.4,0);
\coordinate (a3) at (-1.2,0);
\coordinate (a4) at (0,0);
\coordinate (a5) at (1.2,0);
\coordinate (a6) at (2.4,0);
\coordinate (a2) at (0,1.05);
\coordinate (a7) at (3.6,0);
\coordinate (a8) at (4.8,0);

\draw[dynkin edge] (a1) --(a3) -- (a4) -- (a5) -- (a6) -- (a7)--(a8);
\draw[dynkin edge] (a4) -- (a2);

\node[dynkin node, label={[root label]below:$\bf 3875$}] at (a1) {};
\node[dynkin node, label={[root label]above:$\bf 147250$}] at (a2) {};
\node[dynkin node, label={[root label]above:$\bf 6696000$}] at (a3) {};
\node[dynkin node, label={[root label]below:$\bf 6899079264$}] at (a4) {};
\node[dynkin node, label={[root label]above:$\bf 146325270$}] at (a5) {};
\node[dynkin node, label={[root label]below:$\bf 2450240$}] at (a6) {};
\node[dynkin node, label={[root label]above:$\bf 30380$}] at (a7) {};
\node[dynkin node, label={[root label]below:$\bf 248$}] at (a8) {};

\end{tikzpicture}
\caption{Dynkin diagram for $\E_8$. The numbers denote the dimension of the fundamental weights associated with each node.}
\label{fig:E8-Dynkin}
\end{figure}%

The special tuning 
\eqref{eq:special tuning} 
\begin{align}
\{ M_0, \mathsf{M}_{i};\, M_1, M_2, M_3
\} = \{ \sqrt{q}, 1,\, -1, 1, \, -1;\, M_1, M_2, M_3\}\ ,
\end{align}
where the instanton factor $M_0$ and four masses for the $\SU(2)+7\F$ theory are respectively tuned to be $M_0=\sqrt{q}$ where $q$ is the instanton factor for $\SU(2)+6\F$. 
The tuning then yields

\begin{align}\label{eq:tuned-cha}
    \chi_1^{\E_8}&\to\big(\chi_4^{\tF_4}\big)^2+\chi_4^{\tF_4}-3\chi_1 ^{\tF_4}-2\chi_3^{\tF_4}+3 \ ,\notag\\
    \chi_2^{\E_8}&\to\big(\chi_4^{\tF_4}\big)^2
+2\chi_1^{\tF_4}\chi_4^{\tF_4}
-3\chi_4^{\tF_4}
-3\chi_1^{\tF_4}
-2\chi_2^{\tF_4}
-4\chi_3^{\tF_4} \ ,\notag\\
\chi_3^{\E_8}&\to2\big(\chi_4^{\tF_4}\big)^3
+4\chi_1^{\tF_4}\big(\chi_4^{\tF_4}\big)^2
-8\big(\chi_4^{\tF_4}\big)^2
-5\chi_1^{\tF_4}\chi_4^{\tF_4}
+6\chi_2^{\tF_4}\chi_4^{\tF_4}
-4\chi_3^{\tF_4}\chi_4^{\tF_4}\notag\\
&
\quad+6\chi_4^{\tF_4}
-7\big(\chi_1^{\tF_4}\big)^2
-3\big(\chi_3^{\tF_4}\big)^2
+16\chi_1^{\tF_4}
+11\chi_2^{\tF_4}
-8\chi_1^{\tF_4}\chi_3^{\tF_4}
+17\chi_3^{\tF_4}\ ,\notag\\
\chi_4^{\E_8}&\to\big(\chi_4^{\tF_4}\big)^5
+7\chi_1^{\tF_4}\big(\chi_4^{\tF_4}\big)^4
-8\big(\chi_4^{\tF_4}\big)^4
+8\chi_2^{\tF_4}\big(\chi_4^{\tF_4}\big)^3
-4\chi_3^{\tF_4}\big(\chi_4^{\tF_4}\big)^3
+\big(\chi_4^{\tF_4}\big)^3\notag\\
&
\quad-27\big(\chi_1^{\tF_4}\big)^2\big(\chi_4^{\tF_4}\big)^2
-4\big(\chi_3^{\tF_4}\big)^2\big(\chi_4^{\tF_4}\big)^2
+47\chi_1^{\tF_4}\big(\chi_4^{\tF_4}\big)^2
+12\chi_2^{\tF_4}\big(\chi_4^{\tF_4}\big)^2\notag\\
&
\quad-28\chi_1^{\tF_4}\chi_3^{\tF_4}\big(\chi_4^{\tF_4}\big)^2
+30\chi_3^{\tF_4}\big(\chi_4^{\tF_4}\big)^2
-11\big(\chi_4^{\tF_4}\big)^2
-2\big(\chi_1^{\tF_4}\big)^2\chi_4^{\tF_4}
+3\big(\chi_3^{\tF_4}\big)^2\chi_4^{\tF_4}\notag\\
&
\quad-\chi_1^{F_4}\chi_4^{F_4}
-20\chi_1^{F_4}\chi_2^{F_4}\chi_4^{F_4}
+24\chi_2^{F_4}\chi_4^{F_4}
-2\chi_1^{F_4}\chi_3^{F_4}\chi_4^{F_4}
-16\chi_2^{\tF_4}\chi_3^{\tF_4}\chi_4^{\tF_4}
+2\chi_3^{\tF_4}\chi_4^{\tF_4}\notag\\
&
\quad+\chi_4^{\tF_4}
+26\big(\chi_1^{\tF_4}\big)^3
+8\big(\chi_3^{\tF_4}\big)^3
-61\big(\chi_1^{\tF_4}\big)^2
+\big(\chi_2^{\tF_4}\big)^2
+37\chi_1^{\tF_4}\big(\chi_3^{\tF_4}\big)^2
-37\big(\chi_3^{\tF_4}\big)^2\notag\\
&
\quad+29\chi_1^{\tF_4}
-20\chi_1^{\tF_4}\chi_2^{\tF_4}
+23\chi_2^{\tF_4}
+56\left(\chi_1^{\tF_4}\right)^2\chi_3^{\tF_4}
-98\chi_1^{\tF_4}\chi_3^{\tF_4}
-18\chi_2^{\tF_4}\chi_3^{\tF_4}
+27\chi_3^{\tF_4}
+4\ ,\notag\\
\chi_5^{\E_8}&\to3\big(\chi_4^{\tF_4}\big)^4
+3\chi_1^{\tF_4}\big(\chi_4^{\tF_4}\big)^3
-7\big(\chi_4^{\tF_4}\big)^3
-13\chi_1^{\tF_4}\big(\chi_4^{\tF_4}\big)^2
-3\chi_2^{\tF_4}\big(\chi_4^{\tF_4}\big)^2
-15\chi_3^{\tF_4}\big(\chi_4^{\tF_4}\big)^2\notag\\
&
\quad+6\big(\chi_4^{\tF_4}\big)^2
-6\big(\chi_1^{\tF_4}\big)^2\chi_4^{\tF_4}
+19\chi_1^{\tF_4}\chi_4^{\tF_4}
-2\chi_2^{\tF_4}\chi_4^{\tF_4}
-6\chi_1^{\tF_4}\chi_3^{\tF_4}\chi_4^{\tF_4}
+14\chi_3^{\tF_4}\chi_4^{\tF_4}
-11\chi_4^{\tF_4}\notag\\
&
\quad+13\big(\chi_1^{\tF_4}\big)^2
+19\big(\chi_3^{\tF_4}\big)^2
-11\chi_1^{\tF_4}
+6\chi_1^{\tF_4}\chi_2^{\tF_4}
-7\chi_2^{\tF_4}
+32\chi_1^{\tF_4}\chi_3^{\tF_4}
+6\chi_2^{\tF_4}\chi_3^{\tF_4}
-17\chi_3^{\tF_4}\ ,\notag\\
\chi_6^{\E_8}&\to2\big(\chi_4^{\tF_4}\big)^3
-4\chi_1^{\tF_4}\big(\chi_4^{\tF_4}\big)^2
+6\big(\chi_4^{\tF_4}\big)^2
-3\chi_1^{\tF_4}\chi_4^{\tF_4}
-2\chi_2^{\tF_4}\chi_4^{\tF_4}
-4\chi_3^{\tF_4}\chi_4^{\tF_4}
+2\chi_4^{\tF_4}\notag\\
&
\quad+7\big(\chi_1^{\tF_4}\big)^2
+\big(\chi_3^{\tF_4}\big)^2
-16\chi_1^{\tF_4}
-3\chi_2^{\tF_4}
+8\chi_1^{\tF_4}\chi_3^{\tF_4}
-13\chi_3^{\tF_4}
+6\ ,\notag\\
\chi_7^{\E_8}&\to-3\big(\chi_4^{\tF_4}\big)^2
-\chi_1^{\tF_4}\chi_4^{\tF_4}
+3\chi_4^{\tF_4}
+5\chi_1^{\tF_4}
+\chi_2^{\tF_4}
+7\chi_3^{\tF_4}
-3\ ,\notag\\
\chi_8^{\E_8}&\to\chi^{\tF_4} _1-\chi^{\tF_4} _4-2\ ,
\end{align}
where $\chi^{\tF_4}_i,i=1,2,3,4$, are the characters of the 52-, 1274-, 273-, 26-dimensional representations of $\tF_4$, respectively. It follows that $g_2^{\E_8}$ and $g_3^{\E_8}$ in \eqref{eq:E8} reduce to $g_2^{6\F/\Z_2}$ and $g_3^{6\F/\Z_2}$ for the $\SU(2)+6\F/\Z_2$ theory in \eqref{eq:sw-su26f-f4}. This, therefore, confirms our claim \eqref{eq:proposal}.

We note that the special tuning freezes $\SO(8)$ inside $E_8$, and the remaining symmetry gives rise to an $F_4$, 
\begin{align}
      \E_8 \supset \SO(16)~\xrightarrow{\text{$\D_4$-tuning}}  ~ \SO(8) \subset F_4\ .
\end{align}


\section{Conclusion and discussion}\label{sec:conclusion}
In this paper, we computed the Seiberg–Witten curves for the $\Z_2$ S-folds of five-dimensional rank-1 superconformal theories, $\SU(2)_0$, $\SU(2)+2\F$, $\SU(2)+4\F$, and $\SU(2)+6\F$. We developed two independent systematic methods for calculating the Seiberg–Witten curves based on (generalized) toric diagrams. 

In the first, we lifted the $\Z_2$ S-fold of the 5-brane web to the M-theory  torus. On the SW curve of the rank-1 $\SU(2)+2n\F$ theory, it acts as the involution $(t,w)\mapsto (t^{-1},w^{-1})$. We constructed the quotient curves explicitly by implementing a two-to-one map, which identifies pairs of points on each fiber while the Coulomb branch remains untouched. 

In the second, we followed the 5-brane web prescription of \cite{Kim:2021fxx}, where the 7-brane configuration $\mathcal{S}_{\D_4}={\bf A^4 BC}$ is inserted at the fixed point of the $\Z_2$ symmetry, and then obtained the S-fold curves from the SW curves of the $\E_{5+n}$ theories by the special tuning \eqref{eq:tuning-condi} and \eqref{eq:instTuning}, {\it i.e.,} the four D7-brane fugacities (masses) are set to $(1,-1,1,-1)$, accounting for the inserted 7-branes being stuck at the fixed point, and the instanton fugacity to $\sqrt{q}$, required in the S-folded prepotential. For all four theories, $\SU(2)+2n\F$ $(n=0,1,2,3)$, two constructions give the identical Seiberg–Witten curve when expressed as the Weierstrass normal forms. 

The flavor symmetries of four $\Z_2$ S-folded theories are summarized as:

\begin{table}[ht]
    \centering
    \small
    \renewcommand{\arraystretch}{1.3}
    \begin{tabular}{c||c|c|c|c}
        \hline
        S-folds
        & $\mathrm{SU}(2)_0/\Z_2$
        & $\mathrm{SU}(2)+2\F/\Z_2$
        & $\mathrm{SU}(2)+4\F/\Z_2$
        & $\mathrm{SU}(2)+6\F/\Z_2$ \\ \hline
Flavor symmetry
        & $\SU(2)$
        & $\SU(3)$
        & $\SO(7)$
        & $F_4$ \\ \hline
    \end{tabular}
\end{table}
\FloatBarrier
\noindent
The resulting flavor symmetries can be understood as follows. 
\begin{itemize}
    \item From the perspective of the two-to-one map, the $\Z_2$ S-fold acts on the flavor symmetry by the outer automorphism group of the affine Lie algebra. (See the flavor symmetry for the discrete gauging $\Z_2$ covering~\cite{Xie:2022lcm}.) More explicitly, for the parent theory $\mathcal{T}$ has an $\E_n$ flavor symmetry, and the $\Z_2$ affine folding through the orbit-sum yields:
    \begin{equation}
    \begin{aligned}
        \mathrm{SU}(2)_0/\Z_2 &: \; \widehat{E}_1~\xrightarrow{\text{$\Z_2$}}~\widehat{A}_1 \ ,\cr
        \mathrm{SU}(2)_0+2\F/\Z_2 &: \; \widehat{E}_3~\xrightarrow{\text{$\Z_2$}}~\widehat{A}_2 \ ,\cr
        \mathrm{SU}(2)+4\F/\Z_2 &: \; \widehat{E}_5~\xrightarrow{\text{$\Z_2$}}~\widehat{B}_3 \ ,\cr
        \mathrm{SU}(2)+6\F/\Z_2 &: \; \widehat{E}_7~\xrightarrow{\text{$\Z_2$}}~\widehat{F}_4 \ .       
    \end{aligned}
     \end{equation}
    \item From the perspective of the inserted 7-brane $\mathcal{S}_{\D_4}$, The $\Z_2$ S-folding of $\mathcal{T}=\SU(2)+2n\F$ gives rise to the $E_{n+5}$ theory.
\begin{equation}
        \begin{aligned}
\mathrm{SU}(2)_0/\Z_2 &:  ~~{E_5}~\xrightarrow{\text{$\D_4$-tuning}}~{\mathrm{SU}(2)} \ ,\\
\mathrm{SU}(2)+2\F/\Z_2 &: ~~{E_6}~\xrightarrow{\text{$\D_4$-tuning}}~ {\mathrm{SU}(3)} \ , \\
\mathrm{SU}(2)+4\F/\Z_2 &: ~~{E_7}~\xrightarrow{\text{$\D_4$-tuning}}~{\mathrm{SO}(7)} \ ,  \\
\mathrm{SU}(2)+6\F/\Z_2 &: ~~{E_8}~\xrightarrow{\text{$\D_4$-tuning}}~{F_4} \ . 
\end{aligned}
\end{equation}
Here, the $\D_4$-tuning refers to the special mass holonomy tuning \eqref{eq:tuning-condi} and \eqref{eq:instTuning}.
\end{itemize}

We note that our analysis is related to, but should be distinguished from, other operations with base changes on rank-1 Seiberg--Witten geometries. Among others~\cite{Argyres:2010py,Xie:2022lcm, Sakai:2023fnp,Argyres:2023tfx,Arias-Tamargo:2023duo, Furrer:2024zzu, Arias-Tamargo:2026rrf}, we list the differences in the Kodaira fiber structure as follows:
\begin{table}[htbp]
    \centering
    \small
    \renewcommand{\arraystretch}{1.2}
    \setlength{\tabcolsep}{4pt}

    \refstepcounter{table}\label{tab:comparison}
    \begin{tabular*}{\linewidth}{
        @{\extracolsep{\fill}}lcccc@{}
    }
        \toprule
        \multicolumn{1}{@{}l}{\textbf{Parent theory} $\mathcal{T}$}
        & $\mathrm{SU}(2)_0$
        & $\mathrm{SU}(2)+2\F$
        & $\mathrm{SU}(2)+4\F$
        & $\mathrm{SU}(2)+6\F$
        \\

        \multicolumn{1}{@{}l}{~~~with~Kodaira fibers $(F_\infty; {F_{\rm bulk}})$}
        & $(I_8;\,I_1^4)$
        & $(I_6;\,I_1^6)$
        & $(I_4;\,I_1^8)$
        & $(I_2;\,I_1^{10})$
        \\
        \midrule

        $\Z_2$ Covering~\cite{Xie:2022lcm}
        & $(I_4;\,I_0^*I_1^2)$
        & $(I_3;\,I_0^*I_1^3)$
        & $(I_2;\,I_0^*I_1^4)$
        & $(I_1;\,I_0^*I_1^5)$
        \\

        $\Z_2$ CB folding~\cite{Furrer:2024zzu}
        & $(I_4^*;\,I_1^2)$
        & $(I_3^*;\,I_1^3)$
        & $(I_2^*;\,I_1^4)$
        & $(I_1^*;\,I_1^5)$
        \\

        \textbf{S-fold $\mathcal{T}/\Z_2$}
        & $(I_4;\,I_2^4)$
        & $(I_3;\,I_2^4I_1)$
        & $(I_2;\,I_2^4I_1^2)$
        & $(I_1;\,I_2^4I_1^3)$
        \\
        \bottomrule
    \end{tabular*}
\end{table}

It would be interesting to generalize this analysis to other $\Z_n$ S-folds and also to extend it to other physical observables. 
For instance, an immediate extension of this work is to compute the Seiberg–Witten curves for the 5d $\Z_{3,4,6}$ S-folds in a similar setup based on 5-brane webs~\cite{Kim:2026Z346}. There are also a few interesting directions to pursue along the S-folds: (i) The first one is to compute the Seiberg–Witten curves for the $\Z_n$ S-folds of higher-rank theories. It was proposed \cite{Kim:2021fxx} that two characteristically different S-folds are possible from the point of view of the fixed point of $\Z_n$ symmetry. One is the case where the fixed point lies on the face of a 5-brane web, and the other is the case where it lies on an edge or a vertex.  The former is the $\Z_2$ S-folds of the $\SU(2N)_0$ theories and the latter is the $\Z_2$ S-folds of the $\SU(2N+1)_0$ theories and $T_4$, $T_5$ theories. Another one is to compute more refined observables. As our analysis is based on 5-brane webs and the Nekrasov partition can also be computed using a 5-brane web through the topological vertex formulation, it would be interesting to study how the 7-brane insertion is implemented in an Omega deformed background.


\acknowledgments{We thank Yuzhu Cheng, Hee-Cheol Kim, and Kimyeong Lee for useful discussions. We would like to thank 
Mini Advanced Workshop on ``Supersymmetric Quantum Field Theories Across Dimensions 2026,'' at USTC, ``Aspects of Supersymmetric Quantum Field Theory 2025" held in Seoul National University, Theoretical Physics Group at Shing-Tung Yau Center of Southeast University, KIAS, and BIMSA where this work was presented. SK is supported by National Natural Science Foundation of China grant No. 12250610188 and also by the Brain Pool Program funded by the Ministry of Science and ICT through the National Research Foundation of Korea (RS-2025-25457100). SL is also supported by National Natural Science Foundation of China grant No. 12250610188. MS is supported in part by the National Natural Science Foundation of China (Grant No. 12475105). FY is supported by start-up research grant RK21121 from Huzhou Normal University.}

\bigskip

\appendix

\section{The procedure for transforming to Weierstrass normal form}\label{app:WNF}

This appendix explains how the Seiberg–Witten curves obtained from 5-brane webs are brought into the Weierstrass normal form. In Appendix~\ref{app:WNF}, we review the standard method of \cite{Huang:2013yta} and write down the formula for the special cases, for which we actually compute in this paper. This formula is used in the derivations of \eqref{eq:g2-su2-0}, \eqref{eq:su2+2Fz2main}, \eqref{eq:SU24F-Z2}, \eqref{eq:sw-su26f-f4}, and Appendix ~\ref{sec:physmass-curves}. In Appendix~\ref{app:coodTrans}, we apply this method to the derivation of \eqref{eq:sw-su26f-f4} from \eqref{eq:SW6f-reduction-maintext}, where the curve is of high degree and requires a reduction to a cubic form before the standard method can be applied.

\subsection{An overview of the existing method}\label{app:WNF2}

We begin by reviewing the standard procedure for bringing a genus-one algebraic curve into the Weierstrass normal form,
\begin{align}\label{eq:WS-form}
    y^2=4x^3-g_2x-g_3\ .
\end{align}
For a (non-)toric 5-brane web, the corresponding Seiberg–Witten curve is often given by a cubic curve in the projective space $\mathbb{P}^2=\{[x:y:z]\}$
\begin{align}\label{eq:s1x3}
    0 &=s_1 x^3 + s_2 x^2 y + s_3 x y^2 + s_4 y^3 + s_5 x^2 z + s_6 x y z + s_7 y^2 z + s_8 x z^2 + s_9 y z^2 + s_{10} z^3 \ .
\end{align}
If the Seiberg–Witten curve is given in this form, it is straightforward to bring it into the Weierstrass normal form by setting $z=1$ without loss of generality and then applying eq.(A.8) in \cite{Huang:2013yta}.

In the main text, we use eq.(A.8) in \cite{Huang:2013yta} in the following two cases. 
In the case
\begin{align}
s_1=s_4=s_5=s_7=s_{10}=0,
\end{align}
the corresponding Weierstrass normal form is given by \eqref{eq:WS-form} with
\begin{align}
    g_2&=\frac{s_6^4}{12}-\frac{2}{3} s_3 s_8 s_6^2-\frac{2}{3} s_2 s_9 s_6^2+\frac{4}{3} s_3^2
   s_8^2+\frac{4}{3} s_2^2 s_9^2-\frac{4}{3} s_2 s_3 s_8 s_9\notag\\
   g_3&=\frac{s_6^6}{216}-\frac{1}{18} s_3 s_8 s_6^4-\frac{1}{18} s_2 s_9 s_6^4+\frac{2}{9} s_3^2
   s_8^2 s_6^2+\frac{2}{9} s_2^2 s_9^2 s_6^2+\frac{1}{9} s_2 s_3 s_8 s_9
   s_6^2 \, \notag\\
   &\quad-\frac{8}{27} s_3^3 s_8^3-\frac{8}{27} s_2^3 s_9^3+\frac{4}{9} s_2^2 s_3 s_8
   s_9^2+\frac{4}{9} s_2 s_3^2 s_8^2 s_9 \ .
\end{align}
We use this to derive the Seiberg–Witten curves for 
$\mathrm{SU}(2)_0/\Z_2$,
$\mathrm{SU}(2)+2\F/\Z_2$, and
$\mathrm{SU}(2)+4\F/\Z_2$ in the Weierstrass normal form in the main text, which are given in \eqref{eq:g2-su2-0}, \eqref{eq:su2+2Fz2main} and \eqref{eq:SU24F-Z2}, respectively.

In the case 
\begin{align}\label{eq:second-case-s}
s_2=s_3=s_4=s_6=s_9=s_{10}=0, 
\end{align}
the corresponding Weierstrass normal form is given by \eqref{eq:WS-form} with
\begin{align}\label{eq:WS-second-case}
    g_2&=\frac{4}{3} s_5^2 s_7^2-4 s_1 s_7^2 s_8,\notag\\
    g_3&=\frac{4}{3} s_1 s_5 s_7^3 s_8-\frac{8}{27} s_5^3 s_7^3\ .
\end{align}
This is used in the next section to derive the Seiberg–Witten curves for 
$\mathrm{SU}(2)+6 \F/\Z_2$ in the Weierstrass normal form, which is given in \eqref{eq:sw-su26f-f4} in the main text. 

However, in Appendix~\ref{sec:physmass-curves}, the Seiberg–Witten curve that we encounter is typically not a cubic curve but 
a quadratic curve in $t$: $F(t,w)=\mathcal{A}(w)t^2+\mathcal{B}(w)t+\mathcal{C}(w)=0$. For a curve of this form, completing the square gives
\begin{align}\label{eq:complete-square}
    \left(t\mathcal{A}(w)+\frac{\mathcal{B}(w)}{2}\right)^2
=\frac1{4}\Bigg(\mathcal{B}(w)^2-4\mathcal{A}(w)\mathcal{C}(w)\Bigg)\ .
\end{align}
If the right-hand side of \eqref{eq:complete-square}, by a suitable rescaling of the coordinate on the left-side, becomes a quartic polynomial in $w$ as 
\begin{align}
     y^2=l_1w^4+l_2w^3+l_3w^2+l_4w+l_5,\quad y(t,w)\propto t\mathcal{A}(w) +\frac{\mathcal{B}(w)}{2}\ ,
\end{align}
where $l_i,i=1,2,\dots,5$ ($l_5\neq0$),  depend upon the parameters of the theory but are independent of the curve coordinates, the corresponding Weierstrass coefficients are given by
\begin{align}\label{eq:g2-g3-coef}
g_2&=\frac{1}{12}\left(
a_1^4+8a_2a_1^2+16a_2^2-48a_4-24a_1a_3
\right),\\
g_3&=\frac{1}{216}\left(
a_1^6+12a_2a_1^4+48a_2^2a_1^2+64a_2^3
-72a_1^2a_4-36a_1^3a_3
-144a_2a_1a_3+216a_3^2+576a_6
\right)\notag\ ,
\end{align}
with
\begin{align}
a_1&=\frac{l_4}{\sqrt{l_5}},&
a_2&=l_3-\frac{l_4^2}{4l_5},&
a_3&=2\sqrt{l_5}\,l_2,&
a_4&=-4l_5l_1,&
a_6&=a_2a_4.
\end{align}

\subsection{Reduction of a high-degree SW curve}\label{app:coodTrans}

We now explain the derivation of \eqref{eq:sw-su26f-f4} from \eqref{eq:SW6f-reduction-maintext}. The curve \eqref{eq:SW6f-reduction-maintext}, which is the $\SU(2)+6\F/\Z_2$ curve, is of high degree and is not cubic, so the standard method of \cite{Huang:2013yta} reviewed in Appendix~\ref{app:WNF} cannot be applied directly. As we will see, the reduction to a cubic form is achieved by completing the square in one of the variables after simple coordinate transformation. We start with the curve presented in \eqref{eq:SW6f-reduction-maintext}, which is rewritten in the form
\begin{align}\label{eq:SW6f-reduction}
F_{\SU(2)+6\bf{F}/\Z_2}(X,Y)=\sum_{m,n} C_{m,n} X^m Y^n,
\end{align}
with each coefficient being given by
\begin{align}
&C_{0,2} =  - (M-M^{-1})
\left( \chi_4^{\SU(4)} - \chi_{\bar 4}^{\SU(4)} + M - M^{-1} \right) - \chi_6^{\SU(4)} - \mathcal{U},
\cr
&C_{1,1} = 2 \chi_6^{\SU(4)} - 6 (M-M^{-1})^2 +2\mathcal{U},
\cr
&C_{1,2} = 8(M^2-M^{-2})+4(M^{-1} \chi_4^{\SU(4)} - M \chi_{\bar 4}^{\SU(4)}),
\cr
&C_{1,3} = 
(M+M^{-1})
\left( \chi_4^{\SU(4)}+\chi_{\bar 4}^{\SU(4)} - M -M^{-1} \right) - \chi_6^{\SU(4)} + \mathcal{U},
\cr
&C_{2,0} = 
(M-M^{-1})
\left( \chi_4^{\SU(4)} - \chi_{\bar 4}^{\SU(4)} - M+M^{-1} \right) - \chi_6^{\SU(4)} - \mathcal{U},
\cr
&C_{2,1} = 8(M^2-M^{-2})-4(M^{-1} \chi_4^{\SU(4)} - M \chi_{\bar 4}^{\SU(4)}),
\cr
&C_{2,2} = 2 \chi_6^{\SU(4)} - 6 (M+M^{-1})^2 -2\mathcal{U},
\cr
&C_{3,1} = 
- (M+M^{-1})
\left( \chi_4^{\SU(4)}+\chi_{\bar 4}^{\SU(4)} + M+M^{-1} \right) - \chi_6^{\SU(4)} + \mathcal{U},
\end{align}
while $C_{m,n} = 0 $ for $(m,n)$ other than above. 

To reduce the degree of the curve \eqref{eq:SW6f-reduction}, we first perform the coordinate transformation
\begin{align}
    X\to YZ,\quad Y\to Y,
\end{align}
and then divide the resulting equation by $Y^2$. This brings \eqref{eq:SW6f-reduction} into a quadratic equation in $Y$,
\begin{align}
    A Y^2 + B Y + C = 0,
\end{align}
where $A$, $B$, and $C$ are polynomials in $Z$
\begin{align}
    A &= C_{3,1} Z^3 + C_{2,2} Z^2 + C_{1,3} Z, 
\cr
    B &= C_{2,1} Z^2 + C_{1,2} Z, 
\cr
    C &= C_{2,0} Z^2 + C_{1,1} Z + C_{0,2}.
\end{align}
For an equation of this form, we complete the square in $Y$ by introducing a new variable
\begin{align}
    T=AY+\frac{B}{2},
\end{align}
which brings it into the form
\begin{align}
    T^2=\frac14 B^2 - A C .
\end{align}

After absorbing the overall factor $4M^8 (Z-1)^2$ in the right hand side by the redefinition of $T$
\begin{align}
T = 2M^4 (Z-1) t \,,
\end{align}
we obtain the cubic curve of the form
\begin{align}\label{eq:sw-cubic-curve}
    t^2=c_3Z^3+c_2Z^2+c_1Z\ .
\end{align}
The coefficients $c_i$ are written in terms of SO(8) characters as
\begin{align}
c_3&=
(\chi^{\SO(8)}_V+2-\mathcal{U})^2-(\chi^{\SO(8)}_S+\chi^{\SO(8)}_C)^2,\notag\\
c_2&=
- 2(\mathcal{U}+6)^2
+ 2(\chi^{\SO(8)}_V)^2 + 2(\chi^{\SO(8)}_S)^2 + 2(\chi^{\SO(8)}_C)^2 - 16 \chi_{\mathrm{Adj}}^{\SO(8)} + 64, \notag\\
c_1&=
(\chi^{\SO(8)}_V-2+\mathcal{U})^2-(\chi^{\SO(8)}_S-\chi^{\SO(8)}_C)^2,
\end{align}
where, we have combined the characters of $\mathrm{SU(4)} \times \mathrm{U(1)}$ into the characters of SO(8) by using the relation
\begin{align}
\chi^{\SO(8)}_V 
&= M \chi_4^{\SU(4)} + M^{-1} \chi_{\bar 4}^{\SU(4)},
\cr
\chi^{\SO(8)}_S 
&= \chi_6^{\SU(4)} + M^2 + M^{-2 },
\cr
\chi^{\SO(8)}_C
&= M^{-1} \chi_4^{\SU(4)} + M \chi_{\bar 4}^{\SU(4)},
\cr
\chi^{\SO(8)}_{\mathrm{Adj}}
& = \chi_4^{\SU(4)} \chi_{\bar 4}^{\SU(4)}
+ (M^2 + M^{-2}) \chi_6^{\SU(4)}.
\end{align}

Since the resulting curve \eqref{eq:sw-cubic-curve} is now cubic in a form \eqref{eq:s1x3} with \eqref{eq:second-case-s}, we apply the formula \eqref{eq:WS-second-case}. 
Combining the characters of SO(8) in the result into the characters of $\tF_4$ by using the relation
\begin{align}
\chi^{\tF_4}_{1}
&= \chi^{\SO(8)}_{\mathrm{Adj}} + \chi^{\SO(8)}_{V} + \chi^{\SO(8)}_{S} + \chi^{\SO(8)}_{C} \,,
\cr
\chi^{\tF_4}_{2}
&=   \chi^{\SO(8)}_{\mathrm{Adj}}  (\chi^{\SO(8)}_V + \chi^{\SO(8)}_S + \chi^{\SO(8)}_C + 4) + \chi^{\SO(8)}_V \chi^{\SO(8)}_S \chi^{\SO(8)}_C 
\cr
& \quad - (\chi^{\SO(8)}_V)^2 -  (\chi^{\SO(8)}_S)^2 - (\chi^{\SO(8)}_C)^2 + \chi^{\SO(8)}_V \chi^{\SO(8)}_S + \chi^{\SO(8)}_S \chi^{\SO(8)}_C + \chi^{\SO(8)}_C \chi^{\SO(8)}_V 
\cr 
& \quad - \chi^{\SO(8)}_V - \chi^{\SO(8)}_S - \chi^{\SO(8)}_C + 2\,, 
\cr
\chi^{\tF_4}_{3}
&=
2 \chi^{\SO(8)}_{\mathrm{Adj}} + \chi^{\SO(8)}_V + \chi^{\SO(8)}_S + \chi^{\SO(8)}_C + \chi^{\SO(8)}_V \chi^{\SO(8)}_S + \chi^{\SO(8)}_S \chi^{\SO(8)}_C + \chi^{\SO(8)}_C \chi^{\SO(8)}_V + 1 \,,
\cr
\chi^{\tF_4}_{4}
&= \chi^{\SO(8)}_{V} + \chi^{\SO(8)}_{S} + \chi^{\SO(8)}_{C} + 2 \,.
\end{align}
where 
$\chi_i^{\tF_4} (i=1,2,3,4)$, are the characters of 52-,1274-, 273- and 26-dimensional representations of $\tF_4$, respectively, we obtain the final result \eqref{eq:sw-su26f-f4} after rescaling the Coulomb moduli as $\mathcal{U} = \frac12 u$.

\section{5-brane web realization of S-fold curves via mass tuning of \texorpdfstring{$\E_{5,6,7,8}$}{E5,6,7,8} SCFTs}
\label{sec:phy-mass-para}

In this appendix, we present the Seiberg–Witten curves for the $\Z_2$ S-folds of rank-1 SCFTs in a 5-brane web description. This complements the Weierstrass normal form analysis of section~\ref{sec:SW-EN-SCFT}. While section~\ref{sec:SW-EN-SCFT} starts from the Weierstrass normal form of the $\E_{5,6,7,8}$ SCFTs and imposes the special tuning that realizes the $\Z_2$ S-fold, here we start from the Seiberg–Witten curves obtained directly from the corresponding 5-brane webs and impose the same special tuning. Although the two approaches are both based on the special tuning of the $\E_{5,6,7,8}$ SCFTs, they are different in the language used to describe the curves.

In tuning the mass parameters to special values, it is essential to make clear the convention and the physical meaning of the parameters. In section~\ref{sec:phy-mass-def}, we introduce the \emph{physical mass parameterization}. Since the special tuning is described in terms of the physical masses, one needs to know how the coefficients of the Seiberg–Witten curves are expressed in terms of the physical mass parameters in order to determine the result of the tuning.

For the $\E_{5,6,7,8}$ Seiberg–Witten curves, the polynomial structure has been known from previous studies \cite{Bao:2011rc, Bao:2013pwa, Kim:2014nqa}. In these studies, however, the curves are often expressed in terms of parameters in which a different subgroup of the $\E_{5,6,7,8}$ global symmetry is manifest, and the relation to the physical mass parameters, which correspond to the $\SO(2N_f)$ flavor symmetry, is not entirely straightforward%
\footnote{See, however, the discussion on the physical parameters in Appendix A of \cite{Bao:2011rc}.}.  
In section \ref{sec:physmass-curves}, we provide the explicit form of the coefficients in terms of the physical mass parameters in the 5-brane web parameterization, together with the result of substituting the special tuning. 

\subsection{Physical mass parameterization}\label{sec:phy-mass-def}

Recall that determining the Seiberg–Witten curve completely from the dual diagram of a given 5-brane web involves three rescaling degrees of freedom:
an overall rescaling of the curve, and rescalings of the coordinates $t$ and $w$, which correspond to translations of the 5-brane web in the horizontal and vertical directions, respectively.  These do not affect the physics, so we may use them to fix three coefficients $c_{ij}$ in the Seiberg–Witten curve $F(t,w)=0$ arbitrarily. Once these three coefficients are fixed, the remaining coefficients are determined by the physical parameters of the theory.

In a 5-brane web, the heights of the flavor branes are often called the mass parameters, but their values depend on the choice of the reference point from which the heights are measured. 
 We refer to the particular choice of the heights measured from the center-of-mass of the two color branes as the \emph{physical mass parameterization}, since the resulting mass parameters correspond to the physical masses appearing in the gauge theory. In this parameterization, the exponentiated physical mass parameters are defined as $M_i=e^{-\beta m_i}$, where $m_i$ are the physical masses.

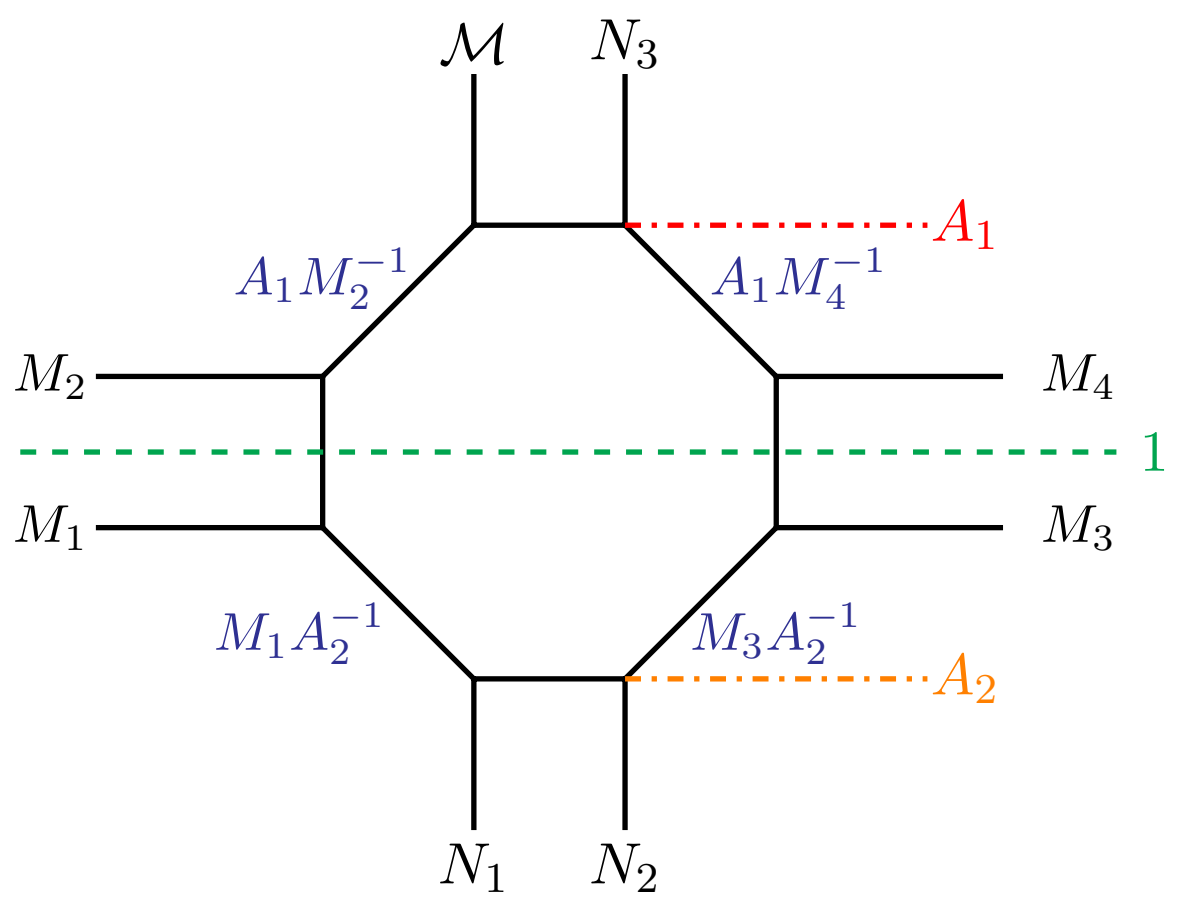
\begin{figure}
\centering
\begin{tikzpicture}[
    scale=0.5,
    line width=1pt,
    mass/.style={font=\small},                 
    param/.style={font=\small,color=Blue}      
]

\begin{scope}
\draw
(0,0)--(2,0)
(0,0)--(-2,2)--(-2,4)
(-5,2)--(-2,2)
(-5,4)--(-2,4)
(-2,4)--(0,6)--(2,6)
(0,6)--(0,8)
(0,-2)--(0,0)
(2,0)--(2,-2)
(2,0)--(4,2)--(4,4)--(2,6)
(2,6)--(2,8)
(4,4)--(7,4)
(4,2)--(7,2);
\end{scope}

\begin{scope}
\node[param] at (-2,5.3) {$A_1M_2^{-1}$};
\node[param] at (4.3,5.3) {$A_1M_4^{-1}$};
\node[param] at (-2.3,0.6) {$M_1A_2^{-1}$};
\node[param] at (4,0.6) {$M_3A_2^{-1}$};
\end{scope}

\begin{scope}
\draw[dash dot,color=red] (2,6)--(6,6);
\node[color=red] at (6.5,6) {$A_1$};
\draw[dash dot,color=orange] (2,0)--(6,0);
\node[color=orange] at (6.5,0) {$A_2$};
\end{scope}

\begin{scope}
\node[mass] at (-5.6,2) {$M_1$};
\node[mass] at (-5.6,4) {$M_2$};
\node[mass] at (8,2) {$M_3$};
\node[mass] at (8,4) {$M_4$};
\end{scope}

\begin{scope}
\draw[dashed,color=Green] (-6,3)--(8.5,3);      
\node[color=Green] at (9,3) {$1$};
\end{scope}

\begin{scope}
\node at (0,-2.5) {$N_1$};
\node at (2,-2.5) {$N_2$};
\node at (0,8.4) {$\mathcal{M}$};
\node at (2,8.4) {$N_3$};
\end{scope}

\end{tikzpicture}
\caption{5-brane web for the $\SU(2)+4\F$ gauge theory, which
corresponds to the $\E_5$ SCFT at the UV fixed point.  The physical mass
parameters $M_i$ and the composite parameters $A_1M_2^{-1}$ etc.\ are
indicated.}
\label{fig:su24fphy}
\end{figure}

As an example, let us consider the $\SU(2)+4\F$ gauge theory shown in figure \ref{fig:su24fphy}. A rescaling of the coordinate $w$ corresponds to shifting the reference position. In the physical mass parameterization, this reference line is chosen to coincide with the green line, which represents the center-of-mass of the two color branes. 
We normalize this center-of-mass position to $w=1$.  The physical mass parameterization is then expressed as the condition
\begin{align}\label{eq:A1A2}
    A_1A_2=1\ , 
\end{align}
where $A_1$ is the exponentiated parameter associated with the height of the upper color D5-brane measured from the reference point $w=1$, while $A_2$ is defined analogously for the lower color D5-brane.  
More generally, throughout this paper, when we use the physical mass parameterization to determine the Seiberg–Witten curves of the $\SU(2)$ gauge theories with $N_f<8$ flavors, we impose the condition \eqref{eq:A1A2}.  In section~\ref{sec:physmass-curves}, we will see that this condition is translated to 
a specific relation among the parameters of the web for each of the $\E_{5,6,7,8}$ SCFTs, which we call the \emph{physical mass condition}.

The physical mass parameterization is particularly convenient for the comparison with the $\Z_2$ S-fold curves, since the center-of-mass of the two color branes is also the natural center of the $\Z_2$ S-fold action.  In particular, it is used both in the mass-tuning approach of this appendix and in the $\Z_2$ quotient approach of section~\ref{sec:2-1}.  Consequently, the mass parameters appearing in the two approaches can be naturally identified with each other.

The same parameterization also allows us to compare the two mass-tuning approaches: the one based on the Weierstrass form in section~\ref{sec:SW-EN-SCFT} and the one based on the 5-brane web in this appendix.  Under this parameterization, the $N_f(=5,6,7)$ physical masses $m_i$ are naturally identified with the Cartan parameters of the $\SO(2N_f)$ flavor symmetry, so that the corresponding $\SO(2N_f)$ fugacities in the Lie-algebra characters can be directly identified with the exponentiated physical mass parameters $M_i$.%
\footnote{The $\E_5$ case requires special care.  For $\E_{6,7,8}$, the characters are most conveniently expressed in terms of the Cartan parameters of the subgroup $\SO(2N_f)\times\U(1)$, where the Cartan parameters of $\SO(2N_f)$ are directly identified with the physical masses. For $\E_5$, on the other hand, it is standard to use the $\SO(10)$ character expressions, in which the $\SO(8)\subset\SO(10)$ is embedded nontrivially.  As a result, the parameters natural from the $\SO(10)$ viewpoint are related to the $\SO(8)$ physical masses by a triality transformation, and one must implement this transformation in order to match the standard $\SO(10)$ character expressions.}  
This is precisely the identification used in section~\ref{sec:SW-EN-SCFT}, where the tuning of a subgroup $\SO(2N_f)\subset \E_{N_f+1}$ in the Weierstrass form is performed.  In the 5-brane web approach of this appendix, the same tuning corresponds to tuning the physical mass parameters $m_i$, which makes the comparison between the two constructions transparent.

\subsection{\texorpdfstring{Tuning of Seiberg–Witten curves of $\E_{5,6,7,8}$ SCFTs}{Tuning of Seiberg–Witten curves of E{5,6,7,8} SCFTs}}\label{sec:physmass-curves}

In this subsection, we present the Seiberg–Witten curves of the $\E_{5,6,7,8}$ SCFTs in the physical mass parameterization, obtained from the corresponding 5-brane webs.  For each theory, the physical mass condition is imposed as a specific relation among the parameters of the web, which is the concrete realization of \eqref{eq:A1A2}. Then, imposing the tuning condition \eqref{eq:tuning-condi} on four physical mass parameters and identifying the instanton parameter as in \eqref{eq:instTuning}, we obtain the Seiberg–Witten curves of the $\Z_2$ S-fold theory directly in the 5-brane web language.

\subsubsection*{\texorpdfstring{$\E_5$ to $\SU(2)_0/\Z_2$}{E5 to SU(2)0/Z2}}

Let us first consider the $\E_5$ SCFT. Its 5-brane web under the physical mass parameterization is given by figure~\ref{fig:su24fphy}, and the proper imposition of the physical mass condition leads to
\begin{align}\label{eq:phy-cond-e5}
    \frac{\mathcal{M}M_1M_2}{N_1}=\frac{M_3M_4N_2}{N_3}=1\ .
\end{align}
This relation serves as a guideline for selecting appropriate coefficients $c_{ij}$. These four $M_i$ are the physical mass parameters.

Under the parameterization \eqref{eq:phy-cond-e5}, the $\E_5$ Seiberg–Witten curve is given by
\begin{align}\label{eq:sw-e5-mass}
    F_{\E_5}(t,w)&=t^2 \big(w-M_3\big) \big(w-M_4\big)-\left(\mathcal{M}+\frac{1}{\mathcal{M}}\right) t w^2-M_3 M_4
   \left(M_1 M_2 \mathcal{M}+\frac{1}{M_3 M_4 \mathcal{M}}\right) t\notag\\
   &\quad+\big(w-M_1\big)
   \big(w-M_2\big)-t U w=0\ ,
\end{align}
where 
\begin{align}
\mathcal{M}=(M_0^2M_1M_2M_3M_4)^{-1/4}
\end{align}
should be understood as the function of the instanton parameter $M_0$ of the $\E_5$ theory.

We now impose the special tuning \eqref{eq:tuning-condi} and use the relation between instanton factors \eqref{eq:instTuning}:
\begin{align}\label{eq:tuning-e5}
    M_1=-M_2=M_3=-M_4=1\ ,\qquad M_0=\sqrt{q}\ .
\end{align}
Substituting \eqref{eq:tuning-e5} into \eqref{eq:sw-e5-mass}, we obtain
\begin{align}\label{eq:sw-e5-tuned-explicit}
    F_{\E_5}^{\text{tuned}}(t,w)
    &=t^2(w^2-1)-\left(q^{-1/4}+q^{1/4}\right)t w^2
    -\left(q^{-1/4}+q^{1/4}\right)t
    +(w^2-1)-t U w\ .
\end{align}
This is the Seiberg–Witten curve for the $\SU(2)_0/\Z_2$ theory obtained from the 5-brane web of $\E_5$ SCFT.
Bringing the curve \eqref{eq:sw-e5-tuned-explicit} into the Weierstrass normal form using \eqref{eq:g2-g3-coef} and then identifying $U\to u$, we find agreement with \eqref{eq:g2-su2-0}, as expected.

\subsubsection*{\texorpdfstring{$\E_6$ to $\SU(2)+2\F/\Z_2$}{E6 to SU(2)+2F/Z2}}

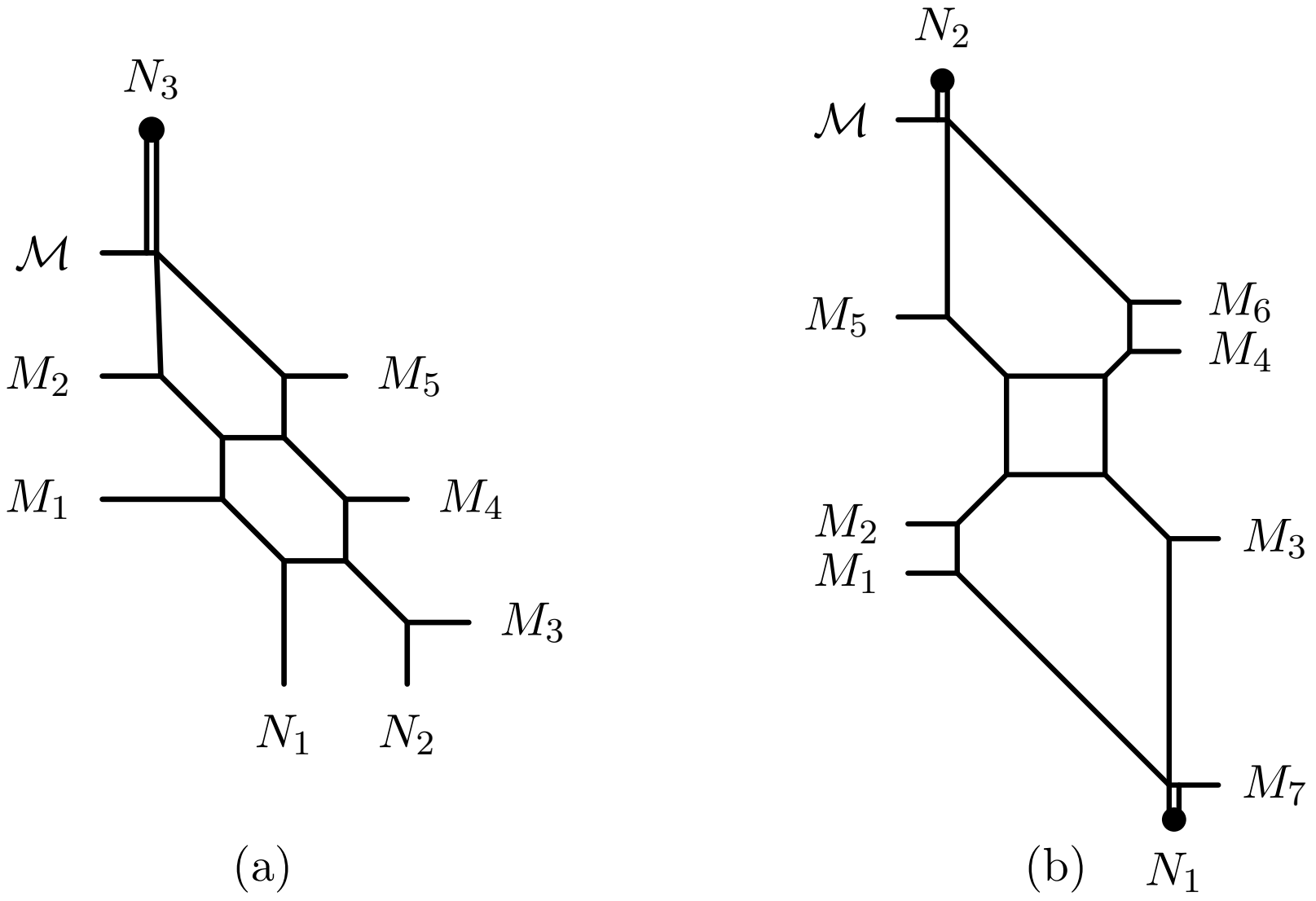
\begin{figure}
\centering
\begin{tikzpicture}[
    scale=0.5,
    line width=1.25pt,
    line cap=round,
    line join=round
]

\coordinate (A1) at (-0.08,5);
\coordinate (A2) at ( 0.08,5);

\coordinate (B) at (0.15,3);
\coordinate (C) at (2.15,3);

\coordinate (D) at (1.15,2);
\coordinate (E) at (2.15,2);

\coordinate (F) at (1.15,1);
\coordinate (G) at (3.15,1);

\coordinate (H) at (2.15,0);
\coordinate (I) at (3.15,0);

\coordinate (J) at (4.15,-1);

\draw (A1) -- (-0.08,7);
\draw (A2) -- ( 0.08,7);

\draw (-0.8,5) -- (A2);
\draw (-0.8,3) -- (B);
\draw (-0.8,1) -- (F);

\draw (A2) -- (B);
\draw (C)  -- (E);
\draw (D)  -- (F);
\draw (G)  -- (I);

\draw (D) -- (E);
\draw (H) -- (I);

\draw (A2) -- (C);
\draw (B)  -- (D);
\draw (E)  -- (G);
\draw (F)  -- (H);
\draw (I)  -- (J);

\draw (C) -- (3.15,3);
\draw (G) -- (4.15,1);
\draw (J) -- (5.15,-1);

\draw (H) -- (2.15,-2);
\draw (J) -- (4.15,-2);

\node[anchor=east] at (-1.0,5) {$\mathcal{M}$};
\node[anchor=east] at (-1.0,3) {$M_{2}$};
\node[anchor=east] at (-1.0,1) {$M_{1}$};

\node[anchor=west] at (3.35,3)  {$M_{5}$};
\node[anchor=west] at (4.35,1)  {$M_{4}$};
\node[anchor=west] at (5.35,-1) {$M_{3}$};

\node[anchor=south] at (0,7.2) {$N_{3}$};
\node[anchor=north] at (2.15,-2.2) {$N_{1}$};
\node[anchor=north] at (4.15,-2.2) {$N_{2}$};

\fill[black] (0,7) circle[radius=6pt];

\node at (1.8,-5) {(a)};

\end{tikzpicture}
\qquad\qquad
\begin{tikzpicture}[
    scale=0.4,
    line width=1.2pt,
    line cap=round,
    line join=round
]

\draw (-1,-1) -- (-1,1);
\draw ( 1,-1) -- ( 1,1);
\draw (-1, 1) -- ( 1,1);
\draw (-1,-1) -- ( 1,-1);

\draw (-1,1) -- (-2.2,2.2);
\draw (-2.2,2.2) -- (-2.2,7);
\draw (-2.4,6.2) -- (-2.4,7);

\draw (1,1) -- (1.5,1.5);
\draw (1.5,1.5) -- (1.5,2.5);
\draw (1.5,2.5) -- (-2.2,6.2);

\draw (-2.2,2.2) -- (-3.2,2.2);
\draw (-2.2,6.2) -- (-3.2,6.2);
\draw (1.5,1.5) -- (2.5,1.5);
\draw (1.5,2.5) -- (2.5,2.5);

\draw (-1,-1) -- (-2,-2);
\draw (-2,-2) -- (-2,-3);
\draw (-2,-3) -- (2.3,-7.3);

\draw (1,-1) -- (2.3,-2.3);
\draw (2.3,-2.3) -- (2.3,-8);
\draw (2.5,-7.3) -- (2.5,-8);

\draw (-2,-2) -- (-3,-2);
\draw (-2,-3) -- (-3,-3);
\draw (2.3,-2.3) -- (3.3,-2.3);
\draw (2.3,-7.3) -- (3.3,-7.3);

\node[anchor=east] at (-3.2,-3)   {$M_{1}$};
\node[anchor=east] at (-3.2,-2)   {$M_{2}$};
\node[anchor=east] at (-3.4, 2.2) {$M_{5}$};
\node[anchor=east] at (-3.4, 6.2) {$\mathcal{M}$};

\node[anchor=west] at (3.4,-7.3) {$M_{7}$};
\node[anchor=west] at (3.4,-2.3) {$M_{3}$};
\node[anchor=west] at (2.7, 1.5) {$M_{4}$};
\node[anchor=west] at (2.7, 2.5) {$M_{6}$};

\node[anchor=north] at (2.4,-8.3) {$N_{1}$};
\node[anchor=south] at (-2.3,7.3) {$N_{2}$};

\fill[black] (2.4,-8) circle[radius=7pt];
\fill[black] (-2.3,7) circle[radius=7pt];

\node at (0,-9) {(b)};

\end{tikzpicture}
\caption{(a) $\SU(2)+5\F$ gauge theory under the physical mass parameterization, where $M_i,i=1,\dotsc,5$, are the physical mass parameters and $\mathcal{M}$ depends on the instanton parameter $M_0$ of the $\E_6$ theory as $\mathcal{M}=M_0\sqrt{M_1M_2M_3M_4M_5}$; (b) $\SU(2)+6\F$ gauge theory under the physical mass parameterization, where $M_i,i=1,\dotsc,6$, are the physical mass parameters, $M_7$ and $\mathcal{M}$ are introduced as  auxiliary mass parameteres for convenince, with $\mathcal{M}$ depending on the instanton parameter $M_0$ of the $\E_7$ theory as $\mathcal{M}=M_0\sqrt{M_1M_2M_3M_4M_5M_6}$.}
\label{fig:E67case}
\end{figure}

For the $\E_6$ SCFT, the 5-brane web under the physical mass parameterization is given by figure~\ref{fig:E67case}(a), where $M_i$ ($i=1,\dots,5$) are the physical mass parameters and $M_0$ is the instanton parameter of the $\E_6$ theory, with $\mathcal{M}=M_0\sqrt{M_1M_2M_3M_4M_5}$. 
Under this parameterization, the physical mass condition is
\begin{align}\label{eq:phy-cond-e6}
    \frac{N_1}{N_3}=M_1M_2\ ,\qquad \frac{N_2}{N_3}=\frac{\mathcal{M}}{M_3M_4M_5}\ ,
\end{align}
and the $\E_6$ Seiberg–Witten curve reads
\begin{align}\label{eq:sw-e6-mass}
    F_{\E_6}(t,w)&=(w-M_1)(w-M_2)(w-\mathcal{M})+t^2\big(w-M_3\big)(w-M_4)(w-M_5)\notag\\
    &\quad+\big(M_1M_2M_3M_4M_5+\mathcal{M}\big)t
    +Utw+tw^2\Big(\sum_{i=1}^5M_i+\mathcal{M}\Big)-2tw^3=0\ .
\end{align}

We now impose the special tuning \eqref{eq:tuning-condi} and use the relation between instanton factors \eqref{eq:instTuning}:
\begin{align}\label{eq:tuning-e6}
    M_1=-M_2=M_3=-M_4=1\ ,\qquad M_0=\sqrt{q}\ ,
\end{align}
while $M_5$ remains untuned. Substituting \eqref{eq:tuning-e6} into \eqref{eq:sw-e6-mass}, we obtain
\begin{align}\label{eq:sw-e6-tuned-explicit}
    F_{\E_6}^{\text{tuned}}(t,w)
    &=(w^2-1)(w-q^{1/2}M_5^{1/2})+t^2(w^2-1)(w-M_5)\notag\\
    &\quad+(M_5+q^{1/2}M_5^{1/2})t+Utw+tw^2(M_5+q^{1/2}M_5^{1/2})-2tw^3\ .
\end{align}
This is the Seiberg–Witten curve for the $\SU(2)+2\F/\Z_2$ theory obtained from the 5-brane web of $\E_6$ SCFT. 
Bringing the curve \eqref{eq:sw-e6-tuned-explicit} into the Weierstrass normal form using \eqref{eq:g2-g3-coef}, we find agreement with \eqref{eq:su2+2Fz2main}, as expected, where we relabel $M_5 \to M_1$ and then rescale
\begin{align}
    U\to\sqrt{(M_1q^{1/3})}u,\quad g_2\to (M_1q^{1/3})^{-2}g_2,\quad g_3\to (M_1q^{1/3})^{-3}g_3\ .
\end{align}

\subsubsection*{\texorpdfstring{$\E_7$ to $\SU(2)+4\F/\Z_2$}{E7 to SU(2)+4F/Z2}}

For the $\E_7$ SCFT, the 5-brane web under the physical mass parameterization is given by figure~\ref{fig:E67case}(b), where $M_i$ ($i=1,\dots,6$) are the physical mass parameters and $M_0$ is the instanton parameter. For convenience, we also introduce the auxiliary masses $\mathcal{M}=M_0\sqrt{M_1M_2M_3M_4M_5M_6}$ and $M_7$ which are not but related to the physical mass parameters and, in particular, the instanton mass parameter. 
Under this parameterization, the physical mass condition is
\begin{align}\label{eq:phy-cond-e7}
    \frac{N_1}{N_2}=\frac{\mathcal{M}}{M_3M_4M_6}=\frac{M_1M_2M_5}{M_7}\ ,
\end{align}
and the $\E_7$ Seiberg–Witten curve reads
\begin{align}\label{eq:sw-e7-mass}
    F_{\E_7}(t,w)&=(w-M_1)(w-M_2)(w-M_5)(w-\mathcal{M})+t^2(w-M_3)(w-M_4)(w-M_6)(w-M_7)\notag\\
    &\quad-2tw^4+tw^3\Big(\sum_{i=1}^7M_i+\mathcal{M}\Big)+Utw^2+c_{11}tw-2M_7\mathcal{M}t\ ,
\end{align}
where
\begin{align}\label{eq:c11-e7}
    c_{11}&=\frac{\mathcal{M}}{M_3M_4M_6}(M_7 M_3 M_4+M_7 M_6 M_4+M_3 M_6 M_4+M_7 M_3 M_6)\notag\\
    &\quad+\frac{M_7}{M_1M_2M_5}(M_1 M_2 M_5+M_1 \mathcal{M} M_5+M_2 \mathcal{M} M_5+M_1 M_2\mathcal{M})\ .
\end{align}

We now impose the special tuning \eqref{eq:tuning-condi} and use the relation between instanton factors \eqref{eq:instTuning}:
\begin{align}\label{eq:tuning-e7}
    M_1=-M_2=M_3=-M_4=1\ ,\qquad M_0=\sqrt{q}\ ,
\end{align}
where $M_5, M_6$ are the remaining untuned physical mass parameters. With this tuning, $\mathcal{M}$ and $M_7$ are determined as
\begin{align}\label{eq:tuning-e7-2}
\mathcal{M} = (q M_5 M_6) ^{\frac12}, \qquad 
M_7 = q^{-\frac12} (M_5 M_6) ^{\frac12} .
\end{align}
Substituting \eqref{eq:tuning-e7}, \eqref{eq:tuning-e7-2} into \eqref{eq:sw-e7-mass}, we obtain
\begin{align}\label{eq:sw-e7-tuned-explicit}
    F_{\E_7}^{\text{tuned}}(t,w)
    &=(w^2-1)(w-M_5)(w-q^{1/2}(M_5M_6)^{1/2})\notag\\
    &\quad+t^2(w^2-1)(w-M_6)(w - q^{-\frac12} (M_5 M_6) ^{\frac12})\notag\\
    &\quad-2tw^4+tw^3\Big(M_5+M_6+q^{-\frac12} (M_5 M_6) ^{\frac12}+q^{1/2}(M_5M_6)^{1/2}\Big)\notag\\
    &\quad+Utw^2+c_{11}^{\text{tuned}}tw-2 (M_5M_6) t\ ,
\end{align}
where $c_{11}^{\text{tuned}}$ is given explicitly by
\begin{align}\label{eq:c11-tuned-e7}
    c_{11}^{\text{tuned}}
    =M_5+M_6+\left(q^{1/2}+q^{-1/2}\right)(M_5M_6)^{1/2}\ .
\end{align}
This is the Seiberg–Witten curve for the $\SU(2)+4\F/\Z_2$ theory obtained directly from the 5-brane web. 
Bringing the curve \eqref{eq:sw-e7-tuned-explicit} into the Weierstrass normal form using \eqref{eq:g2-g3-coef}, we find agreement with \eqref{eq:SU24F-Z2}, as expected, where we relabel $M_5 \to M_1, M_6 \to M_2$, and then rescale 
\begin{align}
U\to \sqrt{M_1M_2}u,\quad g_2\to(M_1M_2)^{-2}g_2,\quad g_3\to (M_1M_2)^{-3}g_3 \ .    
\end{align}

\subsubsection*{\texorpdfstring{$\E_8$ to $\SU(2)+6\F/\Z_2$}{E8 to SU(2)+6F/Z2}}

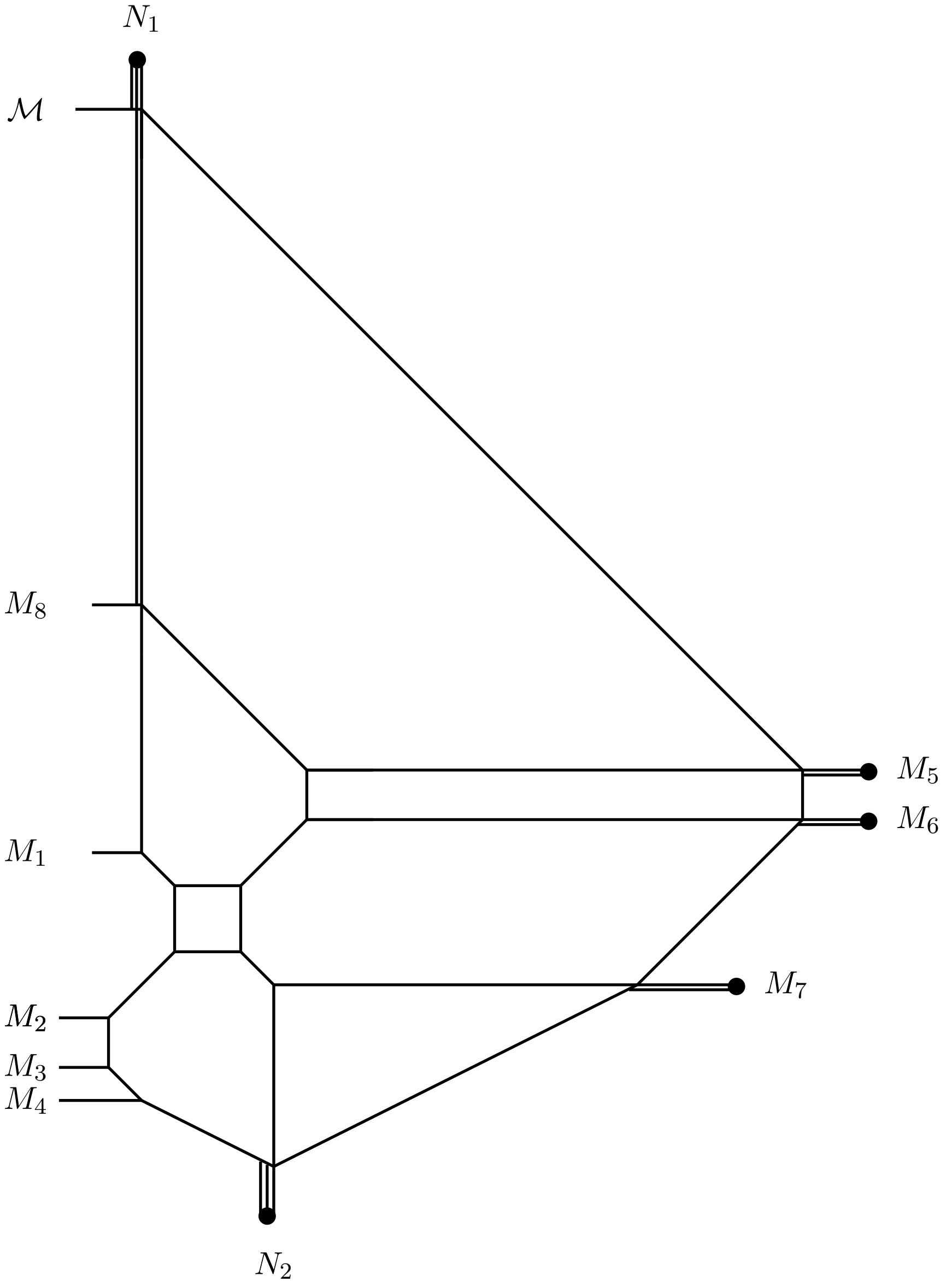
\begin{figure}
    \centering
    \begin{tikzpicture}[scale=0.2,line width=1pt]
        \draw(0,0)--(0,4);
        \draw(0,0)--(4,0);
        \draw(4,0)--(4,4);
        \draw(0,4)--(4,4);
        \draw(0,4)--(-2,6);
        \draw(-5,6)--(-2,6);
        \draw(-2,6)--(-2,18);
        \draw(4,4)--(8,8);
        \draw(12,8)--(8,8);
        \draw(8,8)--(8,11);
        \draw(8,11)--(12,11);
        \draw(8,11)--(-2,21);
        \draw(-5,21)--(-2,21);
        \node at(-9,21){$M_8$};
        \draw(-4,-4)--(0,0);
        \draw(-7,-4)--(-4,-4);
        \draw(-4,-4)--(-4,-7);
        \draw(-4,-7)--(-7,-7);
        \draw(-4,-7)--(-2,-9);
        \draw(-2,-9)--(-7,-9);
        \draw(38,11)--(-2,51);
        \draw(-2,51)--(-2,18);
        \draw(-2.3,21)--(-2.3,51);
        \draw(-2,-9)--(6,-13);
        \draw(4,0)--(6,-2);
        \draw(6,-2)--(34,-2);
        \draw(6,-2)--(6,-16);
        \draw(5.6,-12.9)--(5.6,-16);
        \draw(5.2,-12.7)--(5.2,-16);
        \draw(27.5,-2.3)--(34,-2.3);
        \draw(28,-2)--(38,8);
        \draw(38,8)--(38,11);
        \draw(-2,51)--(-6,51);
        \draw(-2,48)--(-2,54);
        \draw(-2.3,51)--(-2.3,54);
        \node at (-9,51){$\mathcal{M}$};
        \draw(-2.6,51)--(-2.6,54);
        \draw(38,10.7)--(42,10.7);
        \draw(8,11)--(42,11);
            \draw(37.7,7.7)--(42,7.7);
            \draw(8,8)--(42,8);
            \draw(6,-13)--(28,-2);
            \node at (-9,6){$M_1$};
            \node at(-9,-4){$M_2$};
            \node at(-9,-7){$M_3$};
            \node at(-9,-9){$M_4$};
            \node at(-9,-4){$M_2$};
            \node at(45,11){$M_5$};
            \node at(45,8){$M_6$};
            \node at(37,-2){$M_7$};
            \node at (6,-19){$N_2$};
            \node at (-2,56.5){$N_1$};
            \fill[black] (-2.26,54) circle[radius=15pt];
             \fill[black] (5.6,-16) circle[radius=15pt];
             \fill[black] (34,-2.1) circle[radius=15pt];
             \fill[black] (42,7.9) circle[radius=15pt];
             \fill[black] (42,10.9) circle[radius=15pt];
    \end{tikzpicture}
    \caption{5-brane web for SU(2) with $N_f=7$ flavors gauge theory and $M_i,i=1,\dotsc,7$, are the physical mass parameters, and $\mathcal{M}=M_0^2$, with $M_0$ being the instanton parameter of the $\E_8$ theory.}\label{fig:su27}
\end{figure}

For the $\E_8$ SCFT, the 5-brane web under the physical mass parameterization is given by figure~\ref{fig:su27}, where $M_i$ ($i=1,\dots,7$) are the physical mass parameters, and auxiliary masses $\mathcal{M}$ and $M_8$ depend on the instanton parameter $M_0$ as
\begin{align}\label{eq:e8-M}
    \mathcal{M}=M_0^2\ ,\qquad M_8=M_0\sqrt{M_1M_2M_3M_4M_5M_6M_7}\ ,
\end{align}
respectively. 
In this case, the physical mass condition reads
\begin{align}\label{eq:phy-cond-e8}
    \frac{N_2}{N_1}=\sqrt{\frac{M_1M_2M_3M_4}{M_5M_6M_7}}M_0\ .
\end{align}
From this condition, we find
\begin{align}\label{eq:sw-e8-mass}
    F_{\E_8}(t,w)=-t^{3}Q(w)^2+t^2Q(w)S(w)+tR(w)+P(w)=0\ ,
\end{align}
where
\begin{align}\label{eq:e8-coeffts}
    Q(w)&=\prod_{i=5}^7(w-M_i)\ ,\notag\\
    S(w)&=3w^3-\sigma_1w^2+\sigma_2w-3M_8\ ,\notag\\
    R(w)&=-3w^6+2\sigma_1 w^5-\sigma_3 w^4-U w^3-\sigma_4 w^2+2\sigma_5 w-3M_8^2\ ,\notag\\
    P(w)&=(w-\mathcal{M})(w-M_8)\prod_{i=1}^4(w-M_i)\ ,
\end{align}
with
\begin{align}\label{eq:e8-sigma}
\sigma_1&=\sum_{i=1}^8M_i+\mathcal{M}\ ,\notag\\
    \sigma_2&=1+M_8\Big(\frac{1}{\mathcal{M}}+\sum_{i=1}^7\frac1{M_i}\Big)\ ,\notag\\
    \sigma_3&=1+\sum_{1\leq i<j\leq 7}M_iM_j
+\mathcal{M}\sum_{i=1}^{7}M_i
+M_8\bigg[
\mathcal{M}+\frac{1}{\mathcal{M}}
+\sum_{i=1}^{7}\Big(M_i+\frac{1}{M_i}\Big)
\bigg]\ ,\notag\\
    \sigma_4&=\bigg(\prod_{i=1}^{7}M_i\bigg)
\bigg[
\mathcal{M}+\sum_{i=1}^{7}\frac{1}{M_i}
+\mathcal{M}\sum_{1\leq i<j\leq 7}\frac{1}{M_iM_j}
\bigg]
+M_8\bigg[
\mathcal{M}+\frac{1}{\mathcal{M}}
+\sum_{i=1}^{7}\Big(M_i+\frac{1}{M_i}\Big)
\bigg]\ ,\notag\\
    \sigma_5&=M_8+\bigg(\prod_{i=1}^{7}M_i\bigg)
\bigg(1+\mathcal{M}\sum_{i=1}^{7}\frac{1}{M_i}\bigg)\ .
\end{align}

We now impose the special tuning \eqref{eq:tuning-condi} and use the relation between instanton factors \eqref{eq:instTuning}:
\begin{align}\label{eq:tuning-e8}
    M_1=-M_2=M_3=-M_4=1\ ,\qquad M_0=\sqrt{q}\ ,
\end{align}
where $M_5, M_6, M_7$ are the remaining untuned physical mass parameters. 
With this tuning, $\mathcal{M}$ and $M_8$ are determined as
\begin{align}\label{eq:e8-tuned-M}
    \mathcal{M}=q\ ,\qquad M_8=q^{1/2}(M_5M_6M_7)^{1/2}\ .
\end{align}
Substituting \eqref{eq:tuning-e8} and \eqref{eq:e8-tuned-M} into \eqref{eq:sw-e8-mass}, we obtain
\begin{align}\label{eq:sw-e8-tuned-explicit}
    F_{\E_8}^{\text{tuned}}(t,w)
    &=-t^{3}Q(w)^2 +t^2Q(w) S_{\text{tuned}}(w)+tR_{\text{tuned}}(w)+P_{\text{tuned}}(w)\ ,
\end{align}
where
\begin{align}\label{eq:tuning-cof}
Q(w)&=\prod_{i=5}^7(w-M_i)\ ,\notag\\
S_{\text{tuned}}(w)
&=3w^3-\sigma_1^{\text{tuned}}w^2+\sigma_2^{\text{tuned}}w-3q^{1/2}(M_5M_6M_7)^{1/2}\ ,\notag\\
R_{\text{tuned}}(w)
&=-3w^6+2\sigma_1^{\text{tuned}}w^5-\sigma_3^{\text{tuned}}w^4-Uw^3-\sigma_4^{\text{tuned}}w^2+2\sigma_5^{\text{tuned}}w-3q(M_5M_6M_7)\ ,\notag\\
P_{\text{tuned}}(w)
&=(w-q)\bigl(w-q^{1/2}(M_5M_6M_7)^{1/2}\bigr)(w^2-1)^2\ ,
\end{align}
with
\begin{align}
\sigma_1^{\text{tuned}}&=M_5+M_6+M_7+q^{1/2}(M_5M_6M_7)^{1/2}+q\ ,\notag\\
\sigma_2^{\text{tuned}}&=1+q^{1/2}(M_5M_6M_7)^{1/2}
\left(\frac1q+\frac1{M_5}+\frac1{M_6}+\frac1{M_7}\right)\ ,\notag\\
\sigma_3^{\text{tuned}}&= - 1+(M_5M_6+M_5M_7+M_6M_7)
+q(M_5+M_6+M_7)\notag\\
&\quad+q^{1/2}(M_5M_6M_7)^{1/2}
\left[q+\frac1q+M_5+M_6+M_7+\frac1{M_5}+\frac1{M_6}+\frac1{M_7}\right]\ ,\notag\\
\sigma_4^{\text{tuned}}&=M_5M_6M_7\Bigg[\frac1{M_5}+\frac1{M_6}+\frac1{M_7}
+q\left(\frac1{M_5M_6}+\frac1{M_5M_7}+\frac1{M_6M_7}-1\right)\Bigg]\notag\\
&\quad+q^{1/2}(M_5M_6M_7)^{1/2}
\left[q+\frac1q+M_5+M_6+M_7+\frac1{M_5}+\frac1{M_6}+\frac1{M_7}\right]\ ,\notag\\
\sigma_5^{\text{tuned}}&=q^{1/2}(M_5M_6M_7)^{1/2}
+M_5M_6M_7\left[1+q\left(\frac1{M_5}+\frac1{M_6}+\frac1{M_7}\right)\right]\ .
\end{align}
This is the Seiberg–Witten curve for the $\SU(2)+6\F/\Z_2$ theory obtained directly from the 5-brane web.

Next, we would like to bring the curve \eqref{eq:sw-e8-tuned-explicit} into the Weierstrass normal form using eq.(A.8) in \cite{Huang:2013yta}. Multiplying both sides of \eqref{eq:sw-e8-tuned-explicit} by $-Q(w)$, redefining $z:=Q(w)t$, and performing a coordinate transformation $z=wx+S_{\text{tuned}}/3$, we obtain
\begin{align}\label{eq:cubic-curve-wx}
    w^3x^3+w\Bigg(-R_{\text{tuned}}-\frac{S_{\text{tuned}}^2}3\Bigg)x-\frac{2S_{\text{tuned}}^3}{27}-\frac{S_{\text{tuned}}R_{\text{tuned}}}{3}-Q(w)P_{\text{tuned}}=0\ .
\end{align}
Inserting \eqref{eq:tuning-cof} into \eqref{eq:cubic-curve-wx} and dividing both sides of the resulting curve by $w^3$
give rise to the following cubic curve
\begin{align}
\label{eq:cub-wx-tuned}
    x^3+(\alpha_2w^2+\alpha_1w+\alpha_0)x+\beta_3w^3+\beta_2w^2+\beta_1w+\beta_0=0\ ,
\end{align}
where the coefficients $\alpha_i$ and $\beta_j$ are uniquely determined by the above substitution, and their explicit expressions are omitted for brevity. Applying the general formula in Eq.~(A.8) of \cite{Huang:2013yta} to \eqref{eq:cub-wx-tuned}, relabeling $M_5\to M_1$, $M_6\to M_2$, and $M_7\to M_3$, performing the rescalings
\begin{align}
    U\to\sqrt{qM_1M_2M_3}\,u,\quad
    g_2\to(qM_1M_2M_3)^{-2}g_2,\quad
    g_3\to(qM_1M_2M_3)^{-3}g_3\ ,
\end{align}
and finally shifting $u\to u-60$, we find agreement with \eqref{eq:sw-su26f-f4}, as expected.

\section{\texorpdfstring{Characters of $\E_5$, $\E_6$, $\E_7$, and $\E_8$}{Characters of E5, E6, E7, and E8}}\label{app:Characters}

For ease of reference, we collect the characters of $\E_5,\E_6,\E_7$ and $\E_8$ that appear in the Seiberg–Witten curves presented in the main text. We use the Mathematica package
Susyno \cite{Fonseca:2011sy} to re-express these characters in terms of those of the following subgroups: 
\begin{align}
   \E_5&\supset\SO(8)\times \U(1)_I\ ,\notag\\
    \E_6&\supset \SO(10)\times \U(1)_I\ ,\notag\\
    \E_7&\supset\SO(12)\times \SU(2)\supset\SO(12)\times \U(1)_I\ ,\notag\\
    \E_8&\supset\SO(16) \supset\SO(14)\times \U(1)_I\ .
\end{align}
These embeddings give the expression for the characters of $\E_5,\E_6,\E_7$ and $\E_8$ in terms of the instanton parameter $M_0$ and the physical mass parameters $M_i$. 
We also derive the values of each characters under the special tuning of masses, which are given in \eqref{eq:e5-tuned-cha}, \eqref{eq:e6-after-char}, \eqref{eq:E7-tun-cha} and \eqref{eq:tuned-cha}, respectively.

\subsection*{$\E_5$ characters}

The characters of the 10-, $\overline{16}$-, 45-, 120-, and $16$-dimensional representations of the $\E_5=\SO(10)$ are given, respectively by, 
\begin{align}\label{eq:E5-to-SO8xU1}
   \chi_1^{\E_5}&=\chi_{\mathbf{v}}^{\SO(8)}+M_0+M_0^{-1}\, \notag\\ 
   \chi_2^{\E_5}&=\chi_{\mathbf{s}}^{\SO(8)}M_0^{-1/2}+\chi_{\mathbf{c}}^{\SO(8)}M_0^{1/2}\ ,\notag\\
   \chi_3^{\E_5}&=\chi_{\mathbf{adj}}^{\SO(8)}+1+\chi_{\mathbf{v}}^{\SO(8)}M_0+\chi_{\mathbf{v}}^{\SO(8)}M_0^{-1}\ ,\notag\\
   \chi_4^{\E_5}&=\chi_{\mathbf{s}}^{\SO(8)}\chi_{\mathbf{c}}^{\SO(8)}+\chi_{\mathbf{adj}}^{\SO(8)}M_0+\chi_{\mathbf{adj}}^{\SO(8)}M_0^{-1}\ ,\notag\\
   \chi_5^{\E_5}&=\chi_{\mathbf{s}}^{\SO(8)}M_0^{1/2}+\chi_{\mathbf{c}}^{\SO(8)}M_0^{-1/2}\ ,
\end{align}
where $M_0$ denotes the $\U(1)_I$ fugacity, with $+1$  charge, and the characters of $\SO(8)$ are given in terms of physical mass parameters $M_i,i=1,\dotsc,4$ as
\begin{align}
\chi_{\mathbf{v}}^{\SO(8)} 
&= \sum_{i=1}^4 (M_i + M_i^{-1}),
\cr
\chi_{\mathbf{s}}^{\SO(8)} 
&= \frac12 \left( \prod_{i=1}^4 (M_i^{\frac12}+M_i^{-\frac12}) 
+ \prod_{i=1}^4 (M_i^{\frac12} - M_i^{-\frac12}) \right),
\cr
\chi_{\mathbf{c}}^{\SO(8)} 
&= \frac12 \left( \prod_{i=1}^4 (M_i^{\frac12}+M_i^{-\frac12}) 
- \prod_{i=1}^4 (M_i^{\frac12} - M_i^{-\frac12}) \right),
\cr
\chi_{\mathbf{adj}}^{\SO(8)}
&= \sum_{i<j} ( M_i + M_i^{-1})( M_j + M_j^{-1}) + 4.
\end{align}
Imposing the tuning condition $M_1=-M_2=M_3=-M_4=1$ given in \eqref{eq:tuning-condi} on these four physical mass parameters $M_i$, we find the values of the characters of $\SO(8)$ reduces to
\begin{align}
\chi_{\mathbf{v}}^{\SO(8)} \to 0, 
\qquad
\chi_{\mathbf{s}}^{\SO(8)} \to 0,
\qquad
\chi_{\mathbf{c}}^{\SO(8)} \to 0,
\qquad
\chi_{\mathbf{adj}}^{\SO(8)}
& \to -4.
\end{align}
Substituting these into \eqref{eq:E5-to-SO8xU1}, we find \eqref{eq:e5-tuned-cha}.

It is also possible to write down $\E_5=\SO(10)$ characters more directly without decomposing into $\SO(8) \times \U(1)$. However, in this case, we need to take into account the non-trivial map between $\SO(8)$ fugacities identified as physical masses and the $\SO(10)$ fugacities, which map involves the triality of $\SO(8)$:
\begin{align}\label{eq:map-e5}
    z_1&= \sqrt{M_1}\sqrt{M_2} \sqrt{M_3}\sqrt{ M_4},\quad z_2=\frac{\sqrt{M_1}\sqrt{M_2}}{\sqrt{M_3}\sqrt{M_4}},\quad z_3=\frac{\sqrt{M_1}\sqrt{M_3}}{\sqrt{M_2}\sqrt{M_4}},\quad\notag\\
    z_4&=\frac{\sqrt{M_1}\sqrt{M_4}}{\sqrt{M_2}\sqrt{M_3}},\quad z_5=M_0.
\end{align}
With these $\SO(10)$ fugacities, the $\SO(10)$ characters in \eqref{eq:E5-to-SO8xU1} can also be expressed as
\begin{align}\label{eq:sw-e5-cha}
    \chi_1^{\E_5}
    &=\sum_{i=1}^5\left(z_i+\frac1{z_i}\right)\ ,\notag\\
    \chi_2^{\E_5}&=\frac12\left[
    \prod_{i=1}^5\left(\sqrt{z_i}+\frac1{\sqrt{z_i}}\right)
    -\prod_{i=1}^5\left(\sqrt{z_i}-\frac1{\sqrt{z_i}}\right)
    \right] \ ,\notag\\
    \chi_3^{\E_5}
    &=5+\sum_{1\le i<j\le5}
    \left(z_i+\frac1{z_i}\right)
    \left(z_j+\frac1{z_j}\right)\ ,\notag\\
    \chi_4^{\E_5}
    &=\sum_{1\le i<j<k\le5}
    \left(z_i+\frac1{z_i}\right)
    \left(z_j+\frac1{z_j}\right)
    \left(z_k+\frac1{z_k}\right)
    +4\sum_{i=1}^5\left(z_i+\frac1{z_i}\right)\ ,\notag\\
    \chi_5^{\E_5}
    &=\frac12\left[
    \prod_{i=1}^5\left(\sqrt{z_i}+\frac1{\sqrt{z_i}}\right)
    +\prod_{i=1}^5\left(\sqrt{z_i}-\frac1{\sqrt{z_i}}\right)
    \right]\ .
\end{align}
Imposing the tuning condition \eqref{eq:tuning-condi} on these four physical mass parameters $M_i$, due to \eqref{eq:map-e5} we obtain
\begin{align}\label{eq:z1-tuning}
    z_1=-1,\quad z_2=1,\quad z_3=-1,\quad z_4=1,\quad z_5=M_0
\end{align}
where particularly we have chosen
\begin{align}
    \sqrt{M_2}=\sqrt{M_4}=i\ .
\end{align}
By substituting \eqref{eq:z1-tuning} into \eqref{eq:sw-e5-cha}, 
we find the values of the $\E_5$ characters \eqref{eq:e5-tuned-cha} under the special tuning.

\subsection*{$\E_6$ characters}
Under the embedding $\E_6\supset \SO(10)\times\U(1)_I$,
the $\E_6$ characters decompose as follows:
\begin{align}\label{eq:E6cha}
    \chi_1^{\E_6}&=M_0^{4/3}+\chi_1M_0^{-2/3}+\chi_5M_0^{1/3}\ ,\notag\\
\chi_2^{\E_6}&=\chi_2M_0^{-1}+\chi_5M_0+\chi_3+1\ ,\notag\\
\chi_3^{\E_6}&=\chi_3M_0^{-4/3}+\chi_1\chi_5M_0^{-1/3}
+\chi_1M_0^{2/3}+\chi_4M_0^{2/3}+\chi_5M_0^{5/3}\ ,\notag\\
\chi_4^{\E_6}&=\chi_4M_0^{-2}+\chi_4M_0^2
+\chi_5\chi_3M_0^{-1}+\chi_3\chi_2M_0
+\chi_3+\chi_1\chi_4\ ,\notag\\
\chi_5^{\E_6}&=\chi_2M_0^{-5/3}+\chi_3M_0^{4/3}
+\chi_1M_0^{-2/3}+\chi_4M_0^{-2/3}
+\chi_1\chi_2M_0^{1/3}\ ,\notag\\
\chi_6^{\E_6}&=M_0^{-4/3}+\chi_1M_0^{2/3}
+\chi_2M_0^{-1/3}\ .
\end{align}
Here, $M_0$ denotes the U(1)$_I$ fugacity. The characters $\chi_i^{\E_6},i=1,2,\dotsc,6$, correspond respectively to  the  $27$-, $78$-, $\overline{351}$-,  $2925$-, $351$-, and $\overline{27}$-dimensional representations of $\E_6$, while $\chi_i,i=1,2,\dotsc,5$, denote the characters of the 10-, $\overline{16}$-, 45-, 120-, $16$-dimensional representations of $\SO(10)$, respectively. Each $\SO(10)$ characters can be written down explicitly in terms of physical mass parameters $M_i$ as
\begin{align}\label{eq:SO10-characters}
\chi_1
    &=\sum_{i=1}^5\left(M_i+\frac1{M_i}\right)\ ,\notag\\
\chi_2
    &=\frac12\left[
    \prod_{i=1}^5\left(\sqrt{M_i}+\frac1{\sqrt{M_i}}\right)
    -\prod_{i=1}^5\left(\sqrt{M_i}-\frac1{\sqrt{M_i}}\right)
    \right] \ ,\notag\\
\chi_3
    &=5+\sum_{1\le i<j\le5}
    \left(M_i+\frac1{M_i}\right)
    \left(M_j+\frac1{M_j}\right)\ ,\notag\\
\chi_4
    &=\sum_{1\le i<j<k\le5}
    \left(M_i+\frac1{M_i}\right)
    \left(M_j+\frac1{M_j}\right)
    \left(M_k+\frac1{M_k}\right)
    +4\sum_{i=1}^5\left(M_i+\frac1{M_i}\right)\ ,\notag\\
\chi_5
    &=\frac12\left[
    \prod_{i=1}^5\left(\sqrt{M_i}+\frac1{\sqrt{M_i}}\right)
    +\prod_{i=1}^5\left(\sqrt{M_i}-\frac1{\sqrt{M_i}}\right)
    \right]\ .
\end{align}
These give how we express $\E_6$ characters in terms of the instanton parameter $M_0$ and the physical mass parameters $M_i$. 

Imposing the tuning condition $M_1=-M_2=M_3=-M_4=1$ given in \eqref{eq:tuning-condi} on these four physical mass parameters and by relabeling $M_5 \to M_1$, the values of these characters reduces to
\begin{align}
\chi_1 \to M_1 + M_1^{-1},
\quad
\chi_2 \to 0 \ ,
\quad
\chi_3 \to -3 \ , 
\quad
\chi_4 \to -4(M_1 + M_1^{-1})\ ,
\quad
\chi_5\to 0 \ .
\end{align}
By substituting these into \eqref{eq:E6cha}, we find the values of the $\E_6$ characters \eqref{eq:e6-after-char} under the special tuning.

\subsection*{$\E_7$ characters}
For $\E_7$, we denote by $\chi_i^{\E_7}$, $i=1,2,\dotsc,7$, the characters  of the representations of dimensions 133, 912, 8645, 365750, 27664, 1539,  and 56, respectively.
Under the embedding $\E_7\supset\SO(12)\times \SU(2)$ , their decomposition into $\SO(12)\times\SU(2)$  characters are given as follow:
\begin{align}\label{eq:E7cha}
    \chi_1^{\E_7}&=\chi_0^2+\chi _5 \chi_0+\chi _2-1\ ,\notag\\
    \chi_2^{\E_7}&=\chi _6 \chi_0^2+\chi _1 \chi_0+\chi _3 \chi_0+\chi _1 \chi _5-2 \chi _6\ ,
    \notag\\
   \chi_3^{\E_7}&=\chi _5 \chi_0^3+\chi _2 \chi_0^2+\chi _4 \chi_0^2+\chi_0^2+\chi _2 \chi _5 \chi_0-2 \chi _5 \chi_0+\chi _5^2-2 \chi _2+\chi _1 \chi _3-3 \chi _4-2\ ,\notag\\
   \chi_4^{\E_7}&=\chi _4 \chi_0^4+\chi _3 \chi _6 \chi_0^3+\chi _1 \chi _3 \chi_0^2+\chi _2 \chi _4
   \chi_0^2-4 \chi _4 \chi_0^2+\chi _1 \chi _3 \chi _5 \chi_0-3 \chi _3 \chi _6 \chi_0+\chi _1^2-\chi _2^2 \notag\\
   &+\chi _3^2+\chi _2 \chi _5^2+\chi _6^2-2 \chi _2+\chi _1^2 \chi_4-4 \chi _2 \chi _4+\chi _4-\chi _1 \chi _5 \chi _6-1 \ ,\notag\\
   \chi_5^{\E_7}&=\chi _3 \chi_0^3+\chi _2 \chi _6 \chi_0^2+\chi _1 \chi _2 \chi_0-3 \chi _3 \chi_0+\chi _1 \chi _4 \chi_0+\chi _3 \chi _5+\chi _1^2 \chi _6-3 \chi _2 \chi _6 \ ,\notag\\
   \chi_6^{\E_7}&=\chi _1^2+\chi _6 \chi_0 \chi _1+\chi _2 \chi_0^2-2 \chi _2+\chi _4 \ ,\notag\\
   \chi_7^{\E_7}&=\chi _6+\chi _1 \chi_0\ ,
\end{align}
where $\chi_0$ is the character of the fundamental representation of $\SU(2)$ group defined by
\begin{align}
    \chi_0=M_0+\frac1{M_0}
\end{align}
and $\chi_i,i=1,2,\dotsc,6$, are respectively the characters of $12$-, $66$-, $220$-,  $495$-, $\overline{32}$-, and $32$-dimensional representations of the $\SO(12)$ group. Each $\SO(12)$ characters can be written down explicitly in terms of physical mass parameters $M_i$ as
\begin{align}
\chi_1
    &=\sum_{i=1}^6 \left(M_i+\frac1{M_i}\right)\ ,
\cr
\chi_2
    &=\sum_{1\le i<j\le6}
    \left(M_i+\frac1{M_i}\right)
    \left(M_j+\frac1{M_j}\right)
    + 6 \ ,
\cr
\chi_3
    &=\sum_{1\le i<j<k\le6}
    \left(M_i+\frac1{M_i}\right)
    \left(M_j+\frac1{M_j}\right)
    \left(M_k+\frac1{M_k}\right)
    +5\sum_{i=1}^6\left(M_i+\frac1{M_i}\right)\ ,
\cr
\chi_4
    &=\sum_{1\le i<j<k<\ell\le6}
    \left(M_i+\frac1{M_i}\right)
    \left(M_j+\frac1{M_j}\right)
    \left(M_k+\frac1{M_k}\right)
    \left(M_{\ell}+\frac1{M_{\ell}}\right)
\cr
& \qquad \qquad 
    +4\sum_{1\le i<j\le6}
    \left(M_i+\frac1{M_i}\right)
    \left(M_j+\frac1{M_j}\right)
    + 15 \ ,
\cr
\chi_5
    &=\frac12\left[
    \prod_{i=1}^6\left(\sqrt{M_i}+\frac1{\sqrt{M_i}}\right)
    +\prod_{i=1}^6\left(\sqrt{M_i}-\frac1{\sqrt{M_i}}\right)
    \right]\ ,
\cr
\chi_6
    &=\frac12\left[
    \prod_{i=1}^6\left(\sqrt{M_i}+\frac1{\sqrt{M_i}}\right)
    -\prod_{i=1}^6\left(\sqrt{M_i}-\frac1{\sqrt{M_i}}\right)
    \right] \ ,
\end{align}

Imposing the tuning condition $M_1=-M_2=M_3=-M_4=1$ given in \eqref{eq:tuning-condi} on these four physical mass parameters and by relabeling $M_5, M_6 \to M_1, M_2$, the values of these characters reduces to
\begin{align}
\chi_1 &\to M_1 + \frac1{M_1} + M_2 + \frac1{M_2},
\cr
\chi_2 &\to \left( M_1 + \frac1{M_1} \right) \left(M_2 + \frac1{M_2} \right)-2  \ ,
\cr
\chi_3 &\to -3 \left( M_1 + \frac1{M_1} + M_2 + \frac1{M_2} \right) \ , 
\cr
\chi_4 &\to -4 \left( M_1 + \frac1{M_1}\right) \left(M_2 + \frac1{M_2} \right)-1\ ,
\cr
\chi_5 &\to 0 \ ,
\cr
\chi_6 &\to 0 \ .
\end{align}
By substituting these into \eqref{eq:E7cha}, we find the values of the $\E_7$ characters \eqref{eq:E7-tun-cha} under the special tuning. 

\subsection*{$\E_8$ characters}
For $\E_8$, the dimensions of the representations associated with  the characters of  $\chi^{\E_8}_i$, $i=1,2,\dotsc,8$ are 3875, 147250, 6696000, 6899079264, 146325270, 2450240, 30380, 248, respectively.
We here use the embedding $ E_8\supset \SO(16)\supset\SO(14)\times\U(1)_I$.
Mathematically,  the eight characters of $\E_8$ decompose as follows
\begin{align}\label{eq:chie8}
    \chi_1^{\E_8}&=\chi_1^2+\chi_8 \chi_1-\chi_2+\chi_4-\chi_7-1 \ ,\nonumber\\
    \chi_2^{\E_8}&=\chi_7 \chi_1^2+\chi_1^2+\chi_3 \chi_1+\chi_5 \chi_1+\chi_8^2-3 \chi_2-3 \chi
   _4-2 \chi_6-2 \chi_2 \chi_7-\chi_7+\chi_3 \chi_8-2  \ ,\nonumber\\ 
    \chi_3^{\E_8}&=\chi_8 \chi_1^3+\chi_2 \chi_1^2+\chi_4 \chi_1^2+\chi_6 \chi_1^2-\chi_7 \chi
   _1^2-\chi_3 \chi_1-2 \chi_2 \chi_8 \chi_1+\chi_4 \chi_8 \chi_1-\chi_7 \chi
   _8 \chi_1-2 \chi_2^2\notag\\
   &\quad+\chi_7^2+\chi_2 \chi_8^2-\chi_2-2 \chi_2 \chi_4+\chi_3
   \chi_5-3 \chi_2 \chi_6-\chi_6+\chi_2 \chi_7-2 \chi_4 \chi_7+\chi_3 \chi_8\ ,\nonumber\\
    \chi_4^{\E_8}&=\chi_6 \chi_1^4+\chi_1^4+\chi_3 \chi_1^3+\chi_5 \chi_1^3+\chi_5 \chi_7 \chi
   _1^3-\chi_7 \chi_8 \chi_1^3+\chi_7^2 \chi_1^2+\chi_8^2 \chi_1^2-5 \chi_2 \chi
   _1^2-2 \chi_4 \chi_1^2+\chi_3 \chi_5 \chi_1^2\nonumber\\
   &\quad-4 \chi_2 \chi_6 \chi_1^2+\chi
   _4 \chi_6 \chi_1^2-2 \chi_6 \chi_1^2-\chi_4 \chi_7 \chi_1^2-\chi_6 \chi_7
   \chi_1^2-2 \chi_7 \chi_1^2-\chi_1^2-\chi_3 \chi_7^2 \chi_1-\chi_2 \chi_3
   \chi_1\nonumber\\
   &\quad+2 \chi_3 \chi_1-3 \chi_2 \chi_5 \chi_1-2 \chi_4 \chi_5 \chi_1+\chi_5
   \chi_1+3 \chi_3 \chi_6 \chi_1-\chi_5 \chi_6 \chi_1-\chi_3 \chi_7 \chi_1-3
   \chi_2 \chi_5 \chi_7 \chi_1\nonumber\\
   &\quad-2 \chi_5 \chi_7 \chi_1+2 \chi_7^2 \chi_8 \chi
   _1+\chi_2 \chi_8 \chi_1+\chi_4 \chi_8 \chi_1+\chi_3 \chi_5 \chi_8 \chi
   _1-\chi_6 \chi_8 \chi_1+3 \chi_2 \chi_7 \chi_8 \chi_1-\chi_7 \chi_8 \chi
   _1\nonumber\\
   &\quad-\chi_8 \chi_1-\chi_2^3-2 \chi_7^3+\chi_3 \chi_8^3+4 \chi_2^2+\chi_2 \chi
   _3^2+\chi_3^2-\chi_2 \chi_4^2+\chi_2 \chi_5^2+2 \chi_5^2+\chi_6^2+\chi_2^2
   \chi_7^2-2 \chi_2 \chi_7^2\nonumber\\
   &\quad+2 \chi_4 \chi_7^2-\chi_7^2-2 \chi_2 \chi_8^2+\chi
   _2 \chi_4 \chi_8^2-\chi_2 \chi_7 \chi_8^2-\chi_7 \chi_8^2+5 \chi_2-2 \chi
   _2^2 \chi_4+2 \chi_2 \chi_4+\chi_3 \chi_5+\chi_2^2 \chi_6\nonumber\\
   &\quad+\chi_3^2 \chi_6+6
   \chi_2 \chi_6-4 \chi_2 \chi_4 \chi_6-2 \chi_4 \chi_6+\chi_6+\chi_2^2 \chi
   _7+\chi_3^2 \chi_7-\chi_4^2 \chi_7+3 \chi_2 \chi_7+2 \chi_2 \chi_4 \chi_7\nonumber\\
   &\quad+4
   \chi_4 \chi_7+2 \chi_3 \chi_5 \chi_7+3 \chi_2 \chi_6 \chi_7+4 \chi_6 \chi
   _7+4 \chi_7-2 \chi_2 \chi_3 \chi_8-3 \chi_3 \chi_8-2 \chi_3 \chi_4 \chi
   _8+\chi_2 \chi_5 \chi_8\nonumber\\
   &\quad-\chi_5 \chi_8-3 \chi_3 \chi_6 \chi_8-\chi_2 \chi_3
   \chi_7 \chi_8-\chi_3 \chi_7 \chi_8-\chi_5 \chi_7 \chi_8\ ,\nonumber\\
    \chi_5^{\E_8}&=\chi_5 \chi_1^3-\chi_6 \chi_1^2+\chi_4 \chi_7 \chi_1^2-\chi_1^2+\chi_3 \chi_4
   \chi_1-3 \chi_2 \chi_5 \chi_1-\chi_5 \chi_1+\chi_3 \chi_6 \chi_1-\chi_3
   \chi_7 \chi_1-\chi_5 \chi_7 \chi_1 \nonumber\\
   &\quad+\chi_7 \chi_8 \chi_1+\chi_3^2-3 \chi
   _4^2+\chi_5^2-\chi_7^2+\chi_4 \chi_8^2-\chi_8^2+2 \chi_2-2 \chi_2 \chi_4+\chi
   _4+3 \chi_3 \chi_5+\chi_2 \chi_6\nonumber\\
   &\quad-3 \chi_4 \chi_6+3 \chi_6+\chi_3^2 \chi
   _7+\chi_2 \chi_7-3 \chi_2 \chi_4 \chi_7+\chi_4 \chi_7-\chi_7+\chi_2 \chi_5
   \chi_8-\chi_3 \chi_7 \chi_8+2 \ ,\nonumber\\
    \chi_6^{\E_8}&=\chi_4 \chi_1^2-\chi_5 \chi_1+\chi_3 \chi_7 \chi_1+\chi_3^2-\chi_7^2+\chi_2-2
   \chi_2 \chi_4+\chi_2 \chi_6+\chi_6-\chi_2 \chi_7-2 \chi_4 \chi_7+\chi_5
   \chi_8 \ ,\nonumber\\
    \chi_7^{\E_8}&=\chi_1 \chi_3-\chi_4+\chi_6+\chi_2 \chi_7-\chi_7 \ ,\nonumber\\ 
    \chi_8^{\E_8}&=\chi_2+\chi_7\ ,
\end{align}
where $\chi_i$, $i=1,2,\dotsc,8$, denote the characters of SO(16), whose dimensions are respectively given by
\begin{align}
    \chi=\{16,120,560,1820,4368,8008,\overline{128},128\}\ .
\end{align}
Under our physical mass parameterization, the fugacities of the $\E_8$ group are identified with
\begin{align}
    \{M_0^2,M_1,M_2,M_3,M_4,M_5,M_6,M_7\}
\end{align}
where $M_0$ is the instanton parameter of the $\E_8$ SCFT. These fugacities of the $\SO(16)$ can be written in terms of the instanton parameter and physical mass parameters according to the embedding $\SO(16)\supset\SO(14)\times \U(1)_I$ as
\begin{align}\label{eq:SO16-SO14xU1}
\chi_1 & = M_0^2 + M_0^{-2} + \chi^{\SO(14)}_{\mathbf{14}}\ ,
\cr
\chi_2 & = 1 + (M_0^2 + M_0^{-2}) \chi^{\SO(14)}_{\mathbf{14}} + \chi^{\SO(14)}_{\mathbf{91}}, \,
\cr
\chi_3 & = \chi^{\SO(14)}_{\mathbf{14}} + (M_0^2 + M_0^{-2}) \chi_{\mathbf{91}}^{\SO(14)} +\chi_{\mathbf{364}}^{\SO(14)},\,
\cr
\chi_4 &=\chi^{\SO(14)}_{\boldsymbol{91}}+(M_0^2+M_0^{-2}) 
\chi^{\SO(14)}_{\boldsymbol{364}} +\chi^{\SO(14)}_{\boldsymbol{1001}},
\cr
\chi_5&=\chi^{\SO(14)}_{\boldsymbol{364}}+(M_0^2+M_0^{-2}) \chi^{\SO(14)}_{\boldsymbol{1001}} +\ \chi^{\SO(14)}_{\boldsymbol{2002}}\ ,
\cr
\chi_6&=\chi^{\SO(14)}_{\boldsymbol{1001}}
+(M_0^2+M_0^{-2}) \chi^{\SO(14)}_{\boldsymbol{2002}}
+\chi^{\SO(14)}_{\boldsymbol{3003}}\ ,
\cr
\chi_7&= M_0 \chi^{\SO(14)}_{\boldsymbol{64}} + M_0^{-1}\chi^{\SO(14)}_{\mathbf{\overline{64}}}
\cr
\chi_8&=M_0\chi^{\SO(14)}_{\mathbf{\overline{64}}}+M_0^{-1} \chi^{\SO(14)}_{\boldsymbol{64}} \ .
\end{align}
Here, the characters of $\SO(14)$ are given explicitly in terms of the physical mass parameters $M_i$ as
\begin{align}
\chi^{\SO(14)}_{\boldsymbol{14}} 
&= \sum_{i=1}^{7} \left( M_i + \frac1{M_i} \right)
\cr
\chi^{\SO(14)}_{\boldsymbol{91}} 
&= \sum_{1\le i_1<i_2\le7} 
\left( \prod_{k=1}^2 \left(M_{i_k}+\frac1{M_{i_k}}\right) \right)
    + 7 \ ,
\cr
\chi^{\SO(14)}_{\boldsymbol{364}} 
&= \sum_{1\le i_1 < i_2 < i_3 \le 7}
\left( \prod_{k=1}^3 \left(M_{i_k}+\frac1{M_{i_k}}\right) \right)
    +6\sum_{i=1}^7\left(M_i+\frac1{M_i}\right)\ ,
\cr
\chi^{\SO(14)}_{\boldsymbol{1001}} 
&= \sum_{1\le i_1 < i_2 < i_3 < i_4 \le 7}
\left( \prod_{k=1}^4 \left(M_{i_k}+\frac1{M_{i_k}}\right) \right)
    + 5\sum_{1\le i<j\le7}
    \left( \prod_{k=1}^2 \left(M_{i_k}+\frac1{M_{i_k}}\right) \right)
    + 21 \ ,
\cr
\chi^{\SO(14)}_{\boldsymbol{2002}} 
&= \sum_{1\le i_1 < i_2 < i_3 < i_4 < i_5 \le7}
\left( \prod_{k=1}^5 \left(M_{i_k}+\frac1{M_{i_k}}\right) \right)
\cr
& \qquad 
    + 4\sum_{1\le i_1 < i_2 < i_3 \le 7}
\left( \prod_{k=1}^3 \left(M_{i_k}+\frac1{M_{i_k}}\right) \right)
    + 15 \sum_{i=1}^{7} \left( M_i + \frac1{M_i} \right) \ ,
\cr
\chi^{\SO(14)}_{\boldsymbol{3003}} 
&= \sum_{1\le i_1 < i_2 < i_3 < i_4 < i_5 < i_6 \le7}
\left( \prod_{k=1}^6 \left(M_{i_k}+\frac1{M_{i_k}}\right) \right)
\cr
& \qquad 
    + 3\sum_{1\le i_1 < i_2 < i_3< i_4 \le 7}
\left( \prod_{k=1}^4 \left(M_{i_k}+\frac1{M_{i_k}}\right) \right)
\cr
& \qquad 
    + 10 \sum_{1\le i<j\le7}
    \left( \prod_{k=1}^2 \left(M_{i_k}+\frac1{M_{i_k}}\right) \right)
    + 35\ ,
\cr
\chi^{\SO(14)}_{\boldsymbol{64}} 
&= \frac12 \left( \prod_{i=1}^7 \left( \sqrt{M}_i + \frac1{\sqrt{M}_i}\right) + \prod_{i=1}^7 \left( \sqrt{M}_i - \frac1{\sqrt{M}_i}\right) \right) , \
\cr
\chi^{\SO(14)}_{\boldsymbol{\overline{64}}} 
&= \frac12 \left( \prod_{i=1}^7 \left( \sqrt{M}_i + \frac1{\sqrt{M}_i}\right) -\prod_{i=1}^7 \left( \sqrt{M}_i - \frac1{\sqrt{M}_i}\right) \right) \ .
\end{align}

Imposing the tuning condition $M_1=-M_2=M_3=-M_4=1$ given in \eqref{eq:tuning-condi} on these four physical mass parameters and by relabeling $M_5, M_6,M_7 \to M_1, M_2, M_3$, the values of these characters reduces to
\begin{align}\label{eq:SO14-ch-red}
\chi^{\SO(14)}_{\boldsymbol{14}} 
&\to \sum_{i=1}^{3} \left( M_i + \frac1{M_i} \right)
\cr
\chi^{\SO(14)}_{\boldsymbol{91}} 
&\to \sum_{1\le i_1<i_2\le3} 
\left( \prod_{k=1}^2 \left(M_{i_k}+\frac1{M_{i_k}}\right) \right)
    - 1 \ ,
\cr
\chi^{\SO(14)}_{\boldsymbol{364}} 
&\to \prod_{i=1}^3 \left(M_{i_i}+\frac1{M_{i_i}}\right) 
    -2\sum_{i=1}^3\left(M_i+\frac1{M_i}\right)\ ,
\cr
\chi^{\SO(14)}_{\boldsymbol{1001}} 
&\to  -3\sum_{1\le i<j\le7}
    \left( \prod_{k=1}^2 \left(M_{i_k}+\frac1{M_{i_k}}\right) \right)
    -3\ ,
\cr
\chi^{\SO(14)}_{\boldsymbol{2002}} 
&\to -2\prod_{i=1}^3 \left(M_{i_i}+\frac1{M_{i_i}}\right)
    - \sum_{i=1}^{3} \left( M_i + \frac1{M_i} \right) \ ,
\cr
\chi^{\SO(14)}_{\boldsymbol{3003}} 
&\to 2\sum_{1\le i<j\le7}
    \left( \prod_{k=1}^2 \left(M_{i_k}+\frac1{M_{i_k}}\right) \right)
    +3 \ ,
\cr
\chi^{\SO(14)}_{\boldsymbol{64}} 
&\to 0\ ,
\cr
\chi^{\SO(14)}_{\boldsymbol{\overline{64}}} 
&\to 0\ .
\end{align} 
Combining \eqref{eq:SO16-SO14xU1}, \eqref{eq:SO14-ch-red}, and \eqref{eq:chie8}, we can write down the characters of $\E_8$ under the special tuning. These agree with \eqref{eq:tuned-cha}.

\bigskip
\bibliographystyle{JHEP}
\bibliography{ref}
\end{document}